\documentclass[12pt]{iopart}
\usepackage{amsmath}
\usepackage{amsthm}
\usepackage{amssymb}
\usepackage{mathtools}
\usepackage{amsfonts}
\usepackage{tikz}
\usepackage{bbold}
\usepackage{graphicx}
\usepackage{bm}
\usepackage{xcolor}
\usepackage{float}
\usepackage{hyperref}
\usepackage{etoolbox}
\usepackage{cite}
\usepackage[normalem]{ulem}
\usepackage{everypage}
\AddEverypageHook{\setcounter{footnote}{0}}

\definecolor{caribbeangreen}{rgb}{0.0, 0.8, 0.6}
\newcommand{\curt}[1]{{\sf \color{caribbeangreen} (CvK: #1)}}

\newcommand{\eqnref}[1]{Eq.~\ref{#1}}
\newcommand{\secref}[1]{Sec.~\ref{#1}}
\newcommand{\figref}[1]{Fig.~\ref{#1}}

\newcommand{\appref}[1]{App.~\ref{#1}}
\newcommand{\st}[1]{\sout{#1}}

\newtheorem{thm}{Theorem}[section]
\newtheorem*{remark}{Remark}
\theoremstyle{definition}
\newtheorem{definition}{Definition}
\DeclareMathOperator*{\Motimes}{\text{\raisebox{0.25ex}{\scalebox{0.6}{$\bigotimes$}}}}

\DeclarePairedDelimiter\bra{\langle}{\rvert}
\DeclarePairedDelimiter\ket{\lvert}{\rangle}
\DeclarePairedDelimiterX\braket[2]{\langle}{\rangle}{#1 \delimsize\vert #2}

\newcommand{\comm}[2]{\left[#1, #2\right]}
\newcommand{\order}[1]{\mathcal{O}(#1)}
\newcommand{\abs}[1]{\lvert #1 \rvert}

\makeatletter
\newrobustcmd{\fixappendix}{%
  \patchcmd{\l@section}{1.5em}{7em}{}{}%
  \patchcmd{\l@subsection}{2.3em}{7em}{}{}%
}
\makeatother

\begin{document}

\title[Operator Spreading in the Memory Matrix Formalism]{Operator Spreading in the Memory Matrix Formalism}
\author{Ewan McCulloch\footnote[1]{ewan.r.mcculloch@gmail.com}, C.W. von Keyserlingk\footnote[2]{c.vonkeyserlingk@bham.ac.uk}}

\address{University of Birmingham, Birmingham B15 2TT, UK}

\begin{abstract}
The spread and scrambling of quantum information is a topic of considerable current interest. Numerous studies suggest that quantum information evolves according to hydrodynamical equations of motion, even though it is a starkly different quantity to better-known hydrodynamical variables such as charge and energy. In this work we show that the well-known memory matrix formalism for traditional hydrodynamics can be applied, with relatively little modification, to the question of operator growth in many-body quantum systems. On a conceptual level, this shores up the connection between information scrambling and hydrodynamics. At a practical level, it provides a framework for calculating quantities related to operator growth like the butterfly velocity and front diffusion constant, and for understanding how these quantities are constrained by microscopic symmetries. We apply this formalism to calculate operator-hydrodynamical coefficients perturbatively in a family of Floquet models. Our formalism allows us to identify the processes affecting information transport that arise from the spatiotemporal symmetries of the model.

\end{abstract}
\maketitle

\makeatletter

\tableofcontents

\clearpage

\section{Introduction}
Under time evolution a wide class of many-body quantum systems tend towards equilibrium, where the final state is well described by a relatively small number of parameters such as temperature or pressure. At this point, information about the local conditions of the initial state is “scrambled”: it can no longer be determined by simple local measurements but is instead encoded in increasingly delicate and complicated observables. The process of scrambling has been the focus of intense study in the fields of black hole physics and holography \cite{Hayden07, Sekino08, Lashkari2013, Shenker2014a, Shenker2014b, Shenker2015, Maldacena2016, Hartman2013, Liu14a, Liu14b, Mezei16, Blake16}, integrable systems \cite{Dora17, Fagotti08, Gopalakrishnan18, Lin18, Prosen07}, random unitary circuits \cite{Nahum16, Nahum17, RvK17, OTOCDiff1, OTOCDiff2, Brown12, ChanDeLuca1}, quantum field theories \cite{Stanford2016, Asplund15, Banerjee2017, Roberts18, Roberts16, Swingle17, Aleiner16, CalabreseCardy05}, and in the setting of chaotic spin-chains \cite{HyungwonHuse, Bohrdt16, Prosen17, DeChiara06, Hartnoll19, Lieb72, Abanin17, Luitz17, Bertini2018, Zhang20, Xu2019}. A principal reason for the flurry of interest in scrambling is the striking universality observed in the information spreading dynamics of apparently disparate models.

One universal feature of scrambling in ergodic systems is the ballistic growth of operators (with `butterfly velocity' $v_\mathrm{B}$), a feature connected to the universally observed linear growth of quantum entanglement. This result has been demonstrated for random unitary circuits \cite{Nahum17}; in 1D a simple picture emerges where operators grow according to biased diffusion \cite{Nahum17,RvK17}.  This picture is altered in the presence of additional symmetries which give rise to power-law tails in the distributions of operator right end-points \cite{OTOCDiff1,OTOCDiff2}. In all of these cases, it appears that in ergodic systems  quantum information  obeys  a  sort  of  hydrodynamics,  with  an  unusual  conservation  law, which we call “information conservation”. The purpose of the present work is to show that the connection to hydrodynamics is not just an analogy and that a standard hydrodynamical tool -- the so-called memory matrix formalism (MMF) \cite{Forster2018} -- can be applied with only a few technical (but consequential) modifications. Once we postulate a suitable slow manifold (a concept we explain), the ballistic growth of information is inevitable; in the same way that diffusion is inevitable in high temperature systems when the local conserved density is identified as the sole slow variable. Moreover, the formalism provides a framework for the perturbative calculation of information transport coefficients (e.g., the butterfly velocity) in concrete models. This is useful because operator spreading tends to require working at infinite temperature, where quantum field theoretic methods become harder to control. As such, the MMF is one of the only tools available for the analytical calculation of operator transport coefficients (although a similar effective membrane theory has been independently suggested for the related issue of entanglement growth \cite{Zhou20}).

This work is organized as follows. In Section \ref{memory matrix formalism} we extend the MMF to include a slow mode associated with the conservation of quantum information in the setting of translation invariant one-dimensional Hamiltonian systems and in section \ref{MMF for Floquet models} we do the same for translation invariant one-dimensional Floquet models. Our formalism yields a succinct expression for the butterfly velocity, and shows that the biased diffusion of operator fronts observed in ergodic systems is arguably the simplest scenario consistent with the conservation laws. We also explain how microscopic symmetries constrain the butterfly velocity and front diffusion constants ($D$). We demonstrate the usefulness of the MMF in \secref{Minimal model} where we consider a translation invariant Floquet circuit with no additional symmetries and give results for the circuit averaged butterfly velocity and operator front diffusion constant in the limit of large local Hilbert space dimension $q$. In section \ref{MMF calc}, we calculate the $\mathcal{O}(1/q^2)$ corrections to the butterfly velocity and attribute various contributions to the discrete time translation symmetry and to the spatial translation symmetry. Finally, inspired by this calculation we give predictions for $v_B$ and $D$ for a family of Floquet models to order $\mathcal{O}(1/q^2)$. This calculation also serves as a consistency check on our formalism, confirming that the slow manifold we have proposed is sufficiently complete to perform hydrodynamical calculations.

\section{\label{memory matrix formalism}Memory matrix formalism: Operator spreading as a slow mode}
The memory matrix formalism (MMF) is a method for predicting the hydrodynamical properties of many-body systems \cite{Forster2018,Lucas2015,Davison2016,Bentsen19}. The input for the method is a Hamiltonian (or some other dynamics), as well as a guess as to what the likely slow modes are in the system (the local densities of conserved quantities are natural candidates). Under the assumption that all slow modes have been included and that all remaining fast modes decay sufficiently rapidly, the formalism yields predictions for the long-distance behavior of correlation functions involving the slow modes.

In this section, we adapt the formalism to include the slow mode associated with information conservation in the setting of one-dimensional quantum systems with local dynamics, i.e., local Floquet circuits, or systems with local Hamiltonians. In this paper, the ``conservation of quantum information" is equivalent to the statement that Heisenberg evolved operators have a conserved Hilbert-Schmidt norm $\Tr(O^{\dagger}(t)O(t))=\textrm{const}.$, a direct consequence of unitarity.

By averaging over choices of initial operator with the same right endpoint, we express the distribution of operator right endpoints (or the `operator right density' as we refer to it in this paper), as an autocorrelation function of elements in the space of operators on two replicas of the Hilbert space. Using the MMF, we investigate the pole structure of the corresponding spectral function and give Kubo-like formula for the butterfly velocity $v_B$ and an expression for diffusion constant $D$ in terms of the memory matrix.

\subsection{Quantifying operator spreading}\label{quantifying_operator_spreading}
We consider a one-dimensional lattice of $N$ sites with single site Hilbert space $\mathcal{H}_{local}=\mathbb{C}^q$. The space of operators on the full Hilbert space $\mathcal{H}$ is denoted $\mathcal{B}(\mathcal{H})$. A convenient basis for single site operators are the generalised Pauli matrices\footnote[1]{The generalised Pauli matrices are generated by the shift and clock matrices $X$ and $Z$, $\sigma^{\mu} = X^{\mu^{(1)}}Z^{\mu^{(2)}}$. Where $X^q=Z^q=\mathbb{1} \quad \textrm{and} \quad ZX = e^{\frac{2\pi \mathrm{i}}{q}}XZ$.} $\{\sigma^{\mu}\}$, a set of unitary matrices satisfying the orthogonality relation $\Tr(\sigma^{\mu \dagger}\sigma^{\nu})/q = \delta^{\mu,\nu}$. A time evolved operator $O(t)$ can be expressed as a linear combination of strings of generalised Pauli operators,
\begin{equation}
O(t) \equiv U^{\dagger}(t)OU(t) \equiv e^{\mathrm{i}tL}O = \sum_{\mu}C^{O}_{\mu}(t) \sigma^{\mu}.
\end{equation}
The time evolution is generated by a Hamiltonian $H$, $U(t)=\exp(-\mathrm{i}tH)$. In the second equality we have introduced the Liouvillian $L(\cdot)=[H,\cdot \ ]$\footnote{In the context of this paper, a Liouvillian $L$ is a generator of Hamiltonian evolution (in operator space), as opposed to a generator of Markovian dynamics in open quantum systems, as the name often refers.}, and in the last equality only the coefficients $C^{O}_{\mu}(t)$ depend on time. We have normalised $O$ such that the Hilbert-Schmidt norm gives $||O||_{HS}^2 = \dim(\mathcal{H})=q^{N}$, this ensures $\sum_{\mu} |C^O_{\mu}|^2 = 1$. Following the work of \cite{Nahum17,RvK17}, we define the \textit{right density} $\rho_R(x,t)$ (the probability of an operator having right endpoint at position $x$ at time $t$) by
\begin{equation} \label{weight}
\rho_R(x,t) \equiv \sum_{\mu} |C^{O}_{\mu}(t)|^2 \delta(\textrm{Rhs}(\mu)-x),
\end{equation}
where $\textrm{Rhs}(\mu)$ denotes the rightmost site on which the Pauli string has non-trivial support. Summing over the site positions in \eqnref{weight} gives us a conserved quantity,
\begin{equation}\label{conservation of weight}
\sum_x \rho_R(x,t) = \sum_{x,\mu} |C^{O}_{\mu}(t)|^2 \delta(\textrm{Rhs}(\mu)-x) = \sum_{\mu} |C^{O}_{\mu}(t)|^2 = 1.
\end{equation}

\subsection{Operator right density as a correlation function}
The starting point for the MMF is a temporal correlation function of slow variables, typically local charge operators associated to a conserved quantity such as energy or charge. Analogous to the local operators associated with energy or charge, we are able to associate pseudo-local operators $\hat{W}^x$ with the right density $\rho_R$. These operators, which we call the right density operators, will play the role of slow variables in the MMF. One complication is that, rather than being an operator on the original Hilbert space like charge/energy, $\hat{W}^x$ is an operator on operators -- a \textit{super-operator} (or equivalently an operator on two replicas of the original Hilbert space). Thus, in our application of MMF, we will be studying the dynamics of super-operators like $W^x$, eventually writing $\rho_R$ as a temporal correlation function of `vectorised' right density operators $\lvert W^x\rangle $. This serves as the starting point for the MMF. $\hat{W}^x$ is implicitly defined in the equation for $\rho_R(x,t)$ below,
\begin{equation}\label{density}
\rho_R(x,t) \equiv \langle O(t)\rvert \hat{W}^x \lvert O(t)\rangle,\quad \langle a\rvert b\rangle \equiv \Tr(a^{\dagger}b).
\end{equation}
Where the inner-product $\langle a\rvert b\rangle$ is the infinite temperature operator inner-product suitable for $\mathcal{B}(\mathcal{H})$. An explicit definition of $\hat{W}^x$ is given by
\begin{equation}\label{algebraic expression for weight}
\hat{W}^x \equiv \frac{1}{d_\mathcal{H}}\left(\Motimes_{r\leq x-1} \hat{\Lambda}^+\right)\left( \Motimes_x \hat{\Lambda}^0\right)\left( \Motimes_{r \geq x+1} \hat{\Lambda}^-\right),
\end{equation}
where $\hat{\Lambda}^+$ is the identity super-operator, $\hat{\Lambda}^-$ is the projector onto the identity operator and $\hat{\Lambda}^{0}$ is the projector onto the space of non-identity operators. Algebraically, $\hat{\Lambda}^\pm$ are given by
\begin{equation}\label{Lambda projector def}
\hat{\Lambda}^+ \equiv \sum_\mu \frac{\lvert \sigma^\mu \rangle \langle \sigma^{\mu}\rvert}{q}, \quad \hat{\Lambda}^- \equiv \frac{\lvert \mathbb{1}\rangle \langle\mathbb{1}\rvert}{q}, \quad \hat{\Lambda}^0 \equiv \hat{\Lambda}^+ - \hat{\Lambda}^-.
\end{equation}
By writing \eqnref{density} in the language of tensor diagrams, we make the following manipulations to write the density $\rho_R(x,t)$ as an overlap of two `vectors',
\begin{equation}\label{manipulation}
    \langle O(t) \rvert\hat{W}^x\lvert O(t)\rangle\equiv \raisebox{-0.26\totalheight}{\includegraphics[height=0.8cm]{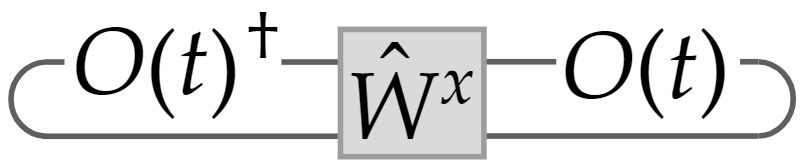}}\equiv\raisebox{-0.36\totalheight}{\includegraphics[height=1.35cm]{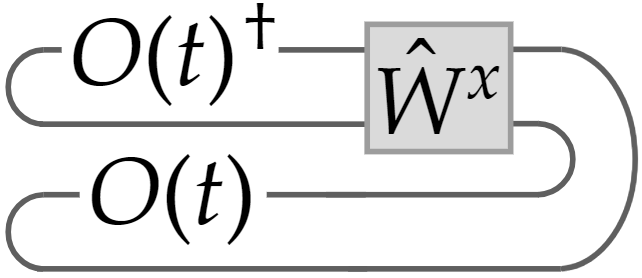}}\equiv\raisebox{-0.39\totalheight}{\includegraphics[height=1.5cm]{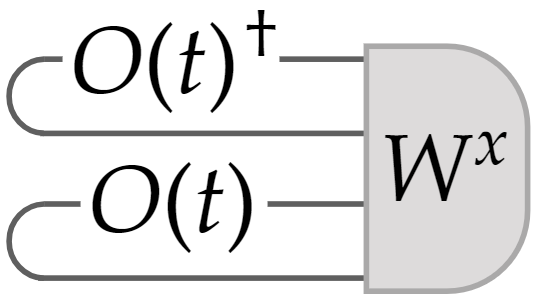}}.
\end{equation}
We view the final expression as the inner-product of two vectors in the vector space $\mathcal{W}\equiv\mathcal{B}(\mathcal{H})\boxtimes \mathcal{B}(\mathcal{H})$ of operators acting on two replicas of the Hilbert space (we have used a boxed tensor product to represent the tensor product between two copies of the operator space). In particular, this allows us to write the density $\rho_R$ as a temporal correlation function of the form $\langle A(t)\rvert B\rangle_{\mathcal{W}}$ as shown below,
\begin{equation}
    \rho_R(x,t)\equiv\langle O(t)\rvert \hat{W}^x\lvert O(t)\rangle \equiv \langle O(t) \boxtimes O(t)^\dagger\rvert W^x\rangle_{\mathcal{W}}, \ \  \langle A\rvert B\rangle_{\mathcal{W}}\equiv\Tr(A^\dagger B), \ A,B \in \mathcal{W}.
\end{equation}
Elements of $\mathcal{W}$ can be written as linear combinations of vectors $\ket{A\boxtimes B}$, which has the following diagrammatic representation,
\begin{equation}\label{leg labelling convention}
	\begin{tikzpicture}[scale = 0.4,baseline=(current  bounding  box.center)]
	\node at (-1.2,0.6) {$\ket{A\boxtimes B} \ \equiv \quad $};
	\node at (1.5,2) {$1$};
	\node at (1.5,1) {$\overline{1}$};
	\node at (1.5,0) {$2$};
	\node at (1.5,-1) {$\overline{2}$};
	\draw[thick,rounded corners=2pt] (2.5,-1) -- (4.8,-1) -- (4.8,0) -- (2.5,0) node [midway,fill=white] {$B$};
	\draw[thick,rounded corners=2pt] (2.5,1) -- (4.8,1) -- (4.8,2) -- (2.5,2) node [midway,fill=white] {$A$};
	\node at (5.5,-1) {,};
	\end{tikzpicture}
\end{equation}
where the \textit{legs} $1,\overline{1},2,\overline{2}$, represent the indices of the operators $A$ and $B$ which act on the first and second replica respectively. An inner-product of two states has the obvious meaning of connecting legs,
\begin{center}
	\begin{tikzpicture}[scale = 0.4]
	\node at (-4,0.6) {$\bra{C\boxtimes D}\ket{A\boxtimes B} \ \equiv \quad $};
	\draw[thick,rounded corners=2pt] (2.5,-1) -- (0.2,-1) -- (0.2,0) -- (2.5,0) node [midway,fill=white] {$D^{\dagger}$};
	\draw[thick,rounded corners=2pt] (2.5,1) -- (0.2,1) -- (0.2,2) -- (2.5,2) node [midway,fill=white] {$C^{\dagger}$};
	\draw[thick,rounded corners=2pt] (2.5,-1) -- (4.8,-1) -- (4.8,0) -- (2.5,0) node [midway,fill=white] {$B$};
	\draw[thick,rounded corners=2pt] (2.5,1) -- (4.8,1) -- (4.8,2) -- (2.5,2) node [midway,fill=white] {$A$};
	\node at (14,0.6) {$\equiv \ \Tr(C^{\dagger}A)\Tr(D^{\dagger}B)\equiv\bra{C}\ket{A}\bra{D}\ket{B}.$};
	\end{tikzpicture}
\end{center}
One also needs to include a rule for moving a symbol around a bend to ensure consistency,
\begin{center}
	\begin{tikzpicture}[scale = 0.4]
	\draw[thick,rounded corners=2pt] (2.3,1) -- (4.8,1) -- (4.8,2) -- (2.3,2) node [midway,fill=white] {$A$};
	\node at (6,1.4) {$\equiv$};
	\draw[thick,rounded corners=2pt] (7,1) -- (9.5,1) node [midway,fill=white] {$A^T$} -- (9.5,2) -- (7,2);
	\node at (10,0.7) {.};
	\end{tikzpicture}
\end{center}
The manipulation of \eqnref{manipulation} has the following consequences for the $\hat{\Lambda}^{\pm}$ and $\hat{\Lambda}^0$,
\begin{equation}\label{Lambda state def}
    \hat{\Lambda}^+ \rightarrow q\ket{+} \equiv \frac{1}{q}\sum_\mu \ket{\sigma^\mu \boxtimes \sigma^{\mu\dagger}}, \quad \hat{\Lambda}^- \rightarrow \ket{-} \equiv \frac{1}{q}\ket{\mathbb{1} \boxtimes \mathbb{1}}, \quad \hat{\Lambda}^0 \rightarrow q\ket{0} \equiv q\ket{+}-\ket{-},
\end{equation}
where we have chosen to normalise the vectors $\ket{+}$ and $\ket{-}$, $\bra{\pm}\ket{\pm}=1$ (this introduces factors of $q$ that differ from \eqnref{Lambda projector def}). The vectors $\ket{\pm}$ take a simple diagrammatic form,
\begin{equation} \label{diagrammatic expression for plus and minus}
	\begin{tikzpicture}[scale = 0.5,baseline=(current  bounding  box.center)]
	\node at (-3.5,0.5) {$\ket{+} = $ \Large $\frac{1}{q}$};
	\node at (-1.2,2) {\color{gray} $1$};
	\node at (-1.2,1) {\color{gray} $\overline{1}$};
	\node at (-1.2,0) {\color{gray} $2$};
	\node at (-1.2,-1) {\color{gray} $\overline{2}$};
	\draw[thick,rounded corners=2pt] (-0.5,-1) -- (0.5,-1) -- (0.5,2) -- (-0.5,2);
	\draw[thick,rounded corners=2pt] (-0.5,0) -- (0,0) -- (0,1) -- (-0.5,1);
	\node at (2.9,0.5) {, \quad $\ket{-} =$ \Large $\frac{1}{q}$};
	\node at (5.4,2) {\color{gray} $1$};
	\node at (5.4,1) {\color{gray} $\overline{1}$};
	\node at (5.4,0) {\color{gray} $2$};
	\node at (5.4,-1) {\color{gray} $\overline{2}$};
	\draw[thick,rounded corners=2pt] (6.2,-1) -- (6.7,-1) -- (6.7,0) -- (6.2,0);
	\draw[thick,rounded corners=2pt] (6.2,1) -- (6.7,1) -- (6.7,2) -- (6.2,2);
	\node at (7.2,0.2) {\color{black} .};
	\end{tikzpicture}
\end{equation}
The vectorised right density super-operators $\ket{W^x}$ are then given by
\begin{equation}
    \ket{W^x}\equiv\frac{1}{d_{>x}}\left(\Motimes_{r\leq x-1} \ket{+}_r\right)\left( \Motimes_x \ket{0}_x\right)\left( \Motimes_{r \geq x+1} \ket{-}_r\right),
\end{equation}
where $d_{>x}$ is the dimensions of the subsystem of sites to the right of site $x$. A closely related super-operator is the \textit{purity super-operator} $\hat{F}^x$, given by
\begin{equation}\label{algebraic expression for F}
    \hat{F}^x \equiv \frac{1}{d_{\leq x}}\hat{\Lambda}^+_{\leq x}\otimes \hat{\Lambda}^-_{>x}, \quad \ket{F^x} \equiv \left(\Motimes_{r\leq x}\ket{+}_r\right)\left(\Motimes_{r>x} \ket{-}_{r}\right),
\end{equation}
where $d_{\leq x}$ is the dimension of the subsystem of sites $r\leq x$. Using this super-operator, we are able to represent the purity $\gamma_{\leq x}(t)$ of the system (partitioned into a subsystem left (inclusive) and right of $x$) as a matrix element,
\begin{equation} \label{purity operator}
d_{\mathcal{H}}\bra{\rho(t)} \hat{F}^x \ket{\rho(t)} d_{\mathcal{H}}\equiv \bra{\rho(t)^\dagger\boxtimes \rho(t)} \ket{F^x}_{\mathcal{W}} \equiv \Tr_{\leq x}(\rho_{\leq x}(t)^2) \equiv \gamma_{\leq x}(t),
\end{equation}
where $\rho(t)$ is the density matrix for the state of the system. 
Up to an overall constant, the purity operators $F^x$ and the integrated right density operators, $\sum_{y\leq x}W^y$, are equivalent. Therefore, the $W^x$ can be written as a difference of (re-scaled) $F^x$'s,
\begin{equation}
W^x = \frac{1}{d_{>x}} F^x - \frac{1}{d_{>x-1}} F^{x-1}. \label{purity-weight relation}
\end{equation}
Time evolution in the doubled operator space $\mathcal{W}$ is generated by a \textit{doubled} Liouvillian $\mathcal{L}\equiv L \boxtimes \mathbb{1} + \mathbb{1}\boxtimes L$, which evolves both replicas of the operator space independently. The time derivative of $W^x(t)\equiv e^{i\mathcal{L}t}(W^x)$ is given by $\partial_t W^x(t)\equiv \mathrm{i}\mathcal{L}(W^x(t))$. We can use this to write down a continuity equation for $W^x$,
\begin{equation}\label{local conservation law}
    \partial_t W^x + \Delta_x(J^x) = 0, \quad \Delta_x(J^x) \equiv J^x - J^{x-1},
\end{equation}
where $J^x=-\mathrm{i}\mathcal{L}(F^x)/d_{>x}$ is the current associated with the operator right density. The currents $J^x$ are pseudo-local super-operators; they look, locally, like $\Lambda^+$ everywhere to the left the \textit{cut} $\{x,x+1\}$ and $\Lambda^-$ everywhere to the right. Local unitary evolution acts trivially in the $+$ and $-$ domains. We show this by considering the action of a (single-site) Liouvillian $\mathcal{L}_{i}$ at site $i$ in the $+$ domain,
\begin{equation}\label{L away from cut}
    \raisebox{-0.45\totalheight}{\includegraphics[height=1.6cm]{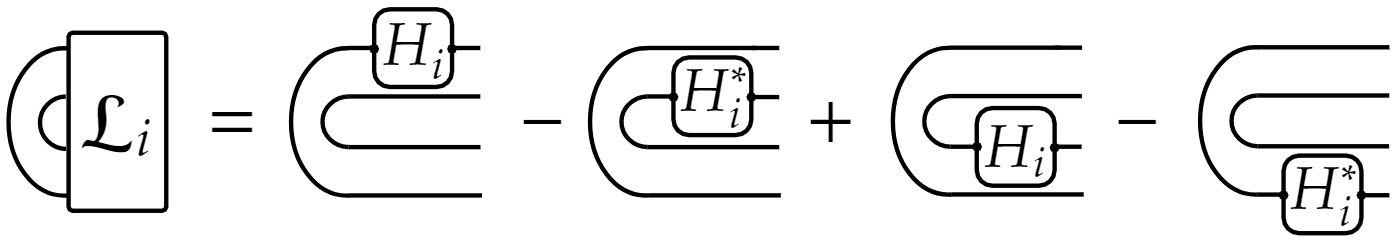}}=0.
\end{equation}
By moving the conjugated Hamiltonians around the `bend' from the barred legs onto the unbarred legs, we find that the first and fourth terms cancel, as do the second and third. This can be easily generalised to $l$-local interactions. By swapping legs $1\leftrightarrow2$ in \eqnref{L away from cut}, we the find the action of $\mathcal{L}$ in the $-$ domain. The region separating the $+$ and $-$ domains can only grow via the local evolution at its edges. In this sense, one can consider \eqnref{local conservation law} to be an equation of local conservation of operator right density.

\subsection{Operator averaging}
Rather than consider the evolution of the right density for a particular operator $O$, we average over the all operators with right endpoint $x$. Doing this in the basis of generalised Pauli matrices, the averaged density $\overline{\rho_R}(x,y,t)$ is given by
\begin{equation}
\overline{\rho_R}(x,y,t) \equiv \frac{q^2}{q^2-1}\frac{1}{d^2_{\leq x}} \sum_{O_x}\bra{O_x^\dagger\boxtimes O_x}e^{-\mathrm{i}t\mathcal{L}}\ket{W^y}_\mathcal{W},
\end{equation}
where we have pulled the time evolution out from the operators $O(t)$. The factor before the sum is normalisation for the average (the reciprocal of the number of linearly independent operators with right endpoint $x$). This is in fact just a temporal-correlation function between right density super-operators,
\begin{equation}
\overline{\rho_R}(x,y,t) = \frac{q^2 d^2_{>x}}{q^2-1}\bra{W^x}e^{-it\mathcal{L}}\ket{W^y}. \label{autocorrelation function}
\end{equation}
The $W^x$ are orthogonal but not normalised with respect to the trace inner-product. The result of this is that the Fourier transformed right-densities $W^k$ are not orthogonal. This lead us to an unusual inner-product $\left(\cdot|\cdot\right)$ with respect to which the $W^x$ are orthonormal,
\begin{equation}\label{inner-product}
\left(A|B\right) \equiv \bra{\Phi\left(A\right)}\ket{B}_\mathcal{W} = \Tr\left(\Phi\left(A\right)^\dagger B\right).
\end{equation}
Where $\Phi$ is given by
\begin{align}
\Phi \equiv \sum_x \frac{1}{\chi_x^{2}}\ket{W^x}\bra{W^x} + Q, \quad \chi_x \equiv \bra{W^x}\ket{W^x}=\frac{q^2-1}{q^2 d^{2}_{>x}}, \implies \Phi\ket{W^x} = \frac{1}{\chi_x} \ket{W^x}.
\end{align}
$Q$ is the projector onto the \textit{fast} subspace $\mathcal{Q}$, the orthogonal complement\footnote{Orthogonal with respect to the inner-product $\bra{\cdot}\ket{\cdot}$, or equivalently $\left(\cdot|\cdot\right)$.} of the \textit{slow} subspace $\mathcal{P} \equiv \textrm{Span}\{W^x\}$. A proof that $\left(\cdot|\cdot\right)$ satisfies the axioms of an inner-product is found in \ref{inner_product_proof}. We must be careful with expression \eqnref{autocorrelation function} as $\mathcal{L}$ is not self-adjoint with respect to this inner-product. Making the operator average implicit, the right density becomes
\begin{equation}
\rho_R(x,y,t) = \left(W^x|e^{-\mathrm{i}t\mathcal{L}}|W^y\right).
\end{equation}
The position dependent re-scaling of the right densities $W^x$ by $\Phi$ reflects the strong entropic bias for operators to grow, i.e., for $y>x$ we find
\begin{equation}\label{operators grow}
    \left(W^x|e^{-\mathrm{i}t\mathcal{L}}|W^y\right)=q^{2(y-x)}\left(W^y|e^{\mathrm{i}t\mathcal{L}}|W^x\right).
\end{equation}
Operators are exponentially more likely to grow than to shrink\footnote{Evolution with $\mathcal{L}$ and $-\mathcal{L}$ can give rise to differing butterfly velocities (in \secref{lightcone} we identify these as $v_R$ and $v_L$, the right/left velocities).}.

\subsection{Spectral and memory function}
Having expressed the right density as a correlation function, we now investigate its late time behavior. A helpful diagnostic in the long time behaviour of a correlation function is the pole structure of the corresponding spectral function,
\begin{equation}
\rho_{x,y}\left(z\right) = \left(W^x\left|\frac{\mathrm{i}}{z-\mathcal{L}}\right|W^y\right)= \int_{0}^{\infty}e^{\mathrm{i}zt}\left(W^x\left|e^{-\mathrm{i}t\mathcal{L}}\right|W^y\right)dt.
\end{equation}
Fourier transforming in space will allow us to express the poles at small $z$ in terms of a wave-number $k$. In particular this allows us to characterise the long-time and long-wavelength behaviour of the averaged density $\rho_R(x,y,t)$. The density super-operators in $k$-space are given by
\begin{equation}\label{momentum space W}
W^k = \frac{1}{\sqrt{N}}\sum_{x} e^{-\mathrm{i}xk} W^x, \quad \left(W^p|W^k\right) = \delta^{k,p}.
\end{equation}
We choose to center the lattice at $x=0$. In The $k$-space, the spectral function is given by
\begin{equation}
\rho_{k,p}\left(z\right) = \left(W^k\left|\frac{\mathrm{i}}{z-\mathcal{L}}\right|W^p\right).
\end{equation}
By taking $\mathcal{P}\equiv\textrm{Span}\{W^k\}$ as the slow space in the memory matrix formalism, we use some formal manipulations of resolvent operators \cite{Forster2018} to express the spectral function as
\begin{equation}
\rho_{k,p}\left(z\right) = \mathrm{i}\left[z\mathbb{1} - \Omega + i\Sigma\left(z\right)\right]^{-1}_{ \ \ k,p},
\end{equation}
where $\Omega_{k,p} \equiv \left(W^k\left|\mathcal{L}\right|W^p\right)$ and
$\Sigma\left(z\right)_{k,p}$ is the memory matrix,
\begin{equation} \label{memory matrix}
\Sigma\left(z\right)_{k,p} = \left(W^k\left|\mathcal{L}Q\frac{\mathrm{i}}{z-\mathcal{L}Q}\mathcal{L}\right|W^{p}\right).
\end{equation}
By writing $\comm{A}{\cdot\ }=l_A-r_A$, where $l_A$ ($r_A$) is left (right) multiplication by $A$, we can express the doubled Liouvillian as $\mathcal{L}=(l_H-r_H)\boxtimes \mathbb{1}+\mathbb{1}\boxtimes (l_H-r_H)$. It is simple to check using \eqnref{algebraic expression for F} and \eqnref{diagrammatic expression for plus and minus} that $\bra{F^x}l_H\boxtimes \mathbb{1}\ket{F^y}=\bra{F^x}r_H\boxtimes \mathbb{1}\ket{F^y}=q^{-|y-x|}\Tr(H)$ (and equivalently for multiplication by $H$ in the second replica). This gives $\bra{F^x}\mathcal{L}\ket{F^y}=0$ and, by linearity, $\Omega_{k,p}=0$.

For translationally invariant systems (in the thermodynamic limit $N\to\infty$), the memory matrix is diagonal, $\Sigma\left(z\right)_{k,p} = \Sigma\left(k,z\right)\delta^{k,p}$ and the inverse of $z\mathbb{1} + i\Sigma\left(k,z\right)$ is readily calculated,
\begin{equation}\label{Hamiltonian spectral func}
\rho_{k,p}(z) = \rho(k,z)\delta^{k,p},\quad \rho(z,k)= \frac{\mathrm{i}}{z + i\Sigma\left(z,k\right)}.
\end{equation}
The long-time and long-wavelength pole structure is found by expanding the memory function $\Sigma\left(z,k\right)$, for small $k$ and $z$. The $k$-space representation of the continuity equation \eqnref{local conservation law} is then given by,
\begin{equation}
\partial_t W^k \equiv \mathrm{i}\mathcal{L}\left(W^k\right) = -(1 - e^{-\mathrm{i}k})J^k,
\end{equation}
where $J^k = \frac{1}{\sqrt{N}}\sum_x e^{-\mathrm{i}kx} J^x$. Putting this back into the memory matrix yields
\begin{align}
\Sigma\left(k,z\right) &= -\mathrm{i}(1-e^{-\mathrm{i}k}) \left(W^k\left|\mathcal{L}Q\frac{\mathrm{i}}{z-\mathcal{L}Q}\right| J^{k}\right)\label{memory matrix k expansion 1} \\
&=- v\left(z\right)\mathrm{i}k - b\left(z\right)k^2 + \cdots. \label{memory matrix k expansion 2}
\end{align}
This will give the pole structure
\begin{equation}
\rho(k,z) \sim \frac{\mathrm{i}}{z - v\left(z\right)k + \mathrm{i}b\left(z\right)k^2 + \cdots}.
\end{equation}

Provided the analyticity of $v(z)$ and $b(z)$ as $-\mathrm{i}z\rightarrow 0^+$, this will be precisely the pole structure associated with a biased diffusion equation. This analyticity condition is met so long as we can assume that the fast variables $J^k$ and $\mathcal{L}\Phi(W^k)$ have rapidly decaying correlations (faster than $1/t$).

\subsection{\label{formal expressions and light-cone}Butterfly velocity and diffusion constant}
In this section we provide formal expression for the butterfly velocity $v_B$ and diffusion constant $D$ and sufficient conditions for the symmetry of the operator growth light-cone, $v_L=v_R$ and $D_L=D_R$.

\subsubsection{Formal expressions for \texorpdfstring{$v_B$}{Lg} and \texorpdfstring{$D$}{Lg}}

Using equation (\ref{memory matrix k expansion 2}), the butterfly velocity $v_B$ is given by
\begin{equation}\label{v(z)}
    v_B = \lim_{z\to i0^+} v(z), \quad v(z)=\lim_{k\to 0}-\mathrm{i}\partial_k\Sigma(k,z).
\end{equation}
We introduce the proxy $\sigma(k,z)$, defined as
\begin{equation}
    \sigma(k,z)=\left(W^k\left|\mathcal{L}\frac{\mathrm{i}}{z-\mathcal{L}}\mathcal{L}\right|W^k\right), \quad \Sigma(k,z)=\frac{\sigma(k,z)}{1+\sigma(k,z)/z}.
\end{equation}
Using $\mathcal{L}(W^k)\sim k$, we conclude that $\lim_{k\to 0}\sigma/k=\lim_{k\to0}\Sigma/k$, provided that we take the $k\to 0$ limit before taking $z\to \mathrm{i}0^+$. Using this, we give a Kubo-like formula for $v_B$,
\begin{equation}\label{Kubo}
v_B = \mathrm{i}\lim_{s \rightarrow 0}\int_0^\infty dt \ e^{-st} \left(W\left|\mathcal{L}\right|J(-t)\right),
\end{equation}
where $J\equiv J^{k=0}$ and $W\equiv W^{k=0}$. Converting to the usual trace inner-product, we have $\left(W\left|\mathcal{L}\right|J(-t)\right)=\bra{\mathcal{L}\Phi(W)}\ket{J(-t)}$. Importantly, $\Phi$ and $\mathcal{L}$ do not in general commute\footnote{An example where $\comm{\Phi}{\mathcal{L}}=0$ is when the Hamiltonian does not couple any sites, i.e., $v_B=0$.}. If they were to commute, we would find $\mathcal{L}(W^{k=0})=0$ and hence $v_B=0$, which would lead to the incorrect conclusion that information propagates diffusively rather than ballistically. Using the biased diffusion ansatz for the pole location,
\begin{equation}\label{ansatz}
    z = v_B k - \mathrm{i} D k^2 + \order{k^3},
\end{equation} 
the diffusion constant is given by
\begin{equation}\label{diffconst}
D = \lim_{z\to \mathrm{i} 0^+}\lim_{k\to0}\left(v_B\partial_z \partial_k \Sigma + \frac{1}{2}\partial_k^2 \Sigma\right).
\end{equation}

\subsubsection{\texorpdfstring{$v_R \neq v_L$}{Lg} and the operator growth light-cone}\label{lightcone}
Generically, the right and left butterfly velocities, $v_R$ and $v_L$, are not equal \cite{Zhang20,Liu2018,stahl2018asymmetric}. In this section we relate $v_R$ and $v_L$ in a way that gives rise to a light-cone structure. We then determine sufficient conditions for a symmetric light-cone, $v_R = v_L$ and symmetric fronts $D_L=D_R$. We need to adapt our notation when talking about both left and left and right density distributions at once, to do this we denote $W_{R}^x$ as the familiar right density super-operators and $W_{L}^x$ to be the left density super-operator. The altered inner-product is also right/left dependent, with $\left(\cdot|\cdot\right)_R$ ($\left(\cdot|\cdot\right)_L$) corresponding to the inner-productive suitable for right (left) endpoint calculations. The right and left density distributions are given by
\begin{align}
\rho_R(x,t) = \left(W_{R}^0\left|e^{-\mathrm{i}t\mathcal{L}}\right| W_{R}^x\right)_R, \quad \rho_L(x,t) = \left(W_{L}^0\left|e^{-\mathrm{i}t\mathcal{L}}\right| W_{L}^x\right)_L.
\end{align}
The equations for $v_L$ and $D_L$ are found to be
\begin{equation}\label{v_L,D_L}
    v^H_L = \lim_{z\to \mathrm{i}0^+}\lim_{k\to 0}i\partial_k\Sigma^H_L(k,z), \quad D^H_L = \lim_{z\to i0^+}\lim_{k\to0}\left(-v_L\partial_z \partial_k + \frac{1}{2}\partial_k^2 \right)\Sigma^H_L(k,z).
\end{equation}
Where we have introduced a superscript $H$ to label which Hamiltonian the systems is being evolved with. In order to convert between these distributions let us introduce the involution $\mathcal{I}$, the spatial inversion symmetry operation, $x\to-x$. It has the following action on the density super-operators and the Liouvillian,
\begin{equation}
\mathcal{I}\ket{W_{L}^x} = \ket{W_{R}^{-x}}, \quad \mathcal{I}\mathcal{L}_H\mathcal{I} = \mathcal{L}_{H_I},
\end{equation}
where $H_I$ is the spatially inversion of the (translationally invariant) Hamiltonian $H$. Using this involution on $\rho_L$ gives the following
\begin{align}
\rho^H_L(x,t) &= \left(W_{L}^0\left|e^{-\mathrm{i}t\mathcal{L}}\right| W_{L}^x\right)_L= \frac{1}{\chi_0} \bra{W_{L}^0}\mathcal{I}^2e^{-\mathrm{i}t\mathcal{L}}\mathcal{I}^2\ket{W_{L}^x}= \frac{1}{\chi_0} \bra{W_{R}^0}e^{-\mathrm{i}t\mathcal{L}_I}\ket{W_{R}^{-x}}\nonumber \\
&= \left(W_{R}^0\left|e^{-\mathrm{i}t\mathcal{L}_I}\right|W_{R}^{-x}\right)_R= \rho^{H_I}_R(-x,t), \label{left-right correspondance}
\end{align}
where $\chi_0 = \bra{W_R^0}\ket{W_R^{0}}=\bra{W_L^{0}}\ket{W_L^{0}}$. We have added labels to make clear under which Hamiltonian the system is being evolved. This relation implies the following,
\begin{equation}
    \Sigma^H_L(k,z)=\Sigma^{H_I}_R(-k,z).
\end{equation}
Using this and expressions for $v_{R/L}$ and $D_{R/L}$ (\eqnref{v(z)}, \ref{diffconst} and \ref{v_L,D_L}), we find the relationship between $v_L$ and $v_R$ and $D_L$ and $D_R$ to be
\begin{equation}\label{left right velocity correspondance}
v^{(H)}_L \equiv v^{(H_I)}_R,\quad D^{(H)}_L \equiv D^{(H_I)}_R.
\end{equation}
This implies a physically obvious result: in an inversion symmetric system the left an right velocities are equal.
A somewhat less obvious result is the following:
\begin{remark}\label{remark}
	For a translationally invariant Hamiltonian $H$, if there exists a transformation $\mathcal{R}$ that performs single site basis rotations, such that $\mathcal{R}^\dagger H \mathcal{R}=-H$ or $\mathcal{R}^\dagger H\mathcal{R}=H^*$, the operator growth light-cone is symmetric, $v_L=v_R$ and $D_L=D_R$.
\end{remark}
The second of these sufficient conditions applies for systems with a time-reversal/anti-unitary symmetry, provided that the symmetry transformation can be achieved with single site transformations. Before we prove this result, we will apply it to several models:

\begin{itemize}
    \item An example where both sufficient conditions are met is the spin-1/2 model $H = -\sum_{<i,j>}JY_i Z_j - \sum_i h X_i$. The choice of $\mathcal{R} = Y^{\otimes N}$ gives $\{\mathcal{R},H\}=0$. Alternatively, if we had chosen $\mathcal{R}$ to be a product of Pauli $X$ operators on every even site and $\mathbb{1}$ on every odd site, we find $\mathcal{R}^\dagger H \mathcal{R} = H^*$, (a time-reversal symmetry). Despite this model lacking inversion symmetry, it has a symmetric operator light-cone, $v_L=v_R$ and $D_L=D_R$.
    \item An example where the second of the sufficient conditions is met is given by another spin-$1/2$ model with two-body interactions and no external field coupling, $H = -\sum_{i,j}\sum_{\alpha,\beta}J_{i,j}^{\alpha,\beta}\sigma^{\alpha}_i\sigma^{\beta}_j$. By choosing $\mathcal{R} = Y^{\otimes N}$, only $\sigma^{\alpha}_i\sigma^{\beta}_j$ terms containing a single $Y$ are sent to their negatives, every other term remains unchanged. The terms that flipped sign make up the imaginary part of $H$, hence $H\to H^*$, so that yet again $v_L=v_R$.
    \item Taking the integrable and non-integrable Hamiltonians of \cite{Zhang20}, in the special case $\lambda=0$ the Hamiltonians lack inversion symmetry but satisfy both of the sufficient conditions above, giving $v_L=v_R$ in agreement with the results in \cite{Zhang20}.
\end{itemize}

To prove this result we make use of the symmetry operation $S= (1\leftrightarrow2)$, which swaps the legs $1$ and $2$ on every site. $S$ has the property $S\ket{W_R^x}=\ket{W^x_L}$ and is a symmetry of $\mathcal{L}$. Therefore, 

\begin{align}\label{light-cone-deriv}
\rho^H_L(x,t) &= \left(W_{L}^0\left|S^2e^{-\mathrm{i}t\mathcal{L}_H}S^2\right| W_{L}^x\right)_L=\frac{1}{\chi_0}\bra{W_{L}^0}Se^{-\mathrm{i}t\mathcal{L}_H}S\ket{W_{L}^x}= \frac{1}{\chi_0}\bra{W_{R}^{x}}e^{-\mathrm{i}t\mathcal{L}_H}\ket{W_{R}^0}\nonumber \\
&= \frac{1}{\chi_0}\bra{W_{R}^{0}}e^{\mathrm{i}t\mathcal{L}_H}\ket{W_{R}^{-x}}^*= \left(W_{R}^0\left|e^{\mathrm{i}t\mathcal{L}_H}\right| W_{R}^{-x}\right)_R^* = \rho^{-H}_R(-x,t),
\end{align}
where we have used the reality of the density. Using this gives
\begin{equation}
    \Sigma^H_L(k,z)=\Sigma^{-H}_R(-k,z) \implies v^{(H)}_L \equiv v^{(-H)}_R,\quad D^{(H)}_L \equiv D^{(-H)}_R.
\end{equation}
This implies the existence of a light cone structure, where the future light cone is a $\pi$ rotation of the past light cone. If, in the final equality of \eqnref{light-cone-deriv}, we had brought the complex conjugation onto each term, we would have instead found
\begin{equation}
\Sigma^H_L(k,z)=\Sigma^{H^*}_R(-k,z) \implies v^{(H)}_L \equiv v^{(H^*)}_R,\quad D^{(H)}_L \equiv D^{(H^*)}_R.
\end{equation}
We can generalise this by considering any transformation which performs on-site basis rotations, such a transformation is an isometry of the density super-operators, $\ket{W^x}$. Letting $\mathcal{R}=\prod_x R_x$ be a product of unitary rotations, we can freely conjugate $H$ by $\mathcal{R}$ at any point in \eqnref{light-cone-deriv} (on-site rotation is an isometry of $W^x$) and maintain equality. This gives
\begin{equation}
v^{H}_L \equiv v^{(\mathcal{R}^\dagger H\mathcal{R})^*}_R \equiv v^{-\mathcal{R}^\dagger H\mathcal{R}}_R,\quad D^{H}_L \equiv D^{(\mathcal{R}^\dagger H\mathcal{R})^*}_R \equiv D^{-\mathcal{R}^\dagger H\mathcal{R}}_R.
\end{equation}

\section{\label{MMF for Floquet models} MMF for Floquet models}
With only minor changes, the MMF generalises to Floquet models. In this section we repeat a number of steps taken in the continuous time case, finding Floquet analogues of the memory matrix and $\Omega$. The main results of this section are \eqnref{Floquet memory matrix} for the Floquet analogue of $\Sigma$ and $\Omega$ and formal expressions \eqnref{formal expression v and D} for $v_B$ and $D$ (and a Kubo-like formula for $v_B$ in \eqnref{kubo2}). Readers only interested in main results should look at these equations and skip the rest of the section. 

Let $U$ be a Floquet unitary, then the Heisenberg evolution of an operator $O$ is given by
\begin{equation}
O(n) = U^{-n}_{\varepsilon} O U^n_{\varepsilon}.
\end{equation}
Writing $U^{\textrm{ad}}= l_{U} r_{U^{-1}}$, where $l_U$ ($r_U$) is left (right) multiplication by $U$, allows us to rewrite the operator evolution as
\begin{equation}
O(n) = (U^{\textrm{ad}})^{-n}(O).
\end{equation}
Evolution of elements in the doubled operator space $\mathcal{W}$ is given by
\begin{equation}
\left(A\boxtimes B\right)(n) = (U^{\textrm{ad}}\boxtimes U^{\textrm{ad}})^{-n} (A\boxtimes B).
\end{equation}
Denoting $\mathcal{U} = U^{\textrm{ad}}\boxtimes U^{\textrm{ad}}$, the evolution of an element $X \in \mathcal{W}$ is given compactly by
\begin{equation}\label{Floquet evolution}
X(n) = \mathcal{U}^{-n} (X).
\end{equation}
Using \eqnref{Floquet evolution}, we define the discrete time derivative $\Delta_t X(t)\equiv X(t)-X(t-1) = -\mathcal{L}X(t)$, where $\mathcal{L}=\mathcal{U}-\mathbb{1}$. The Floquet analogue of the continuous time continuity equation \eqnref{local conservation law} is then given below, where we have analogously defined a current $J$,
\begin{equation}
    \Delta_t W^k(t) = -(1-e^{-\mathrm{i}k})J^k(t), \quad J^x \equiv \frac{1}{d_{>x}}\mathcal{L}(F^x),
\end{equation}
Following the steps in \secref{quantifying_operator_spreading}, the operator averaged right density is given by
\begin{equation}
\rho(x,y,n) = \left(W^{x}\big|\mathcal{U}^{n}\big|W^y\right).
\end{equation}
For translationally invariant Floquet models, the spectral function $\rho(z,k)$ is given by \eqnref{Floquet spectral func}, in analogy to \eqnref{Hamiltonian spectral func} in the Hamiltonian case,
\begin{equation}\label{Floquet spectral func}
    \rho(k,z) = \frac{1}{1-e^{\mathrm{i}z}(1+\Omega+\Sigma)}.
\end{equation}
$\Omega$ and the memory matrix $\Sigma$ are given, in analogy to \eqnref{memory matrix}, by
\begin{equation}\label{Floquet memory matrix}
    \Omega(k) = \left(W^k\left|\mathcal{L}\right|W^k\right), \quad 
    \Sigma(k,z) = \left(W^k\left|\mathcal{L}Q\frac{1}{e^{-\mathrm{i}z}-1-\mathcal{L}Q}\mathcal{L}\right|W^{k}\right).
\end{equation}
Unlike in the Hamiltonian case, $\Omega$ does not in general vanish, as $\mathcal{L}\equiv \mathcal{U}-\mathbb{1}$ is not a (doubled) Liouvillian. The pole of the spectral function encodes the long time/distance hydrodynamical limit, and is given by \eqnref{pole equation},
\begin{equation}\label{pole equation}
z = \mathrm{i} \log(1 + \Omega(k) + \Sigma(k,z)).
\end{equation}
Using the biased diffusion ansatz \eqnref{ansatz}, the butterfly velocity and operator front diffusion constants are formally given by
\begin{equation}\label{formal expression v and D}
    v_B = \lim_{z\to \mathrm{i}0^+}\lim_{k\to0} \mathrm{i}\partial_k \left(\Omega+\Sigma\right), \quad D = -\lim_{z\to \mathrm{i}0^+}\lim_{k\to 0}\left(\partial_z\partial_k + \frac{1}{2}\partial_k^2\right)\left(\Omega+\Sigma\right)-\frac{v_B^2}{2}.
\end{equation}

As in the continuous time case, the memory function is difficult to calculate. Once again, it is convenient to define an auxiliary quantity $\sigma$, defined as
\begin{equation}
	\sigma(k,z) \equiv \left(W^k\left|\mathcal{L}\mathcal{Q}\frac{1}{e^{-\mathrm{i}z}-1-\mathcal{L}}\mathcal{Q}\mathcal{L}\right|W^{k}\right), \quad \Sigma = \frac{\sigma}{1+\frac{\sigma}{e^{-\mathrm{i}z}-1-\Omega}}.
\end{equation}
Using $\mathcal{L}(W^k)=\mathrm{i}(1-e^{-\mathrm{i}k})J^k\sim k$, as in the continuous time case, we deduce both $\Sigma(k,z)\sim k$ and $\sigma(k,z) \sim k$ for small $k$. Once again giving
\begin{equation}\label{proxy}
\lim_{k\to0}\Sigma(k,z)/k = \lim_{k\to0}\frac{\sigma/k}{1+\frac{(\sigma/k)k}{e^{-\mathrm{i}z}-1-\Omega}}=\lim_{k\to0}\sigma(k,z)/k.
\end{equation}
By using this equivalence and by splitting the current $J$ it up into its slow and fast components $J^k_P=P(J^k)$ and $J^k_Q=Q(J^k)$, we find the Kubo-like formula for $v_B$, analogous to \eqnref{Kubo} found in the continuous time case,
\begin{equation}\label{kubo2}
    v_B = -\left(W|J_P\right) - \lim_{s\to 0^+} \sum_{t=0}^\infty e^{-st} \left(W|\mathcal{L}Q|J_Q(-t)\right),
\end{equation}
where $J=J^{k=0}$ and $W=W^{k=0}$.  Although a slightly different object to the Liouvillian used in the continuous time case, it remains true that $\mathcal{L}$ does not commute with $\Phi$. Leading again, inevitably, to the ballistic spreading of operators.

\section{\label{Minimal model} A minimal Floquet model}
As a test of the formalism, we investigate a translationally invariant Floquet model with on-site random unitary scramblers. We take the single site Hilbert space to be dimension $q$ and choose the Floquet unitary to be composed of a layer of nearest neighbour two-site unitary gates, followed by a layer of on-site Haar random scrambling unitary $V$ applied to every site, this ensures that the model has no additional conservation laws beyond those guaranteed by unitarity,
\begin{equation}
U_{\varepsilon} = V^{\otimes N} e^{-\mathrm{i}\varepsilon H}, \quad H = \sum_{j} Z_j Z_{j+1}.
\end{equation}
\begin{figure}[H]
    \centering
    \large $U_\varepsilon \  = \ \  $\raisebox{-0.5\totalheight}{\includegraphics[height=5cm]{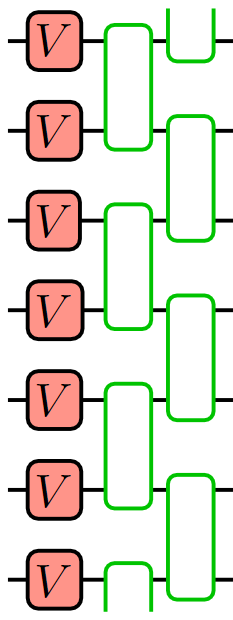}}
    \caption{A Floquet layer with $2$-local gates followed by a layer of single-site scramblers $V$}
    \label{Floquet layer}
\end{figure}

The coupling unitary can be written as a product of commuting two-site unitaries $e^{-\mathrm{i}\varepsilon H}=\prod_j e^{-\mathrm{i}\varepsilon Z_j Z_{j+1}}$. A single gate $e^{-i\varepsilon Z_x Z_{x+1}}=\cos(\varepsilon)\mathbb{1}_x\mathbb{1}_{x+1}-i\sin(\varepsilon)Z_x Z_{x+1}$, straddling sites $x$ and $x+1$, has diagrammatic representation
\begin{equation}\label{Uxxp1}
    e^{-i\varepsilon Z_x Z_{x+1}} \equiv \raisebox{-0.42\totalheight}{\includegraphics[height = 1.2cm]{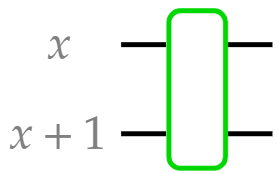}}=\cos(\varepsilon)\raisebox{-0.39\totalheight}{\includegraphics[height = 0.8cm]{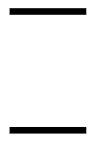}} - i\sin(\varepsilon)\raisebox{-0.39\totalheight}{\includegraphics[height = 0.9cm]{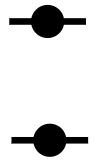}}.
\end{equation}
Where a black dot represent a $Z$ operator, 
\begin{equation}\label{legconvention}
	\centering
	\includegraphics[height = 0.5cm]{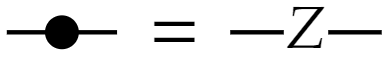}.
\end{equation}
Likewise, the conjugate of the gate $e^{i\varepsilon Z_x Z_{x+1}}$ is given by
\begin{equation}\label{Ustarxxp1}
    e^{i\varepsilon Z_x Z_{x+1}} \equiv \raisebox{-0.42\totalheight}{\includegraphics[height = 1.1cm]{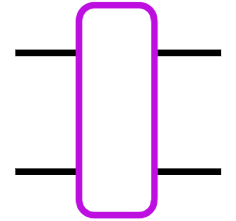}}=\cos(\varepsilon)\raisebox{-0.39\totalheight}{\includegraphics[height = 0.8cm]{Diagrams/Idxxp1.png}} + i\sin(\varepsilon)\raisebox{-0.39\totalheight}{\includegraphics[height = 0.9cm]{Diagrams2/ZZxxp1.png}}.
\end{equation}
We write the Floquet unitary for the doubled operator space as $\mathcal{U} =\mathcal{U}_Z\mathcal{V}$, where $\mathcal{U}_Z=e^{-\mathrm{i}\varepsilon H} \otimes e^{\mathrm{i}\varepsilon H} \otimes e^{-\mathrm{i}\varepsilon H} \otimes e^{\mathrm{i}\varepsilon H}$ contains the 2-local gates and $\mathcal{V}$ is the on-site scrambling unitary (appropriate for the four copies of state space). As in the case with a single replica, We split $\mathcal{U}_Z$ up into a product \textit{bricks} $\mathcal{U}_{x,x+1}$, given by $\mathcal{U}_{x,x+1}=e^{-\mathrm{i}\varepsilon Z_x Z_{x+1}} \otimes e^{\mathrm{i}\varepsilon Z_x Z_{x+1}} \otimes e^{-\mathrm{i}\varepsilon Z_x Z_{x+1}} \otimes e^{\mathrm{i}\varepsilon Z_x Z_{x+1}}$. This replicated gate has the diagrammatic representation
\begin{equation}
    \mathcal{U}_{x,x+1} \equiv \raisebox{-0.42\totalheight}{\includegraphics[height = 1.6cm]{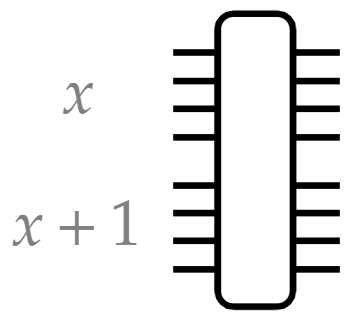}}\equiv\raisebox{-0.42\totalheight}{\includegraphics[height = 1.6cm]{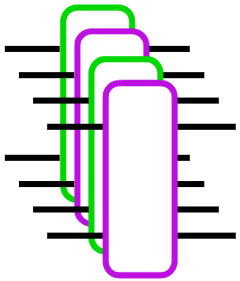}}.
\end{equation}
On each leg (labelled by the replica index $1,\overline{1},2,\overline{2}$ introduced in \eqnref{leg labelling convention} and by site position), the brick has the option of carrying either a $Z$ or a $\mathbb{1}$. If a leg is carrying a non-identity factor $A$, we say that the leg is \textit{decorated} and that the factor $A$ is the decoration. In this spirit, and using equations \ref{Uxxp1} and \ref{Ustarxxp1}, we find the decoration expansion of the brick to be given by
\begin{equation}\label{fullbrickdecomposition}
	\raisebox{-0.13\totalheight}{\includegraphics[height = 5.5cm]{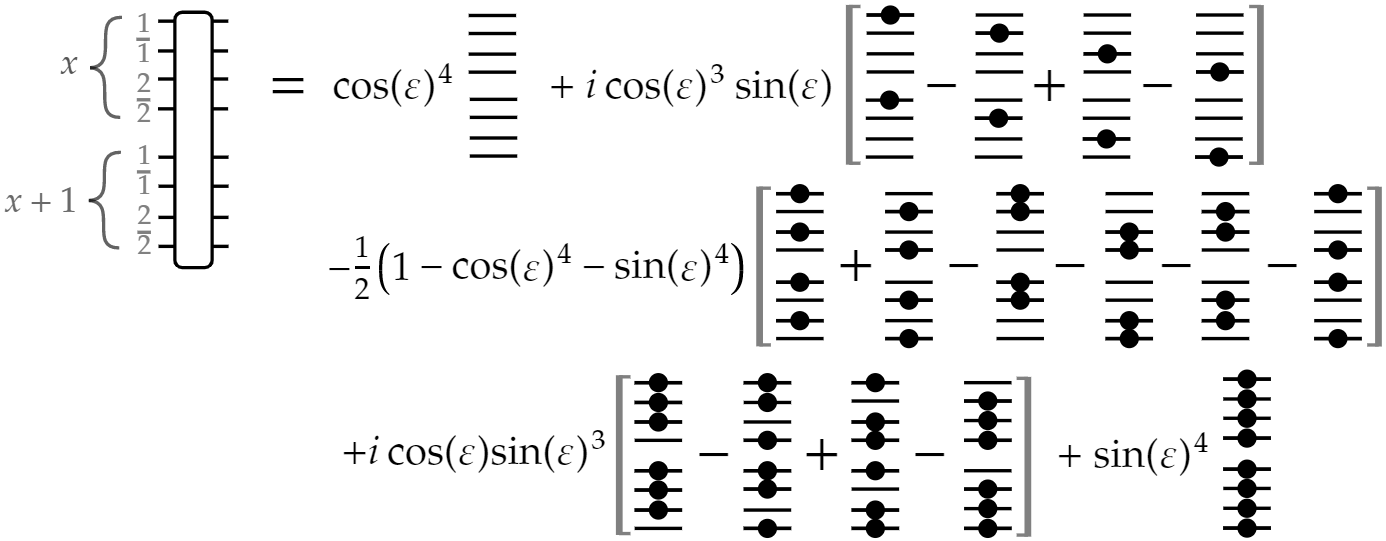}},
\end{equation}
Multiplying $\bra{F^x}$ by a layer of two-site gates yields the following,
\begin{equation}\label{FL}
	\centering
	\includegraphics[height = 2.1cm]{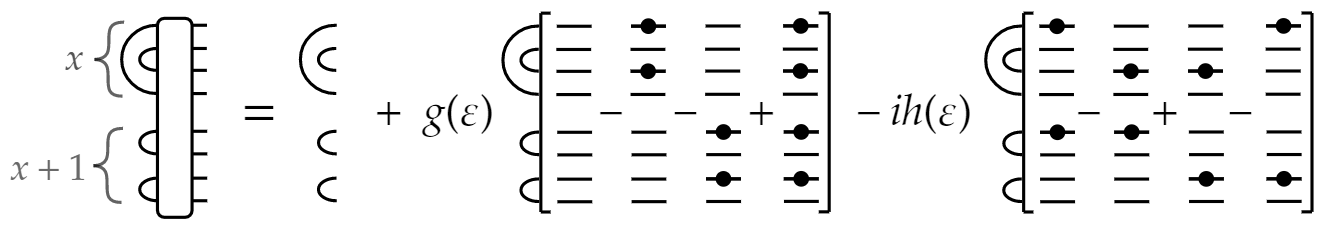},
\end{equation}
where 
\begin{equation}\label{g def}
    g(\varepsilon)\equiv\frac{\cos(4\varepsilon) - 1}{4}, \quad \textrm{and} \quad h(\varepsilon)\equiv\frac{\sin(4\varepsilon)}{4}.
\end{equation}
We have only depicted the sites $x$ and $x+1$ either side of the \textit{cut} (the domain wall between the $+$ and $-$ wiring configurations). Every brick $\mathcal{U}_{r,r+1}$ that does not straddle the cut is `absorbed' into the state using the property
\begin{equation}
    \raisebox{-0.45\totalheight}{\includegraphics[height = 1cm]{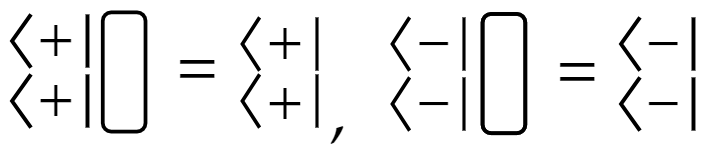}}
    \label{rules_away_from_cut}.
\end{equation}
To see this, we notice that the state $\bra{+}\otimes\bra{+}$ connects the replica $1$ with $\overline{2}$ and $2$ with $\overline{1}$ (see \eqnref{diagrammatic expression for plus and minus}), so that the two copies of $U_{r,r+1}$ each find a copy of $U_{r,r+1}^\dagger$ to yield $\bra{+}\otimes\bra{+}\mathcal{U}_{r,r+1}=\bra{+}\otimes\bra{+}$. The isometry of $\mathcal{U}_{r,r+1}$, $S=(1\leftrightarrow2)$, relates $\bra{+}$ and $\bra{-}$ through $\bra{\pm}S=\bra{\mp}$. Using this isometry, the first equation in \eqnref{rules_away_from_cut} implies the second.

The calculation of $\Omega(k)$ is then straight forward. By utilising translational symmetry and definition \eqnref{momentum space W} for the momentum space density super-operators $W^k$, we write
\begin{equation}
\Omega(k) \equiv \left(W^k\left|(\mathcal{U} - \mathbb{1}) \right|W^{k}\right) = \eta(k)(1-e^{-\mathrm{i}k})\sum_{x} q^x e^{-\mathrm{i}kx}\langle F^0 \left|(\mathcal{U} - \mathbb{1})\right|F^x\rangle.
\end{equation}
where $\eta(k) = \frac{1-q^{-2}e^{\mathrm{i}k}}{1-q^{-2}}$. When the cuts are misaligned ($x\neq0$), \eqnref{rules_away_from_cut} can be used to say $\bra{F^0}(\mathcal{U}-\mathbb{1})\ket{F^x}=\bra{F^0}(\mathcal{U}_{0,1}-\mathbb{1})\ket{F^x}=\bra{F^0}(\mathbb{1}-\mathbb{1})\ket{F^x}=0$. For aligned cuts ($x=0$), \eqnref{FL} yields $\bra{F^0}(\mathcal{U}-\mathbb{1})\ket{F^0}=g(\varepsilon)$.  $\Omega(k)$ is then succinctly given by
\begin{equation}
\Omega(k) = \eta(k)(1-e^{-\mathrm{i}k})g(\varepsilon).
\end{equation}
In \secref{MMF calc} we will see that the circuit averaged\footnote{A circuit average refers to the average of the Haar random unitary $V$} memory matrix is $\order{1/q^2}$. Therefore we can use $\Omega$ alone in \eqnref{formal expression v and D} to obtain a leading order expression for $v_B$ and $D$
\begin{equation}\label{infinite q results}
    v_0(\varepsilon) \equiv \lim_{q\to\infty}\langle v_B(\varepsilon) \rangle = \frac{1-\cos(4\varepsilon)}{4}, \quad \lim_{q\to\infty}\langle D(\varepsilon) \rangle = \frac{v_0(1-v_0)}{2}.
\end{equation}
A Straightforward calculation shows that brick-work circuits with commuting even and odd bricks (within one Floquet layer) have a strict light-cone of $v_{LC}=1$ (as opposed to $v_{LC}=2$ in brick-work circuits with non-commuting even and odd layers). Notably, the $\order{1}$ expression \eqnref{infinite q results} for $D$ vanishes when there is either no ballistic spreading ($v_B=0$) or when operators spread at the light-cone velocity ($v_B=1$), although it should be noted that $v_0$ can never approach this limit in this particular model, $0\leq v_0\leq 1/2$. This is reassuring as an operators spreading at the geometric light-cone velocity must have a front with zero width.

The operator spreading dynamics of this model occupies a markedly different regime than that of holographic/SYK physics which exhibit a sharp front and of random unitary circuits \cite{RvK17, Nahum17}, where the operator front diffusion constant is strongly suppressed at large $q$ and $v_B$ is close to the maximum velocity allowed by any (two-local) brick-work circuit (recent work \cite{Xu2019} investigates the crossover from holographic to random circuit behaviour of the front). In the large $q$ limit of our Floquet circuit, the operator front diffusion constant is roughly the same size as $v_B$, which itself is far from the light-cone velocity.

\section{Corrections from memory effects: a summary}\label{Summary}
We have so far calculated the contributions to information transport (the butterfly constant $v_B$ and operator front diffusion constant $D$) coming from slow processes only. In the remainder of this paper we perturbatively calculate the corrections to the butterfly velocity coming from the fast processes packaged in the memory matrix. The parameter that controls this perturbative expansion is $1/q$, and the corrections are a sum of real-time Feynman-like diagrams. This perturbative expansion encounters technical subtleties at large times where $q\to\infty$ asymptotic methods are insufficient for circuit averaging. Fortunately, we are able to resolve this subtlety by conjecturing that processes contributing to the memory matrix possess a natural exponential decay timescale $\tau(\varepsilon)\sim 1/|g(\varepsilon)|$\footnote{$g(\varepsilon)$ is as defined in \eqnref{g def}.} for the (circuit averaged) real-time memory matrix. We provide analytical evidence backing this conjecture. This allows us to argue that the problematic large-time contributions are in-fact subleading, enabling us to safely compute corrections from memory effects to $\order{1/q^2}$.

In section \ref{theorems}, we present the tools for the computation of memory effects at $\order{1/q^2}$. These tools are $q\to\infty$ scaling results for the Haar average of correlators and products of correlators. For proofs of these results see \cite{mcculloch2021haar}. In \secref{MMF calc}, we use these large $q$ results to compute $\Sigma(t)$, finding that the leading order contributions decay exponentially with a timescale $\tau(\varepsilon)\sim1/|g(\varepsilon)|$. Although only valid for times $t\lesssim q$, this result motivates our conjecture that the exponential decay of $\Sigma(t)$ persists to arbitrarily late times. A detailed discussion of late-time correlation functions and limitation of the large $q$ results of \secref{theorems} is given in \secref{late times}.

\begin{figure}[H]
    \centering
    \includegraphics[height = 6cm]{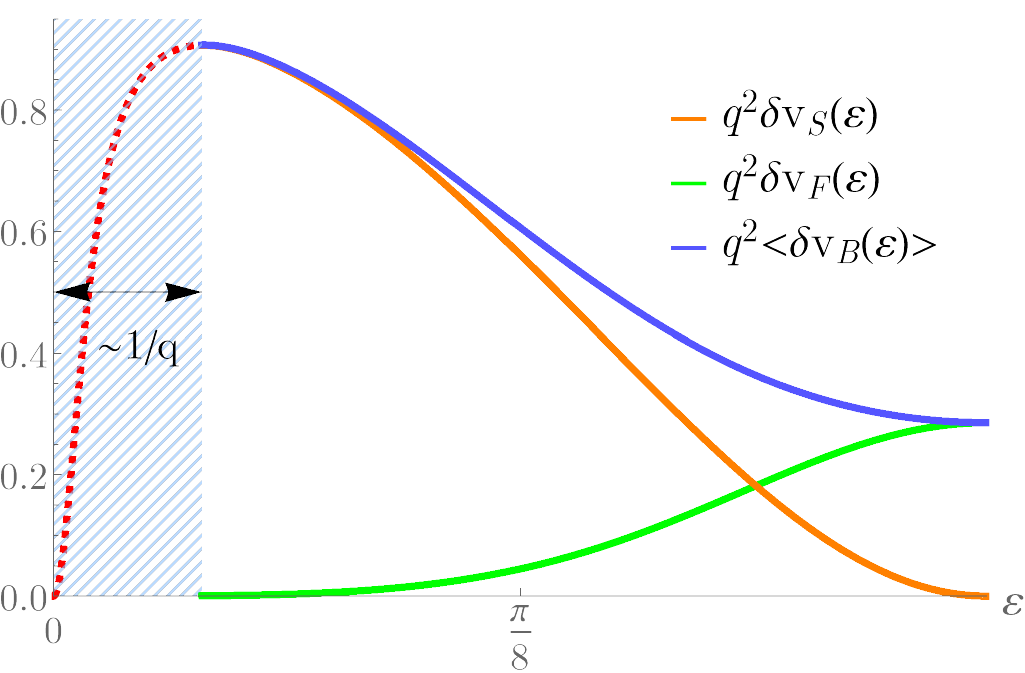}
    \caption{The contributions $\delta v_S(\varepsilon)$ and $\delta v_F(\varepsilon)$ to $\langle \delta v_B\rangle$. As $\varepsilon\to0$, the perturbative expansion in $1/q$ breaks down (see \secref{small epsilon}). $\delta v_B$ must rapidly approach zero, in \ref{independent small epsilon} we argue that this happen over an $\order{1/q}$ window.}
\label{delta_v_B_plot}
\end{figure}

Before going through the calculation of memory corrections in detail, we present the result of section \ref{MMF calc}, in the form of a circuit averaged butterfly velocity,
\begin{equation}\label{full velocity}
    \langle v_B \rangle = v_0(\varepsilon) + \delta v_S(\varepsilon) + \delta v_F(\varepsilon)  + \order{1/q^3}.
\end{equation}
where the memory corrections $\delta v_S(\varepsilon)$ and $\delta v_F(\varepsilon)$ are given at $\order{1/q^2}$ by
\begin{equation}
    \delta v_S(\varepsilon) =\frac{1}{q^2}\frac{1+5s-4s^2}{1-s-3s^2},\quad \delta v_F(\varepsilon) = 2\frac{g^2}{q^2}(\nu(\varepsilon)-f(\varepsilon)),
\end{equation}
with $s(\varepsilon)=\sin(\varepsilon)^2\cos(\varepsilon)^2$ and $\nu(\varepsilon)=[4(1-2s)(1-s(1-2s))]^{-1}$ and where $f(\varepsilon)$ is given to good approximation\footnote{$f(\varepsilon)$ is found by numerically evaluating a sum involving a $5\times5$ transfer matrix in \ref{(2,1) appendix} and the form for the fitting function is motivated in \secref{results}} (\figref{f_plot}) by
    \begin{equation}
        f(\varepsilon) = \frac{1}{7}s(\varepsilon)(1 - 4s(\varepsilon))^2(1 + 6.8 s + 16.1 s^2).
    \end{equation}
It turns out that the contribution $\delta v_S(\varepsilon)$ only arises because of the spatial ($S$) translation symmetry of the model: the scramblers in any particular Floquet layer are identical. On the other hand, $\delta v_F(\varepsilon)$ arises due to both the spatial and Floquet ($F$) symmetries in the problem. To see this, we analyse variations of the model without spatial translation and Floquet symmetry (see \appref{independent scramblers}). In \figref{delta_v_B_plot} we plot the different contributions to $\langle\delta v_B \rangle\equiv\langle v_B \rangle - v_0$.

Spatial disorder is often associated with localization, and a suppression of transport. On the other hand the spatiotemporal randomness in random circuits promotes ergodicity, and the rapid growth of operators. Our results for $v_B$ are more in agreement with the former intuition -- translational symmetry enhances transport -- because the butterfly velocity receives an enhancement $\delta v_S>0$ when spatial translation symmetry is present. Even less obvious is the role of Floquet symmetry; like translational symmetry, we find an enhancement $\delta v_F>0$ to $v_B$. It will be interesting to understand whether these effects are robust at higher orders in perturbation theory, and hold for more general time evolutions.

We now briefly discuss the difficulties with small $\varepsilon$. The $\varepsilon\to0$ limit represents an obvious sanity check on our results, but also represents a challenging limit in a memory matrix calculation. As $\varepsilon\to0$, the memory time must diverge, limiting our ability to truncate the memory effects. This is demonstrated in the failure of \eqnref{full velocity} for $v_B$ to vanish for $\varepsilon=0$ where sites decouple and the butterfly velocity is zero. As the memory time reaches $t\sim q$, or equivalently once $\varepsilon\sim 1/q$, our perturbative expansion in $1/q$ breaks down. Therefore, the corrections $\langle \delta v_B\rangle$ as given in \eqref{full velocity} are valid only for $\varepsilon > \order{1/q}$, below which $\langle \delta v_B \rangle$ must quickly go to zero as shown in \figref{delta_v_B_plot}. We discuss this in detail in \secref{small epsilon}.

\section{Calculating corrections from memory effects}\label{dedicated memory effect section}
In this section we first present the large $q$ scaling results that form the backbone of the perturbative expansion of $\Sigma(t)$, before presenting a intricate booking keeping of the $\order{1/q^2}$ contributions in \secref{MMF calc}.
\subsection{Averaging \texorpdfstring{$n$}{Lg}-point correlation functions and their moments}\label{theorems}
In this section we will present several useful theorems concerning the Haar averages of correlators and products of correlators with random unitary dynamics.
The reader is directed to \cite{mcculloch2021haar} for the proofs of the theorems below. We need only consider correlation functions involving $Z$'s:
\begin{equation}\label{eq:atodefn}
    \langle \mathcal{Z}(\boldsymbol{t})\rangle \equiv \langle Z(t_1) \cdots Z(t_n)\rangle = \frac{1}{q}\Tr[ Z(t_1) \cdots Z(t_n) ],
\end{equation}
where $\boldsymbol{t}=\left(t_1,\cdots,t_n\right)$ and $\mathcal{Z}(\boldsymbol{t})=Z(t_1)\cdots Z(t_n)$ is a product of $n$ `scrambled' Pauli $Z$ matrices $Z(t)=V^t Z V^{-t}$ (with a Haar random unitary $V$). We call correlators of form \eqnref{eq:atodefn} arbitrary-time-ordered (ATO) $n$-point correlation functions as the times $t_i$ are not forced to be ordered. A correlator is trivial if $\mathcal{Z}(\boldsymbol{t})=\mathbb{1}$. For the purposes of this section, we assume that none of the correlators are trivial.

Before we present the theorems, let us introduce what we call the decoration delta constraint, $\delta^{A,B}$. This is zero unless the  two operators $A=Z(a_1)\cdots Z(a_n)$ and $B=Z(b_1)\cdots Z(b_n)$ are equal, $AB^{-1}=\mathbb{1}$, for \emph{all} scrambling unitaries $V$. If the strings $\boldsymbol{a}=(a_1,\cdots,a_n)$ and $\boldsymbol{b}=(b_1,\cdots,b_{n'})$ do not have repeated consecutive times, then the delta constraint simply checks that $\boldsymbol{a}=\boldsymbol{b}$. Otherwise, we need to introduce a concept called the \textit{minimal form} of a operator.
\begin{definition}[Minimal form]\label{minimal form}
With $\mathcal{Z}(\boldsymbol{t})=Z(t_1)\cdots Z(t_n)$ as a product of scrambled Pauli Z operators and $\mathcal{Z}(\boldsymbol{t}')=Z(t'_1)\cdots Z(t'_n)$ as the form reached after exhaustively using the property $Z(t)^2=\mathbb{1}$, we define the minimal form of $\mathcal{Z}(\boldsymbol{t})$ as $\textrm{Min}(\mathcal{Z}(\boldsymbol{t}))\equiv\mathcal{Z}(\boldsymbol{t}')$. We also define the minimal form of the string $\boldsymbol{t}$ as $\textrm{Min}(\boldsymbol{t})\equiv\boldsymbol{t}'$.
\end{definition}
We can then give a more general definition of the delta constraint.
\begin{definition}[Decoration delta constraint]\label{decoration delta constraint}
The delta constraint $\delta^{A,B}$ on operators $A=Z(a_1)\cdots Z(a_n)$ and $B=Z(b_1)\cdots Z(b_n)$ is defined by
\begin{equation}
    \delta^{A,B}=\begin{cases}
    1, \quad \textrm{if } \textrm{Min}(\boldsymbol{a}) = \textrm{Min}(\boldsymbol{b})\\
    0, \quad \textrm{otherwise}.
    \end{cases}
\end{equation}
\end{definition}

\renewcommand{\thethm}{\arabic{thm}}

\begin{thm}\label{productofcorrelators}
	The Haar average of a product of $p$ ATO correlators has the following scaling behaviour,
	\begin{equation}
	    \int dU \langle \mathcal{Z}(\boldsymbol{t}^1)\rangle\cdots \langle\mathcal{Z}(\boldsymbol{t}^p)\rangle = \order{1/q^{2\lfloor p/2 \rfloor}} \quad \textrm{as $q\to\infty$}.
	\end{equation}
\end{thm}

\begin{thm}\label{two-correlators}
    The Haar average of a product of two ATO correlators is given by
    \begin{equation}
        \int dU \langle \mathcal{Z}(\boldsymbol{t}) \rangle \langle \mathcal{Z}(\boldsymbol{t}') \rangle^* = \frac{S(\boldsymbol{t})}{q^2}\sum_{\tau\in\mathbb{Z}}\delta^{\mathcal{Z}(\boldsymbol{t+\tau}),\mathcal{Z}(\boldsymbol{t}')} + \order{1/q^{3}} \quad \textrm{as $q\to\infty$},
    \end{equation}
\end{thm}
where we have used the decoration delta constraint and where the sum over $\tau$ allows for $\boldsymbol{t}$ and $\boldsymbol{t}'$ to differ by a global shift in time. The symmetry factor $S(\boldsymbol{t})$ counts the degree of cyclic symmetry of the list of times $\boldsymbol{t}$, if there exists $n$ cyclic permutations $\alpha$ such that $\alpha(\boldsymbol{t})=\boldsymbol{t}$, then $S(\boldsymbol{t})=n$.

We will often study a special subset of ATO correlators, which we dub \emph{physical OTOCs}. These take the form $\langle Z \Gamma_1 Z(T) \Gamma_{\overline{2}}^\dagger Z \Gamma_2 Z(T) \Gamma_{\overline{1}}^\dagger \rangle$, where $\Gamma_i=Z(1)^{s^i_1}Z(2)^{s^i_2}\cdots Z(T-1)^{s^i_{T-1}}$ for binary strings $\boldsymbol{s}^i=(s^i_1,\cdots,s^i_{T-1})$.

\begin{thm}\label{OTOCtheorem}
	The Haar average of a physical OTOC is given by
	\begin{align}
        \int dV \langle Z \Gamma_1 Z(T) \Gamma_{\overline{2}}^\dagger Z \Gamma_2 Z(T) \Gamma_{\overline{1}}^\dagger \rangle = \frac{1}{q^2}&\left(\delta^{\Gamma_1\Gamma_{\overline{1}}^\dagger\Gamma_{2}\Gamma_{\overline{2}}^\dagger,\mathbb{1}} - \delta^{\Gamma_1,\Gamma_{\overline{2}}}\delta^{\Gamma_2,\Gamma_{\overline{1}}} -\delta^{\Gamma_1,\Gamma_{\overline{1}}}\delta^{\Gamma_2,\Gamma_{\overline{2}}}\right)\nonumber\\
        &+ \order{1/q^3},
    \end{align}
\end{thm}
where we have again used the decoration delta constraint. This result relies on theorem 4 of \cite{mcculloch2021haar} and is obtained in \eqnref{OTOC theorem derived} of \ref{decoration delta constraint appendix}.

\subsection{Computing \texorpdfstring{$\Sigma$}{Lg} at \texorpdfstring{$\mathcal{O}(1/q^2)$}{Lg}}\label{MMF calc}
We now compute the memory matrix at leading order in $1/q$, enabling us to calculate the butterfly velocity to $\mathcal{O}(1/q^2)$. As discussed in \secref{MMF for Floquet models}, we calculate the proxy $\sigma(k,z)$ instead of $\Sigma(k,z)$. 
We will find that the $\order{1/q^2}$ contributions to $\sigma(k,t)$ decays exponentially fast with a decay-rate $\gamma(\varepsilon)\approx 2\abs{g(\varepsilon)}$ set by the interaction strength $\varepsilon$. 

It is convenient to express $\sigma(k,z)$ as
\begin{equation}
    \sigma(k,z)=\eta(k)(1-e^{-\mathrm{i}k})\mathcal{D}(k,z),
\end{equation}
where
\begin{equation}
    \mathcal{D}(k,z) = \sum_{x}q^x e^{-ikx}\mathcal{D}(x,z), \quad \mathcal{D}(x,z)= q^x\bra{F^0}\mathcal{L}\mathcal{Q}\frac{1}{e^{-\mathrm{i}z}-1-\mathcal{L}}\mathcal{Q}\mathcal{L}\ket{F^x}.
\end{equation}
Corrections to $v_B$ from fast processes (i.e., processes counted by $\Sigma$) are then given by
\begin{equation}\label{correction}
    \delta v = \lim_{z\to i0^+}\lim_{k\to0}i\sigma(k,z)/k = -\mathcal{D}(k=0,z=0).
\end{equation}
We will often work with the real time version of $\mathcal{D}$,
\begin{equation}\label{Ddef}
    \mathcal{D}(x,T) = q^x\bra{F^0}\mathcal{L}\mathcal{Q}\mathcal{U}^{T-1}\mathcal{Q}\mathcal{L}\ket{F^x}.
\end{equation}
We separate the scrambling part of each Floquet layer from the two-local bricks as before; $\mathcal{U}=\mathcal{V}\mathcal{U}_{Z}$. Then a product $\mathcal{U}^n$ can be written $\mathcal{U}^n=\mathcal{U}_Z(1)\mathcal{U}_Z(2)\cdots\mathcal{U}_Z(n)\mathcal{V}^n$ where $\mathcal{U}_Z(t)=\mathcal{V}^t \mathcal{U}_Z \mathcal{V}^{-t}$. One consequence is that
\begin{equation}
    \bra{F^0}\mathcal{L}=\bra{F^0}\mathcal{U}-\bra{F^0}\mathbb{1}=\bra{F^0}\mathcal{V}\mathcal{U}_Z-\bra{F^0}=\bra{F^0}\mathcal{L}_Z
\end{equation}
where $\mathcal{L}_Z=\mathcal{U}_Z-\mathbb{1}$. We can similarly show $\mathcal{L}\ket{F^x}=\mathcal{V}\mathcal{L}_Z\ket{F^x}$. Using this and $\left[\mathcal{V},\mathcal{Q}\right]=0$ and simplifying the notation $\mathcal{L}_Z\to L$ and $\mathcal{U}_Z\to U$, $\mathcal{D}(x,T)$ can be written as
\begin{equation}
    \mathcal{D}(x,T) = q^x\bra{F^0}L\mathcal{Q}U(1)\cdots U(T-1)\mathcal{Q}L(T)\ket{F^x},
\end{equation}
Using the decoration expansion for each brick \eqnref{fullbrickdecomposition}, we express $\mathcal{D}(x,T)$ as a sum over decorations $\Gamma$,
\begin{equation}\label{decexpansion}
    \mathcal{D}(x,T)=\sum_\Gamma C_\Gamma \mathcal{D}_\Gamma(x,T), \quad \mathcal{D}_\Gamma(x,T)=q^x\bra{F^0}L\mathcal{Q}\Gamma \mathcal{Q}L(T)\ket{F^x}.
\end{equation}
$\Gamma=\bigotimes_{r}\,\Gamma^r$ is a product of decorations on every site. The decoration on some site $r$ is given by $\Gamma^r=\Gamma^r_1\otimes\Gamma^{r*}_{\overline{1}}\otimes\Gamma^r_2\otimes\Gamma^{r*}_{\overline{2}}$ (i.e., a product of decorations on each leg of the site) and where $\Gamma^r_i=Z(1)^{s^{r,i}_{1}}Z(2)^{s^{r,i}_{2}}\cdots Z(T-1)^{s^{r,i}_{T-1}}$ for a binary string $\boldsymbol{s}^{r,i}=(s^{r,i}_{1},\cdots,s^{r,i}_{T-1})$. We will use the decoration expansion, and a carefully chosen decomposition of the initial states to express contributions as products of ATO'. We will use the theorems \ref{productofcorrelators} - \ref{OTOCtheorem} to see what kinds of decorations can give rise to $O(1/q^2)$ corrections to the circuit averaged $\sigma$, and then evaluate those. It turns out that only certain values of $x$ are relevant, $x=0,1,2$ at this order, as we will see.

Using the inversion symmetry of the Floquet unitary, we find
\begin{equation}\label{x<0}
    \mathcal{D}(-x,T)=q^{-2x} \mathcal{D}(x,T), \quad \textrm{for $x>0$}.
\end{equation}
We conclude that the contributions from $x<0$ are at least a factor $1/q^2$ smaller than the $x> 0$ contributions. In the remainder of this section we will see that the contributions for $x\geq 0$ are no larger than $\order{1/q^2}$, and that therefore $\mathcal{D}(x<0,T)=\order{q^{-4}}$. We will therefore only consider $x\geq 0$ in the following sections.

To evaluate $\mathcal{D}_\Gamma$, we first project $L(T)\ket{F^x}$ onto the fast space.
\begin{equation}\label{projected state}
    \begin{matrix}\mathcal{Q}L(T)\ket{F^x}=g\left[-
    \begin{matrix}
    \\
    \
    \end{matrix}
    \right.\\
    \
    \end{matrix}
    \begin{matrix}
    \ket{\phi_+(T)}\\ 
    \ket{-}\\
    \textcolor{gray}{\textrm{1}}
    \end{matrix}
    \begin{matrix}\ - \
    \begin{matrix}
    \\
    \
    \end{matrix}
    \\
    \
    \end{matrix}
    \begin{matrix}
    \ket{+}\\ 
    \ket{\phi_-(T)}\\
    \textcolor{gray}{\textrm{2}}
    \end{matrix}
    \begin{matrix}\ + \
    \begin{matrix}
    \\
    \
    \end{matrix}
    \\
    \
    \end{matrix}
    \begin{matrix}
    Z(T)^{\otimes 2}\ket{+}\\ 
    Z(T)^{\otimes 2}\ket{-}\\
    \textcolor{gray}{\textrm{3}}
    \end{matrix}
    \begin{matrix}
    \left.
    \begin{matrix}
    \\
    \
    \end{matrix}
    \right]
    \\
    \
    \end{matrix}
    \begin{matrix}
    -\mathrm{i}h
    \begin{matrix}
    \\
    \
    \end{matrix}
    \\
    \
    \end{matrix}
    \begin{matrix}
    K(T)\ket{+}\\ 
    K(T)\ket{-}\\
    \textcolor{gray}{\textrm{4}}
    \end{matrix} \ \
    \begin{matrix}
    \textcolor{gray}{\textrm{site } x}\\ 
    \textcolor{gray}{\textrm{site } x+1}\\
    \
    \end{matrix}
\end{equation}

where we have suppressed sites $r<x$ ($r>x+1$) and carefully chosen an orthogonal decomposition of $QL(T)\ket{F^x}$ in terms of four fast states, numbered from $1$ to $4$. We have used the shorthand $Z^{\otimes 2}$ and $K$ to represent the following,
\begin{equation}\label{Kdef_Zsquareddef}
	\raisebox{-0.45\totalheight}{\includegraphics[height = 1.1cm]{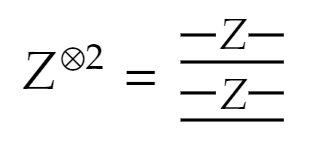}}, \
	\raisebox{-0.45\totalheight}{\includegraphics[height = 1.1cm]{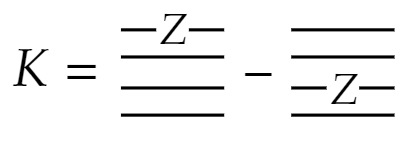}}.
\end{equation}
$Z(T)^{\otimes 2}$ and $K(T)$ are defined identically but with $Z(T)$ in place of $Z$. Finally, $\ket{\phi_+(T)}$ and $\ket{\phi_-(T)}$ are given by
\begin{equation}\label{phi}
    \ket{\phi_+(T)}=Z(T)^{\otimes 2}\ket{+} - \frac{1}{q-q^{-1}}\ket{\perp}, \quad
    \ket{\phi_-(T)}=Z(T)^{\otimes 2}\ket{-} - \frac{1}{q-q^{-1}}\ket{0},
\end{equation}
where $\ket{\perp} = \ket{-}-\frac{1}{q}\ket{+}$. The initial state is easily found using $\bra{F^0}L\mathcal{Q}=\left(\mathcal{Q} L\ket{F^0}\right)^T$. 

The four states numbered in \eqnref{projected state} obey a useful set of identities, which allow us to identify and discard many lower order diagrams and significantly simplify the memory matrix calculation.

\subsubsection{Identities of the \texorpdfstring{$\phi_{\pm}$}{Lg} states}
It will be useful to determine some properties of $\ket{\phi_+}$ and $\ket{\phi_-}$. The isometry $S=(1\leftrightarrow2)$ (swaps legs $1$ and $2$) relates the two vectors, $S\ket{\phi_-}=\ket{\phi_+}$. We can then investigate $\ket{\phi_-}$ only. Firstly, $\ket{\phi_-}$ has no overlap with either $\ket{+}$ or $\ket{-}$.
\begin{align}
    \bra{-}\ket{\phi_-}&=\bra{-}Z^{\otimes 2}\ket{-}-\frac{1}{q-q^{-1}}\bra{-}\ket{0}=\bra{-}Z^{\otimes 2}\ket{-}=\raisebox{-0.38\totalheight}{\includegraphics[height = 1cm]{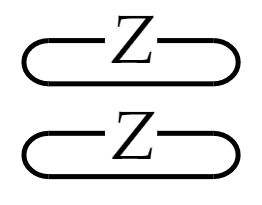}}=0\nonumber\\
    \bra{+}\ket{\phi_-}&=\bra{+}Z^{\otimes 2}\ket{-}-\frac{1}{q-q^{-1}}\bra{+}\ket{0}=\raisebox{-0.4\totalheight}{\includegraphics[height = 1.1cm]{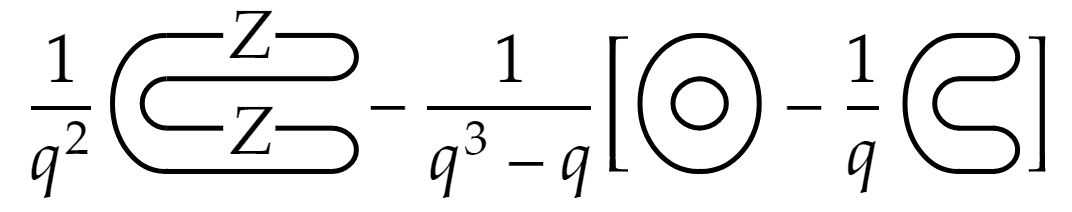}}\nonumber\\
    &=\frac{1}{q}-\frac{1}{q-q^{-1}}(1-q^{-2})=0.\nonumber
\end{align}
Using $S$, we can then write
\begin{equation}\label{nodecphi}
    \bra{\phi_+}\ket{\pm}=\bra{\phi_-}\ket{\pm}=\bra{\pm}\ket{\phi_{+}(T)}=\bra{\pm}\ket{\phi_{-}(T)}=0.
\end{equation}
We next consider what happens when the wires of $\bra{\pm}$ are decorated. Let $\Gamma^r$ be a decoration on the four legs of some site $r$, i.e., $\Gamma^r=\Gamma^r_1\otimes\Gamma^{r*}_{\overline{1}}\otimes\Gamma^r_2\otimes\Gamma^{r*}_{\overline{2}}$, where each $\Gamma^r_i$ take the form $\Gamma_i=Z(1)^{s^i_1}\cdots Z(T-1)^{s^i_{T-1}}$ for some bit-string $\boldsymbol{s}_i$. This is shown graphically below.
\begin{equation}
    \Gamma^r=\raisebox{-0.45\totalheight}{\includegraphics[height=2.6cm]{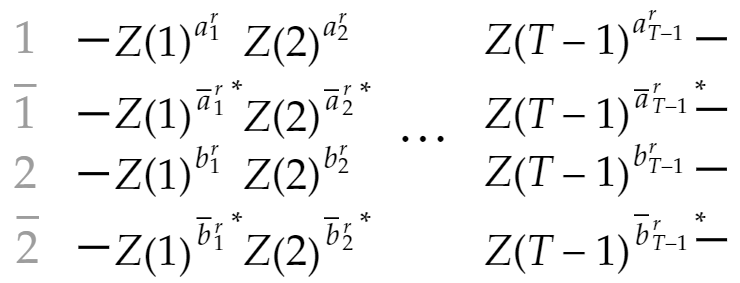}}.
\end{equation}
The overlaps between $\ket{\phi_-(T)}$ and a decorated $\bra{+}$ state is then given by
\begin{align}
    \bra{+}\Gamma^r\ket{\phi_-(T)}&=\raisebox{-0.45\totalheight}{\includegraphics[height = 2cm]{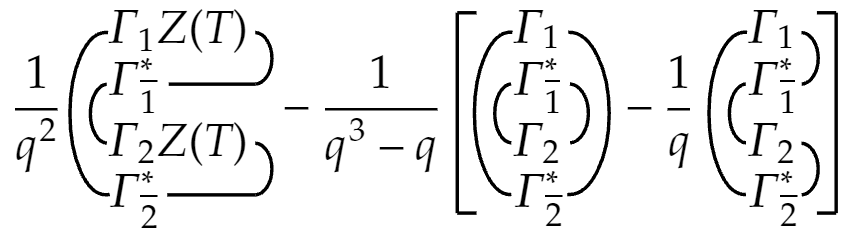}}\nonumber\\
    &=\frac{1}{q}\langle Z(T)\Gamma^{r\dagger}_{\overline{2}}\Gamma^r_1 Z(T)\Gamma^{r\dagger}_{\overline{1}} \Gamma^r_2\rangle\nonumber\\
    & \quad \quad -\frac{1}{q-q^{-1}}\left(\langle \Gamma^{r\dagger}_{\overline{2}}\Gamma^r_1\rangle \langle \Gamma^{r\dagger}_{\overline{1}}\Gamma^r_2\rangle-\frac{1}{q^2}\langle\Gamma^{r\dagger}_{\overline{2}}\Gamma^r_1\Gamma^{r\dagger}_{\overline{1}}\Gamma^r_2\rangle\right).\label{plusphiminus}
\end{align}
Similar identities hold for the overlaps $\bra{+}\ket{\phi_-(T)}$ and $\bra{\phi_-}\ket{\pm}$. These  identities  can  be  summarised  as  follows
\begin{align}
    \bra{-}\Gamma^r\ket{\phi_-(T)}&=\langle Z(T)\Gamma^{r\dagger}_{\overline{1}}\Gamma^r_1\rangle \langle Z(T)\Gamma^{r\dagger}_{\overline{2}}\Gamma^r_2\rangle-\frac{1}{q^2-1}\left(\langle\Gamma^{r\dagger}_{\overline{1}}\Gamma^r_1\Gamma^{r\dagger}_{\overline{2}}\Gamma^r_2\rangle-\langle\Gamma^{r\dagger}_{\overline{1}}\Gamma^r_1\rangle\langle\Gamma^{r\dagger}_{\overline{2}}\Gamma^r_2\rangle\right),\nonumber\\
    \bra{\phi_-}\Gamma^r\ket{+}&=\frac{1}{q}\langle Z\Gamma^r_1\Gamma^{r\dagger}_{\overline{2}} Z(T)\Gamma^r_2\Gamma^{r\dagger}_{\overline{1}}\rangle-\frac{1}{q-q^{-1}}\left(\langle \Gamma^r_1\Gamma^{r\dagger}_{\overline{2}}\rangle \langle \Gamma^r_2\Gamma^{r\dagger}_{\overline{1}}\rangle-\frac{1}{q^2}\langle\Gamma^r_1\Gamma^{r\dagger}_{\overline{2}}\Gamma^r_2\Gamma^{r\dagger}_{\overline{1}}\rangle\right),\nonumber\\
    \bra{\phi_-}\Gamma^r\ket{-}&=\langle Z(T)\Gamma^r_1\Gamma^{r\dagger}_{\overline{1}}\rangle \langle Z(T)\Gamma^r_2\Gamma^{r\dagger}_{\overline{2}}\rangle-\frac{1}{q^2-1}\left(\langle\Gamma^r_1\Gamma^{r\dagger}_{\overline{1}}\Gamma^r_2\Gamma^{r\dagger}_{\overline{2}}\rangle-\langle\Gamma^r_1\Gamma^{r\dagger}_{\overline{1}}\rangle\langle\Gamma^r_2\Gamma^{r\dagger}_{\overline{2}}\rangle\right).\label{minusphiminus}
\end{align}
In general, $\Gamma^{r}$ will insert operators on each of the four legs of the input state. However, sometimes one can utilize the wirings between each of the legs to simplify the resulting expression, an example for $\bra{+}\Gamma^r$ is shown below,
\begin{equation}
    \raisebox{-0.45\totalheight}{\includegraphics[height=2cm]{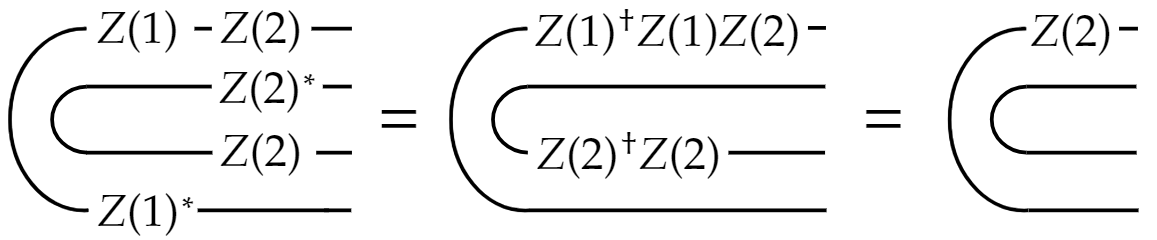}}.
\end{equation}
We say $\Gamma^{r}$ decorates the state if this simplification process cannot be used to remove all four components of $\Gamma^{r}$. In the example above we were able to remove all of the non-identity operators from the $(2,\overline{1})$ wiring but not from the $(1,\overline{2})$ wiring. Notice that in every case in \eqnref{minusphiminus} and in \eqnref{plusphiminus}, if either of the wirings in the $+/-$ states carry a non-identity operator, the overlap with the $\phi_{\pm}$ states vanishes\footnote{This is easily verified in \eqnref{plusphiminus} by substituting either $\Gamma^{r\dagger}_{\overline{2}}\Gamma^r_1=\mathbb{1}$ or $\Gamma^{r\dagger}_{\overline{1}}\Gamma^r_2=\mathbb{1}$}.

Assuming that the decorations non-trivially decorate both wirings of the $+/-$ states, these overlaps can be summarised as follows
\begin{align}
    q\bra{\pm}\Gamma^r\ket{\phi_\mp(T)}&=
    \textrm{OTOC}-\textrm{Corr}\times\textrm{Corr}'+\order{1/q^2}\nonumber\\
    q\bra{\phi_\pm}\Gamma^r\ket{\mp}&=
    \textrm{OTOC}-\textrm{Corr}\times\textrm{Corr}'+\order{1/q^2}\nonumber\\
    \bra{\pm}\Gamma^r\ket{\phi_\pm(T)}&=
    \textrm{Corr}\times\textrm{Corr}'+\order{1/q^2}\nonumber\\
    \bra{\phi_\pm}\Gamma^r\ket{\pm}&=
    \textrm{Corr}\times\textrm{Corr}'+\order{1/q^2}.\label{dec phi overlaps}
\end{align}
Where rather than give the full expressions we have simply presented the types of contributions (i.e., OTOCs, products of non-trivial correlators or terms that are manifestly $\order{1/q^2}$). This is often enough to identify diagrams that contribute to $\mathcal{D}(x,T)$ at $\order{1/q^3}$. In cases that require a more careful analysis we refer to \eqnref{plusphiminus} and \eqnref{minusphiminus}.

These are useful identities because the diagrams contributing to the memory kernel tend to involve products of terms of this form. We will now see how, using our OTOC identities from Sec 5, this result allows us to pinpoint which diagrams are able to contribute at leading order in $\order{1/q^2}$

\subsubsection{\texorpdfstring{$\mathcal{D}^{a,b}(x,T)$}{Lg}}
Casting our attention back to the orthogonal decomposition of the projected vector $\mathcal{Q}L(T)\ket{F^x}$ in \eqnref{projected state} where labelled each of four orthogonal states from $1$ to $4$, we now use a short hand $\{\ket{1,x,T},\cdots,\ket{4,x,T}\}$ to denote each of these states. This is also done for $\bra{F^0}L\mathcal{Q}$. Using this, we define the following quantity,
\begin{equation}\label{Dabdef}
    \mathcal{D}^{a,b}(x,T) \equiv q^x\bra{a,0,0}U(1)U(2)\cdots U(T-1)\ket{b,x,T},
\end{equation}
and also the decoration expansion quantity,
\begin{equation}
    \mathcal{D}^{a,b}_\Gamma(x,T) = q^x\bra{a,0,0}\Gamma\ket{b,x,T}.
\end{equation}
The decoration $\Gamma$ is a product of $T-1$ \textit{decoration layers}, $\Gamma=\Gamma(1)\Gamma(2)\cdots\Gamma(T-1)$, one for each unitary layer $U(t)$ of \eqnref{Dabdef}.

In what follows, we examine $\mathcal{D}^{a,b}$ for all possible pairs $a,b$; some calculations are carried out in \ref{(2,1) appendix} and \ref{Remaining (a,b)}. All contributions are $\leq \mathcal{O} (1/q^3)$, except for the $(2,1)$ and $(4,4)$ terms as summarised in \ref{summary table}.

\subsubsection{\texorpdfstring{$(a,b)=(1,1),(2,2)$}{Lg} are \texorpdfstring{$\mathcal{O}(1/q^3)$}{Lg}}
The arguments used for $(a,b)=(1,1)$ are the same as used for $(2,2)$, for brevity we will only present them for $(1,1)$. We study $x>0$ and $x=0$ separately, writing $\mathcal{D}_\Gamma^{1,1}(x,T)$ as diagram in both cases.
\begin{itemize}
    \item $x>0$:
    \begin{equation}
        \mathcal{D}^{1,1}_\Gamma(x>0,T)= g^2\times \left[ \ \begin{matrix}
        \\
        \vspace{-0.7cm}
        \\
    \textcolor{gray}{\textrm{site } 0}\\ 
    \textcolor{gray}{\textrm{site } 1}\\
    \\ \\ \\ \\
    \end{matrix} \ \ 
    \begin{matrix}
    \ \ \vdots\\
    \ \ \hspace{-5pt}\bra{\phi_+}\\ 
    q\bra{-}\\
    \ \ \vdots\\
    q\bra{-}\\
    \ \ \bra{-}\\
    \ \ \vdots
    \end{matrix}
    \ \fbox{ $\begin{matrix}
    \\
    \\
    \\
    \hspace{-4pt}\Gamma\\
    \\
    \\
    \\
    \end{matrix}$}
    \
    \begin{matrix}
    \hspace{-20pt}\vdots\\
    \hspace{-20pt}\ket{+}\\
    \hspace{-20pt}\ket{+}\\
    \hspace{-20pt}\vdots\\
    \ket{\phi_+(T)}\\ 
    \hspace{-20pt}\ket{-}\\
    \hspace{-20pt}\vdots\\
    \end{matrix} \ \ 
    \begin{matrix}
    \\ \\ \\ \\ \vspace{-0.2cm}
    \\
    \hspace{-10pt}\textcolor{gray}{\textrm{site } x}\\ 
    \hspace{-10pt}\textcolor{gray}{\textrm{site } x+1}\\
    \\
    \end{matrix}\right]\nonumber
    \end{equation}
    Using the overlap identities \eqnref{nodecphi}, \eqnref{plusphiminus} and \eqnref{minusphiminus}, we see that the contribution from site $0$ and $x$ either vanish or have the form
    \begin{equation}
        \left(\textrm{Corr}_0\times\textrm{Corr}'_0+\order{1/q^2}\right)\left(\textrm{OTOC}_x+\textrm{Corr}_x\times\textrm{Corr}'_x+\order{1/q^2}\right).\nonumber
    \end{equation}
    Every other site may contribute either trivial or non-trivial correlators to the product. Therefore, after Haar averaging, theorem \ref{productofcorrelators} of \secref{theorems} gives 
    \begin{equation}
        \int dV \mathcal{D}^{1,1}_\Gamma(x>0,T)=\order{1/q^3}.
    \end{equation}
    
    \item $x=0$:
    \begin{equation}
    \mathcal{D}^{1,1}_\Gamma(0,T)=g^2\times \left[ \
    \ \textcolor{gray}{\textrm{site } 0} \ \ 
    \begin{matrix}
    \ \ \vdots\\
    \ \bra{+}\\
    \bra{\phi_+}\\ 
    \ \bra{-}\\
    \ \ \vdots
    \end{matrix} \ 
    \fbox{ $\begin{matrix}
    \\
    \\
    \hspace{-5pt}\Gamma\\
    \\
    \\
    \end{matrix}$} \ 
    \begin{matrix}
    \hspace{-20pt}\vdots\\
    \hspace{-20pt}\ket{+}\\
    \ket{\phi_+(T)}\\ 
    \hspace{-20pt}\ket{-}\\
    \hspace{-20pt}\vdots
    \end{matrix}\right]\nonumber
    \end{equation}
    On site $x=0$, we have
    \begin{align}
        \bra{\phi_+}\Gamma^0\ket{\phi_+(T)} = \ &\langle Z\Gamma^0_1 Z(T)\Gamma^{0\dagger}_{\overline{2}}\rangle \langle Z\Gamma^0_2 Z(T)\Gamma^{0\dagger}_{\overline{1}}\rangle - \frac{1}{q^2-1}\langle \Gamma^{0\dagger}_{\overline{1}}\Gamma^0_1 Z(T) \Gamma^{0\dagger}_{\overline{2}}\Gamma^0_2 Z(T)\rangle \nonumber\\ 
        &- \frac{1}{q^2-1}\langle Z\Gamma^0_1\Gamma^{0\dagger}_{\overline{1}} Z\Gamma^0_2\Gamma^{0\dagger}_{\overline{2}} \rangle
        + \frac{1}{q^2(1-q^{-2})^2}\langle\Gamma^0_1\Gamma^{0\dagger}_{\overline{1}}\rangle \langle\Gamma^0_2\Gamma^{0\dagger}_{\overline{2}}\rangle \nonumber \\
        &+\frac{1}{q^2-1}\langle Z\Gamma^0_1\Gamma^{0\dagger}_{\overline{2}}\rangle\langle Z \Gamma^0_2\Gamma^{0\dagger}_{\overline{1}}\rangle
        +\frac{1}{q^2-1}\langle \Gamma^{0\dagger}_{\overline{2}}\Gamma^0_1 Z(T)\rangle\langle \Gamma^{0\dagger}_{\overline{1}}\Gamma^0_2 Z(T)\rangle\nonumber\\
        &-\frac{1}{(q^2-1)^2}\langle \Gamma^0_1\Gamma^{0\dagger}_{\overline{1}}\Gamma^0_2\Gamma^{0\dagger}_{\overline{2}}\rangle
        -\frac{1}{(q^2-1)^2}\langle \Gamma^0_1\Gamma^{0\dagger}_{\overline{2}}\Gamma^0_2\Gamma^{0\dagger}_{\overline{1}}\rangle\nonumber\\
        &+\frac{1}{(q^2-1)^2}\langle \Gamma^0_1\Gamma^{0\dagger}_{\overline{2}}\rangle\langle\Gamma^0_2\Gamma^{0\dagger}_{\overline{1}}\rangle.
    \end{align}
    The final three terms are manifestly $\order{1/q^{4}}$. The fifth and sixth terms are of the form $\textrm{Corr}\times\textrm{Corr}'/q^2$, where these correlators are non-trivial. Therefore, using theorem \ref{productofcorrelators}, the Haar average of these terms (possibly multiplied by additional non-trivial correlators from other sites) is $\order{1/q^{4}}$. The second, third and fourth terms all have pre-factors of $1/q^2$; if they are to contribute at this order, the accompanying correlators must be trivial. Using decoration delta constraints, this fact (a consequence of theorem \ref{productofcorrelators}) is written below
    \begin{equation}
        \int dV \frac{1}{q^2}\langle \mathcal{Z}_1\rangle\cdots\langle\mathcal{Z}_m \rangle = \frac{1}{q^2}\prod_i \delta^{\mathcal{Z}_i,\mathbb{1}} + \order{1/q^3}.
    \end{equation}
     All together, in the context of a Haar average, the following replacement is valid up to $\order{1/q^2}$.
    \begin{equation}\label{site0-corr}
        \bra{\phi_+}\Gamma^0\ket{\phi_+(T)} = \langle Z\Gamma^0_1 Z(T)\Gamma^{0\dagger}_{\overline{2}}\rangle \langle Z\Gamma^0_2 Z(T)\Gamma^{0\dagger}_{\overline{1}}\rangle - \frac{1}{q^2}\delta^{ \Gamma^0_{\overline{1}},\Gamma^0_1}\delta^{ \Gamma^0_{\overline{2}},\Gamma^0_2}
        +\order{1/q^3}.
    \end{equation}
    We say that a decoration $\Gamma^r$ leaves a site $r$ undecorated if it contributes only trivial correlators, $\langle \mathbb{1} \rangle$ . In the present case Keeping only $\order{1/q^2}$ contributions forces all sites $r\neq0$ to be left undecorated. This allows us to take the Haar average of \eqnref{site0-corr} directly, using theorem \ref{two-correlators} for the Haar average of a product of two correlators. This gives,
    \begin{equation}
        \int dV \bra{\phi_+}\Gamma^0\ket{\phi_+(T)} = \frac{1}{q^2}\delta^{ \Gamma^0_{\overline{1}},\Gamma^0_1}\delta^{ \Gamma^0_{\overline{2}},\Gamma^0_2} - \frac{1}{q^2}\delta^{ \Gamma^0_{\overline{1}},\Gamma^0_1}\delta^{ \Gamma^0_{\overline{2}},\Gamma^0_2}
        +\order{1/q^3} = \order{1/q^3}.
    \end{equation}
\end{itemize}
The analysis of $x<0$ would replicate that of $x>0$, but with an additional factor of $q^{-2\abs{x}}$. The for all $x$, $\int dV \mathcal{D}^{1,1}_\Gamma(x,T) = \order{1/q^3}$. Alternatively,
\begin{equation}
    \int dV \mathcal{D}^{1,1}(k,T)=\order{1/q^3}.
\end{equation}

\subsubsection{\texorpdfstring{$(a,b)=(4,4)$}{Lg} is \texorpdfstring{$\mathcal{O}(1/q^2)$}{Lg}}\label{(4,4)}
The (4,4) calculation is significantly more difficult; we present the full calculation here, however; readers interested only in the final result should skip to the summary in \secref{summary table}.

We use the decoration expansion once again to rule out contributions from $x>0$ and to identify the relevant contributions from $x=0$.
\begin{itemize}
    \item $x > 0$:
    \begin{equation}
        \mathcal{D}^{4,4}_\Gamma(x>0,T)= -h^2\times \left[ \ \begin{matrix}
        \vspace{-4pt}
        \\
    \textcolor{gray}{\textrm{site } 0}\\ 
    \textcolor{gray}{\textrm{site } 1}\\
    \\ \\ \\ \\
    \end{matrix} \ \ 
    \begin{matrix}
    \ \ \vdots\\
    \ \ \bra{+}K\\ 
    q\bra{-}K\\
    \ \ \vdots\\
    q\bra{-}\\
    \ \ \bra{-}\\
    \ \ \vdots
    \end{matrix}
    \ \ \fbox{ $\begin{matrix}
    \\
    \\
    \\
    \hspace{-5pt}\Gamma\\
    \\
    \\
    \\
    \end{matrix}$} \ \ 
    \begin{matrix}
    \hspace{-20pt}\vdots\\
    \hspace{-20pt}\ket{+}\\
    \hspace{-20pt}\ket{+}\\
    \hspace{-20pt}\vdots\\
    K(T)\ket{+}\\ 
    K(T)\ket{-}\\
    \hspace{-20pt}\vdots
    \end{matrix} \ \ 
    \begin{matrix}
    \vspace{5pt}
    \\ \\ \\ \\
    \hspace{-2pt}\textcolor{gray}{\textrm{site } x}\\ 
    \hspace{-2pt}\textcolor{gray}{\textrm{site } x+1}\\
    \
    \end{matrix}\right]\nonumber
    \end{equation}
    For $x>1$, each of the sites $0$, $1$, $x$ and $x+1$ contribute non-trivial correlators. When $x=1$, sites $0$, $1$ and $2$ all contribute non-trivial correlators. In either case, theorem \ref{productofcorrelators} gives
    \begin{equation}
        \int dV \mathcal{D}^{4,4}_\Gamma(x>0,T)=\order{1/q^3}.
    \end{equation}
    \item $x=0$:

\begin{equation}
    \mathcal{D}^{4,4}_\Gamma(x=0,T)=-h^2\times \left[\ \begin{matrix}
    \textcolor{gray}{\textrm{site } 0}\\ 
    \textcolor{gray}{\textrm{site } 1}\\
    \end{matrix} \ \ 
    \begin{matrix}
    \ \ \ \ \vdots\\
    \ \ \ \bra{+}\\
    \bra{+}K\\ 
    \bra{-}K\\
    \ \ \ \bra{-}\\
    \ \ \ \ \vdots
    \end{matrix} \ 
    \fbox{ $\begin{matrix}
    \\
    \\
    \hspace{-5pt}\Gamma\\
    \\
    \\
    \end{matrix}$} \ 
    \begin{matrix}
    \hspace{-25pt}\vdots\\
    \hspace{-25pt}\ket{+}\\
    K(T)\ket{+}\\ 
    K(T)\ket{-}\\
    \hspace{-25pt}\ket{-}\\
    \hspace{-25pt}\vdots
    \end{matrix}\right]\nonumber
    \end{equation}
    Both sites $0$ and $1$ contribute non-trivial correlators. Keeping only the $\order{1/q^2}$ contributions means selecting decorations on sites $r\neq 0,1$ that give trivial correlators only. For $r>1$ this means selecting decorations such that $\bra{-}\Gamma^r\ket{-}=\langle \Gamma^{r}_1 \Gamma^{r\dagger}_{\overline{1}} \rangle \langle \Gamma^{r}_2 \Gamma^{r\dagger}_{\overline{2}} \rangle=\langle \mathbb{1} \rangle^2$ (for $r<0$ simply switch $-\leftrightarrow+$ and $1\leftrightarrow 2$ in these equations). Choosing only $\Gamma^r$ that leave a site $r<0$ ($r>1$) undecorated (contributing only trivial correlators) is equivalent to the decoration delta constraint $\delta^{\Gamma^r_1,\Gamma^r_{\overline{2}}}\delta^{\Gamma^r_2,\Gamma_{\overline{1}}}$ ($\delta^{\Gamma^r_1,\Gamma^r_{\overline{1}}}\delta^{\Gamma^r_2,\Gamma_{\overline{2}}}$). The implementation of these decoration delta constraints is discussed in \ref{decoration delta constraint appendix}. The result of which is that for sites $r<0$ we sandwiching each decoration layer $\Gamma(t)$ by $\bra{+}$ and $\ket{+}$ and by $\bra{-}$ and $\ket{-}$ for sites $r<0$.
    
    Writing the definition of $K$ in \eqnref{Kdef_Zsquareddef} as $K=Z_1-Z_2$, where the index refers to which leg the $Z$ decorates, we write the following,
    \begin{equation}
        \begin{matrix}
        \bra{+}K\Gamma^0 K(T)\ket{+}\\
        \bra{-}K\Gamma^1 K(T)\ket{-}
        \end{matrix}=\sum_{i,j\in \{1,2\}}
        \begin{matrix}
        \bra{+}Z_i\Gamma^0 Z(T)_i\ket{+}\\ 
        \bra{-}Z_j\Gamma^1 Z(T)_j\ket{-}
        \end{matrix} + 
        \begin{matrix}
        \textrm{ terms with more than two}\\
        \textrm{ non-trivial correlators}.
        \end{matrix}
    \end{equation}
    Because we are selecting only decorations which leave sites $r\neq 0,1$ undecorated, we are able to take the Haar average of this expression in isolation. Terms with more than two non-trivial correlators are $\order{1/q^3}$ or smaller and are therefore discarded, leaving only the Haar average of the first term. Using theorem \ref{two-correlators}, this is given by
    \vspace{-4mm}
    \begin{equation}
        \int dV \sum_{i,j\in \{1,2\}}
        \begin{matrix}
        \bra{+}Z_i\Gamma^0 Z(T)_i\ket{+}\\ 
        \bra{-}Z_j\Gamma^1 Z(T)_j\ket{-}
        \end{matrix} =
        \begin{matrix}
        \\
        
        \frac{1}{q^2}\delta^{\Gamma^0_1,\Gamma^1_{\overline{1}}}\delta^{\Gamma^0_{\overline{2}},\Gamma^1_1}
        \delta^{\Gamma^0_{2},\Gamma^0_{\overline{1}}}\delta^{\Gamma^1_{2},\Gamma^1_{\overline{2}}} 
        + 
        \frac{1}{q^2}\delta^{\Gamma^0_1,\Gamma^1_{\overline{2}}}\delta^{\Gamma^0_{\overline{2}},\Gamma^1_2}
        \delta^{\Gamma^0_{2},\Gamma^0_{\overline{1}}}\delta^{\Gamma^1_{1},\Gamma^1_{\overline{1}}} \\
        + \left(1 \leftrightarrow 2\right) + \order{1/q^3}.
        \end{matrix}
    \end{equation}
    These delta constraints are implemented by sandwiching the decoration layers $\Gamma(t)$ with the appropriate wirings. The four different wirings configurations for sites $0$ and $1$ are shown below.
    \begin{figure}[H]
    \centering
    \raisebox{-0.45\totalheight}{\includegraphics[height = 3.5cm]{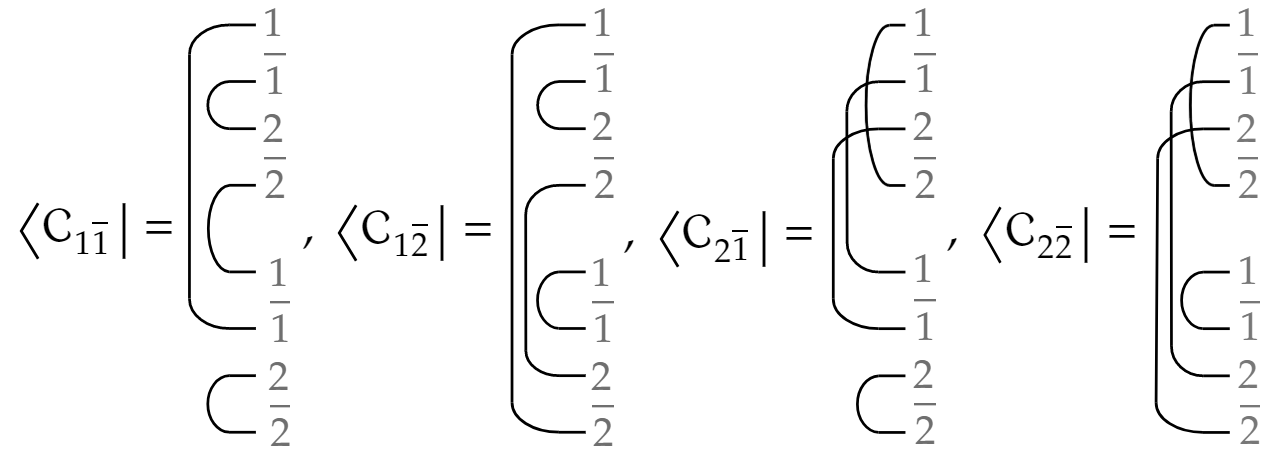}}.
    \label{site_1_projector_insertions}
    \end{figure}
    Counting only the relevant decorations $\Gamma$, $\mathcal{D}^{1,1}_\Gamma(x=0,T)$ given by
    \begin{equation}
        \int dV \mathcal{D}^{4,4}_\Gamma(x=0,T)=-\frac{h^2}{q^2}\sum_{i,j\in \{1,2\}} \raisebox{-0.47\totalheight}{\includegraphics[height=3.8cm]{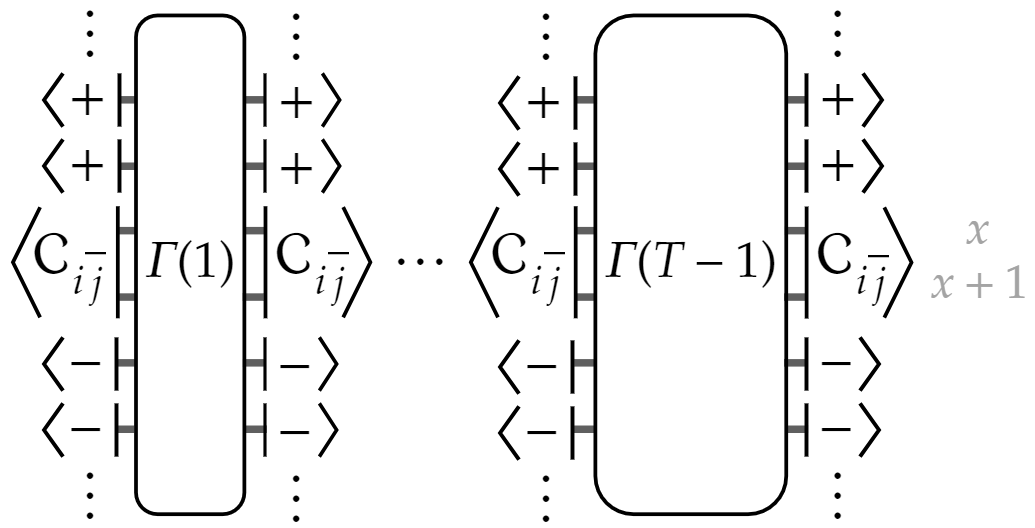}}
    \end{equation}
    We now sum over all decorations $\Gamma$ weighted by the coefficients $C_\Gamma$ appearing in \eqnref{decexpansion}. This converts back into the picture with full unitary layers $U(t)$. Using the property \eqnref{rules_away_from_cut} for 2-local bricks contracted with the states $\bra{\pm}\otimes\bra{\pm}$ or $\ket{\pm}\otimes\ket{\pm}$, $\int dV \mathcal{D}^{1,1}(x=0,T)$ is given by the simplified form
    \begin{align}
    \int dV \mathcal{D}^{4,4}(x=0,T)&=-\frac{h^2}{q^2}\sum_{i,j} \raisebox{-0.45\totalheight}{\includegraphics[height = 2cm]{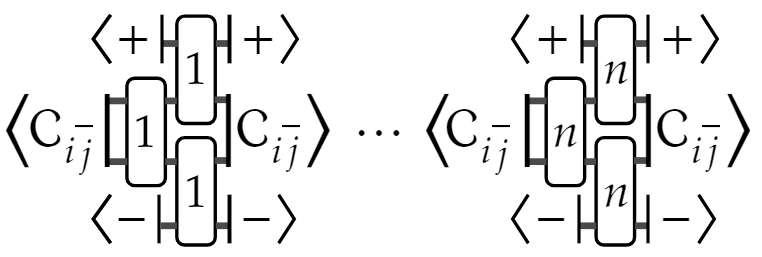}}+\order{1/q^3}\nonumber\\
    &= -\frac{h^2}{q^2}\sum_{i,j} \bra{\mathcal{C}_{i,\overline{j}}}\mathcal{T}\ket{\mathcal{C}_{i,\overline{j}}}^n+\order{1/q^3},
    \end{align}
    where $\mathcal{T}$ is given by 
    \begin{equation}\label{calTdef}
     \raisebox{-0.45\totalheight}{\includegraphics[height = 1.8cm]{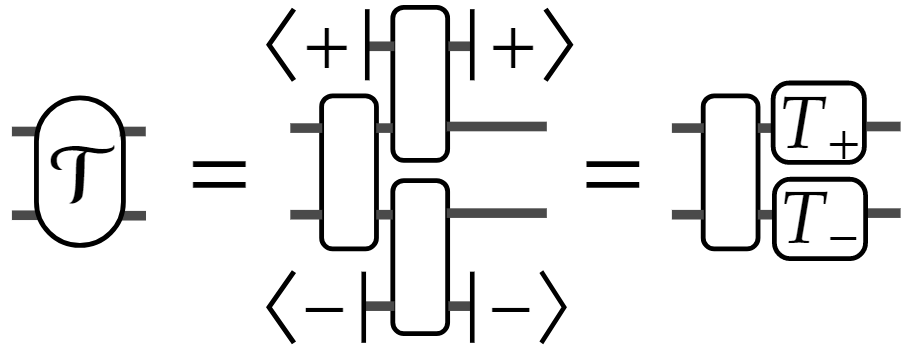}},
    \end{equation}
    and where the tensor contractions $T_+$ and $T_-$ are given algebraically by
    \begin{equation}
        \raisebox{-0.3\totalheight}{\includegraphics[height = 0.97cm]{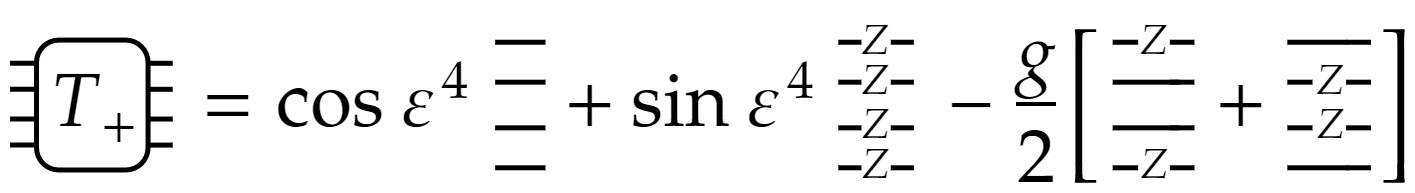}}, \  \raisebox{-0.31\totalheight}{\includegraphics[height = 0.85cm]{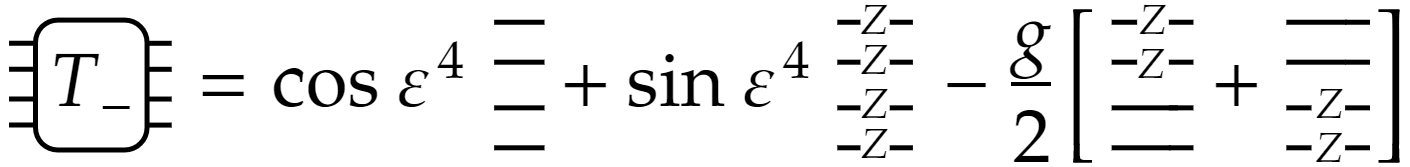}}
    \label{Tplus}
    \end{equation}
    Due to the replica symmetry of the unitary evolution operator, $(1,\overline{1})\leftrightarrow (2,\overline{2})$, the contributions from the $(i,j)=(1,1)$ and $(2,2)$ wirings are identical, as are $(1,2)$ and $(2,1)$ wirings. The unitary $e^{-\mathrm{i}\varepsilon\mathcal{L}}$ also has the property $\textrm{Sw}(1,\overline{1})\textrm{Sw}(2,\overline{2})e^{-\mathrm{i}\varepsilon\mathcal{L}}\textrm{Sw}(1,\overline{1})\textrm{Sw}(2,\overline{2})=(e^{-\mathrm{i}\varepsilon\mathcal{L}})^*$, where $\textrm{Sw}(i,\overline{i})$ swaps unbarred leg $i$ and barred leg $\overline{i}$. This transformation is a symmetry of the $\bra{+}$ and $\bra{-}$ wirings while exchanging the $\mathcal{C}_{1,\overline{1}}$ ($\mathcal{C}_{2,\overline{2}}$) and $\mathcal{C}_{2,\overline{1}}$ ($\mathcal{C}_{1,\overline{2}}$) wirings. Therefore, \begin{equation}\label{104}
        \xi(\varepsilon)\equiv\bra{\mathcal{C}_{1,\overline{1}}}\mathcal{T}\ket{\mathcal{C}_{1,\overline{1}}}=\bra{\mathcal{C}_{2,\overline{2}}}\mathcal{T}\ket{\mathcal{C}_{2,\overline{2}}}=\bra{\mathcal{C}_{1,\overline{2}}}\mathcal{T}\ket{\mathcal{C}_{1,\overline{2}}}^*=\bra{\mathcal{C}_{2,\overline{1}}}\mathcal{T}\ket{\mathcal{C}_{2,\overline{1}}}^*.
    \end{equation}
    Using the decoration decomposition of a brick in \eqnref{fullbrickdecomposition} and the expressions for $T_+$ and $T_-$ in \eqnref{Tplus}, $\xi(\varepsilon)$ is found to be 
    \begin{equation}
        \xi(\varepsilon) = (1+g)^2 -2\mathrm{i}hg.
    \end{equation}
    All together we find,
    \begin{equation}\label{exponential decay}
        \int dV\mathcal{D}^{4,4}(x=0,T)=-\frac{4h^2}{q^2}\Re \{ \xi^{T-1}\} + \order{1/q^3}.
    \end{equation}
\end{itemize}

Summing over $x$, we find the $\order{1/q^2}$ contribution to $\int dV \mathcal{D}^{4,4}(k,T)$ is given by precisely the same quantity. The decay rate is given by $\gamma(\varepsilon)\equiv \ln(\abs{\xi}^{-1})$, which is always close to $2\abs{g(\varepsilon)}$, $2\abs{g(\varepsilon)}\leq \gamma(\varepsilon)\leq 4\ln(2)\abs{g(\varepsilon)}$.

\subsection{Table of results and summary}\label{results}
We summarise the contributions $\int dV \mathcal{D}^{a,b}(k=0,T)$ in the table below, highlighting the only contributions at $\order{1/q^2}$.
\begin{table}[H]\label{summary table}
\centering
\begin{tabular}{||c|| c c c c||} 
 \hline
 $a$ $\backslash$ \ $b$ & $1$ & $2$ & $3$ & $4$ \\ [0.5ex] 
 \hline\hline
 $1$ & \color{red} $q^{-3}$ & \color{red} $q^{-3}$ & \color{red} $q^{-3}$ & \color{red} $q^{-3}$ \color{black} \\ 
 $2$ & \color{green} $q^{-2}$ & \color{red} $q^{-3}$ & \color{red} $q^{-3}$ & \color{red} $q^{-3}$ \color{black} \\
 $3$ & \color{red} $q^{-3}$ & \color{red} $q^{-3}$ & \color{red} $q^{-3}$ & \color{red} $q^{-3}$ \color{black} \\
 $4$ & \color{red} $q^{-3}$ & \color{red} $q^{-3}$ & \color{red} $q^{-3}$ & \color{green} $q^{-2}$ \color{black} \\
 \hline
\end{tabular}
\end{table}
\color{black}

We calculate the $(a,b)=(2,1)$ contribution in \ref{(2,1) appendix}, and in \ref{Remaining (a,b)} we find that the remaining pairs $(a,b)$ contribute at $\order{1/q^3}$ or smaller. Using \eqnref{correction}, we are only required to know $\int dV \mathcal{D}^{a,b}(k=0,z=0)$ to determine the butterfly velocity\footnote{We sum over times up to the cutoff $t_{\varepsilon}(q)$ as discussed in \secref{Summary}, incurring only an error exponentially small in $q$.}. For $(4,4)$ this is given below, with the re-parameterisation $s(\varepsilon)=\sin(\varepsilon)^2\cos(\varepsilon)^2$,
\begin{equation}
    \int dV \mathcal{D}^{(4,4)}(k=0,z=0) =
    -\frac{1}{q^2}\frac{1+5s-4s^2}{1-s-3s^2} + \order{1/q^3}.
\end{equation}

This re-parameterisation is motivated by the following observations about the dependence of $v_B$ and $D$ on $\varepsilon$. Firstly, under the variable shift $\varepsilon\to \pi/2+\varepsilon$, the full coupling unitary then transforms as $e^{-\mathrm{i}\varepsilon H}\to (-\mathrm{i})^{N-1} e^{-\mathrm{i}\varepsilon H}$. The operator dynamics is blind to global phases, meaning that $\varepsilon\to\pi/2+\varepsilon$ is a symmetry of $v_B(\varepsilon)$ and $D(\varepsilon)$. Secondly, by globally swapping leg $1$ with $\overline{1}$ and $2$ with $\overline{2}$, we find $\Sigma_{V,\varepsilon}(k,z)=\Sigma_{V^*,-\varepsilon}(k,z)$. and $\Omega_\varepsilon(k)= \Omega_{-\varepsilon}(k)$, where we have have labelled $\Sigma$ with the scrambler and coupling strength and labelled $\Omega$ with the coupling strength used ($\Omega$ is independent of $V$). By integrating over $V$, we find another symmetry of the circuit averaged butterfly velocity and diffusion constant $\langle v_B(\varepsilon) \rangle$ and $\langle D(\varepsilon) \rangle$, namely $\varepsilon\to -\varepsilon$. Using these symmetries, we determine that $\langle v_B \rangle$ is a function of $s(\varepsilon)$ only. 

For $(a,b)=(2,1)$, we find
\begin{equation}
    \int dV \mathcal{D}^{(2,1)}(k=0,z=0) = \frac{2g^2}{q^2}(f(\varepsilon)-\nu(\varepsilon))
     + \order{1/q^3}.
\end{equation}
where $\nu(\varepsilon)=[4(1-2s)(1-s(1-2s))]^{-1}$ and where $f(\varepsilon)$ is found by numerically evaluating a sum involving a $5\times5$ transfer matrix in \ref{(2,1) appendix} and given to good approximation (\figref{f_plot}) by
    \begin{equation}
        f(\varepsilon) = \frac{1}{7}s(\varepsilon)(1 - 4s(\varepsilon))^2(1+as+bs^2)
    \end{equation}
where $a=6.8$ and $b=16.1$. The factors $\frac{1}{7}s(1-4s)^2$ is obtained analytically by diagonalising the transfer matrix at small $s$ and around the point $s=1/4$. Altogether, we then find
\begin{equation}
    \int dV \mathcal{D}(k=0,z=0) = 2\frac{g^2}{q^2}(f(\varepsilon)-\nu(\varepsilon))
    -\frac{1}{q^2}\frac{1+5s-4s^2}{1-s-3s^2} + \order{1/q^3}.
\end{equation}
so that the correction to the circuit averaged butterfly velocity is
\begin{equation}\label{butterfly_velocity_correction}
    \langle \delta v_B(\varepsilon)\rangle = 
    -\int dV \mathcal{D}(k=0,z=0) = 
    \delta v_F(\varepsilon) + \delta v_S(\varepsilon)  + \order{1/q^3},
\end{equation} 
where
\begin{equation}\label{v contributions}
    \delta v_F(\varepsilon) = 2\frac{g^2}{q^2}(\nu(\varepsilon)-f(\varepsilon)), \quad
    \delta v_S(\varepsilon) =\frac{1}{q^2}\frac{1+5s-4s^2}{1-s-3s^2}.
\end{equation}
As discussed in \secref{Summary}, the correction $\delta v_S(\varepsilon)$ only arises because of the spatial translation symmetry of the model. Whereas $\delta v_F(\varepsilon)$ arises due to both the spatial and Floquet symmetries. To illustrate this point, we carry out a similar calculation for a version of the model with independently distributed scramblers $V_{x,t}$ for each site $x$ and at each layer of unitaries $U_t$. This is done in \ref{independent scramblers} where we find the corrections to the circuit averaged butterfly velocity $\langle \delta v\rangle\sim\order{1/q^3}$, i.e., $v_{S,F}$ both vanish. Turning to the variant of the model with spatial translation symmetry but no time translation symmetry, we find $\langle \delta v_B \rangle = \delta v_S(\varepsilon)  + \order{1/q^3}$. Comparing these result to \eqnref{full velocity} allows us to identify the corrections to $v_B$ with the symmetries of the model.

Away from the weak coupling limit, the averaged butterfly velocity is given by \eqnref{full velocity}. The weak coupling limit is discussed in section \ref{small epsilon}.

\section{Discussion}
In this section we discuss the late time behaviour of the memory function, addressing the limitations of the Haar averaged ATO' results of \secref{theorems}. We also discuss the break down of the perturbative scheme used in \secref{MMF calc} as $\varepsilon\to 0$.

\subsection{Late times}\label{late times}
We have presented a Kubo-like formula for $v_B$ in \eqnref{kubo2}, involving the time integral of a correlator of fast operators, analogous to the $J$-$J$ correlation functions in Kubo formulae for more conventional transport. For $\varepsilon>0$ (more precisely $g(\varepsilon)\neq0$), we find that the leading order correlation functions decay exponentially in time at leading order in $1/q$. Given that the correlation functions are between fast variables, it is plausible that this exponential decay is not simply an artifact of the large $q$ limit, i.e., the correlation functions decay over some time scale which is bounded below by a $q$-independent quantity $\tau(\varepsilon)$, which is positive except at exceptional point where the sites decouple i.e., when $\varepsilon=n\pi/2$ (we discuss the decoupling limit later). Assuming that $t_\varepsilon(q)$ is a (cutoff) time up to which we can safely use the large $q$ correlator theorems of \secref{theorems} to identify the $\order{1/q^2}$ contributions to $\Sigma$, the error incurred by truncating the infinite time evolution is $\order{\exp[- t_\varepsilon(q)/\tau(\varepsilon)]}$. We will now argue that $t_\varepsilon(q)$ increases linearly with $q$. As a result, the error due to truncation decays exponentially in $q$, which is much smaller than the $\mathcal{O}(1/q^2)$ corrections we calculate. This demonstrates the validity of our expansion in $q^{-1}$.

We now give some justification for the existence of a cutoff $t_\varepsilon(q)=cq$ for a $q$-independent constant $c$. We do this in two parts: (1) we argue that the expressions in \secref{theorems} for the Haar average of few correlators are valid for times $t<cq$; (2) we argue that this is enough to guarantee that contributions involving many correlators (more than two) are $\order{1/q^3}$ (for times $t<cq$).
    
In order to do this, we must go beyond the $q\to\infty$ scaling results of \secref{theorems}. In the absence of an analytical understanding of the corrections at finite $q$, we took a numerically Haar average of an assortment of ATO correlators and products of correlators, finding good evidence that the expression for the Haar average of a product of two correlators (theorem \ref{two-correlators}) and of OTOCs (theorem \ref{OTOCtheorem}) experience only $\order{1/q^3}$ errors for times $t<cq$. It turns out that this, along with the use of a Holder's inequality and the monotonicity of the $p$-norm, is enough to bound the Haar average of arbitrarily many correlators as $\order{1/q^3}$. 
    
To show this, we start by bound the expectation of a product of many (more than two) non-trivial correlators as shown below,
\begin{equation}
    \abs{\int dV \prod_{i}\langle \mathcal{Z}(\boldsymbol{t}_i)\rangle}\leq \int dV \prod_{i}\abs{\langle \mathcal{Z}(\boldsymbol{t}_i)\rangle}.
\end{equation}
Then, using the fact $\abs{\langle \mathcal{Z}(\boldsymbol{t}_i)\rangle}\leq 1$, we can bound the average of a product of many correlators by an average over only a few. We choose to highlight three correlators
\begin{equation}\label{3corrs}
    \abs{\int dV \prod_{i}\langle \mathcal{Z}(\boldsymbol{t}_i)\rangle}\leq \int dV \abs{\langle \mathcal{Z}(\boldsymbol{t}_1)\rangle}\abs{\langle \mathcal{Z}(\boldsymbol{t}_2)\rangle}\abs{\langle \mathcal{Z}(\boldsymbol{t}_3)\rangle}.
\end{equation}
Using a generalised Holder's inequality and monotonicity of the $p$-norm, we further bound \eqnref{3corrs} by
\begin{equation}
    \abs{\int dV \prod_{i}\langle \mathcal{Z}(\boldsymbol{t}_i)\rangle}\leq
    \sqrt{
    \int dV \abs{\langle \mathcal{Z}(\boldsymbol{t}_1)\rangle}^2 \int dV\abs{\langle \mathcal{Z}(\boldsymbol{t}_2)\rangle}^2 \int dV\abs{\langle \mathcal{Z}(\boldsymbol{t}_3)\rangle}^2}.
\end{equation}
Then, for times $t<cq$ and using theorem \ref{two-correlators} for the second moment of a correlator, this bound takes the form
\begin{equation}\label{bound for three corr}
    \abs{\int dV \prod_{i}\langle \mathcal{Z}(\boldsymbol{t}_i)\rangle}\leq C \frac{S(\boldsymbol{t}_1) S(\boldsymbol{t}_2) S(\boldsymbol{t}_3)}{q^3},
\end{equation}
for an $\order{1}$ constant $C$. The correlators contributing to $\Sigma$ have $1\leq S(\boldsymbol{t})\leq 2$, so that the right hand-side of \eqnref{bound for three corr} then simplifies to $C'/q^3$ for an $\order{1}$ constant $C'$. For times $t<cq$, any contribution to $\Sigma$ with three of more non-trivial correlators is $\order{1/q^3}$, and the $\order{1/q^2}$ contributions are counted precisely as we have done in \secref{MMF calc} and the \ref{(2,1) appendix}.
    
\subsection{The \texorpdfstring{$\varepsilon\to0$}{Lg} limit}\label{small epsilon}
In the $\varepsilon\to0$ limit, the Floquet unitary is given by a product of single-site unitaries, so that there is no operator growth dynamics at all ($v_B=0$). However, the $\varepsilon \to 0$ limit of the expression \eqnref{full velocity} yields $v_B=1/q^2$. The reason for this failure to predict the correct operator dynamics as $\varepsilon\to0$ lies in the fact that our $1/q$ perturbative scheme breaks down once $\varepsilon$ is as small as $\varepsilon\sim1/q$. Diagrams that we previously dismissed as $\order{1/q^3}$ at strong coupling, can in fact contribute at $\order{1/q^2}$ due to appearance of factors of $\varepsilon\sim 1/q$ in the denominator (after summing over time).

We do not have access to the exact expressions for the Haar average of ATO correlators in the Floquet model, instead relying on $\order{1/q^2}$ results. However, in \ref{independent scramblers}, we study a variant of the model with independently distributed scramblers $V_t$ between time-step for which exact Haar averaged results are possible for certain diagrams. If we were to naively identify the $1/q^2$ contributions before taking a sum over time, as we did in \secref{MMF calc}, we would find the same, incorrect $(a,b)=(4,4)$ contribution and incorrect behaviour of $v_B$ as $\varepsilon\to0$. In \ref{independent small epsilon}, we consider the same family of diagrams, those involving only two non-trivial correlators (whose contribution we dub $\mathcal{D}^{4,4}_2$), but now take the Haar average exactly. In doing so, we find that the previously troublesome $(4,4)$ contribution now vanishes as $\varepsilon\to0$ \eqnref{small epsilon 44},
\begin{equation}
    \langle \mathcal{D}^{4,4}_2(k=0,z=0)\rangle \approx -\frac{8\varepsilon^2}{1+8q^2\varepsilon^2}.
\end{equation}
We suggest then, that there is a region of width $\order{1/q}$ in \figref{delta_v_B_plot}, where $v_B$ rapidly approaches zero.

\section{Conclusion}
In this paper, we re-purposed a hydrodynamic formalism (the memory matrix formalism) for information transport calculations by identifying the conservation of the right density of a Heisenberg time evolved operator as a pseudo-local conservation law. A number of modifications to the existing MMF are necessary: in particular, we are led to use an unusual inner product on our space of observables \eqnref{inner-product}, which  leads to the prediction of ballistic operator growth assuming we have identified a sufficiently complete space of slow operators. We use this new formalism to produce a Kubo formula for the butterfly velocity \eqnref{Kubo} and also find symmetry constraints on the operator growth light-cone (remark \ref{remark}).

In section \ref{Minimal model} we used this formalism to investigate a family of translationally invariant Floquet models, finding leading order expressions for the circuit averaged butterfly velocity $v_B$ and operator front diffusion constant $D$. By leveraging large $q$ random unitary dynamics \cite{mcculloch2021haar}, we found that a simple hierarchy of contributions to the averaged memory matrix $\langle \Sigma \rangle$ emerges, organised by the number of non-trivial correlators contributed. This enabled us to select only the $\order{1/q^2}$ contributions, associated with processes that explore a manageable sub-region of $\mathcal{Q}$ in which only the sites directly either side of the cut (the $+/-$ domain wall) are decorated by non-identity operators. We have then counted all these processes and found that at $\order{1/q^2}$, the memory matrix decays exponentially fast, with an $\order{1}$ decay rate. We used this to calculate corrections to $v_B$, and were able to distinguish the effects of spatiotemporal symmetry on information transport by identifying which processes arise as a consequence of spatial translation and Floquet symmetry.

\section{Further work/outlook}
More exotic information hydrodynamics than biased diffusion is possible in the presence of additional conserved charges. With a U$(1)$ charge, the diffusive conserved components acts as a source of non-conserved operators, giving rise to power-law tails in the spatial distributions of operator weight \cite{OTOCDiff1,OTOCDiff2}. It will be interesting to incorporate additional symmetries, such as a U$(1)$ or fracton symmetry, into the MMF and perform a mode coupling analysis to confirm and perhaps extend existing results. Even in the presence of conservation laws, the diffusive broadening of the operator front appears ubiquitous in chaotic and interacting integrable systems in 1D \cite{Gopalakrishnan18,Gopalakrishnan18b}. What makes this diffusive broadening so universal? Perhaps, by examining MMF expressions for the front diffusion constant, we can say something about the nature or number of additional slows required to find a hydrodynamical equation other than biased diffusion. Other avenues to explore include the formulation of an information mode MMF at finite temperature/chemical potential and, with only minor modifications, calculating purity.
Another potentially fruitful application of formalism is in the setting of perturbed dual unitary circuits, which may serve as a testing ground for perturbative MMF calculations.

\section{Acknowledgements}
EM is supported by EPSRC studentship. C.v.K. is supported by a UKRI Future Leaders Fellowship MR/T040947/1.

\section{References}
\bibliography{global} 
\bibliographystyle{ieeetr}

\appendix
\addtocontents{toc}{\fixappendix}

\section{An unusual inner-product}\label{inner_product_proof}
In this appendix we prove that the inner-product $\left(\cdot|\cdot\right)$ defined in \eqnref{inner-product} satisfies the necessary axioms. We repeat the definition below,
\begin{equation}
\left(A|B\right) \equiv \bra{\Phi\left(A\right)}\ket{B}_\mathcal{W} = \Tr\left(\Phi\left(A\right)^\dagger B\right),
\end{equation}
where the super-operator $\Phi$ is given by
\begin{align}\label{Phi}
\Phi \equiv \sum_x \frac{1}{\chi_x^{2}}\ket{W^x}\bra{W^x} + Q, \quad \chi_x \equiv \bra{W^x}\ket{W^x}, \implies \Phi\ket{W^x} = \frac{1}{\chi_x} \ket{W^x}.
\end{align}
$Q$ is the Hermitian projector onto $\mathcal{Q}$. We must show that this constitutes a bona fide inner product by checking each of the inner product axioms.
\begin{enumerate}
	\item Conjugate symmetry: $\left(A|B\right)^* = \left(B|A\right)$,

	 Exploiting the conjugate symmetry of the inner product $\bra{\cdot}\ket{\cdot}$, we write
	\begin{align}
	\left(A|B\right) = \bra{\Phi\left(A\right)}\ket{B} = \bra{A}\ket{\Phi\left(B\right)} = \bra{\Phi\left(B\right)}\ket{A}^* = \left(B|A\right)^*
	\end{align}
	where we have used the fact that $\Phi$ is Hermitian, this is because $\chi_x$ is real and the projector $\hat{Q}$ is Hermitian.
	\item Linearity in second argument: $\left(A|\beta B + \gamma C\right) = \beta\left(A|B\right) + \gamma\left(A|C\right)$ for scalars $\beta, \gamma$,
	\begin{align}
	\left(A|\beta B + \gamma C\right) &= \bra{\Phi\left(A\right)}\ket{\beta B+\gamma C} = \beta \bra{\Phi\left(A\right)}\ket{B} + \gamma \bra{\Phi\left(A\right)}\ket{ C}\nonumber \\
	&= \beta \left(A|B\right) + \gamma \left(A|C\right)
	\end{align}
	where we have used linearity in the second argument of the inner product $\bra{\cdot}\ket{\cdot}$.
	\item Positive definiteness: $\left(A|A\right) > 0$ for $A \neq 0$,
	\begin{equation}
	\left(A|A\right) = \bra{\Phi\left(A\right)}\ket{A} = \bra{A} \Phi \ket{A} > 0
	\end{equation}
	where we have used the fact that $\Phi$ is a positive definite matrix. This is easily seen by noticing that all $\chi^2_x$ are real and positive.
\end{enumerate}

This confirms that $\left(A|B\right)$ is indeed an inner-product and allows us to consider a new inner-product space in which the weight operators are orthonormal.

\section{Imposing the decorations delta constraints}\label{decoration delta constraint appendix}
In this section we revisit the Haar average identities for ATO correlators and their moments, specifically theorems \ref{two-correlators} and \ref{OTOCtheorem}. The final results (at leading order in $1/q$) involved "delta constraints" (as defined in Def. \ref{decoration delta constraint}) which are zero/one depending on whether or not the decorations were equal (for all scramblers $V$). We have also seen these delta constraints whenever we have demanded that a correlator be trivial (see \eqnref{site0-corr}). In this section, we show how these delta constraints can be imposed by placing the decorations on a contour and inserting projectors at every time step; the insertion of projectors has an appealingly simple graphical interpretation, which facilitates our calculation of $\sigma$ in the main text.

\subsection{The constraint \texorpdfstring{$\delta^{\Gamma_1,\Gamma_{\overline{1}}}$}{Lg}}
The simplest example of a decoration delta constraint to consider is $\delta^{\Gamma_1,\Gamma_{\overline{1}}}$, where the two decorations are time-ordered, i.e., $\Gamma_1=Z(1)^{a_1}Z(2)^{a_2}\cdots Z(n)^{a_n}$ and $\Gamma_{\overline{1}}=Z(1)^{b_1}Z(2)^{b_2}\cdots Z(n)^{b_n}$ for binary strings $\boldsymbol{a}$ and $\boldsymbol{b}$. In this case, the delta constraint checks that $a_i=b_i$ for all $i$. This is equivalent to putting $\Gamma_1$  and $\Gamma_{\overline{1}}$ on a wiring with a single forward and backward contour (labelled $1$ and $\overline{1}$) and then placing a projector between each of the decoration layers are shown below.
    \begin{equation}\label{eq:deltaGG_constr}
    \delta^{\Gamma_1,\Gamma_{\overline{1}}}=\frac{1}{q^n}\raisebox{-0.4\totalheight}{\includegraphics[height = 1.2cm]{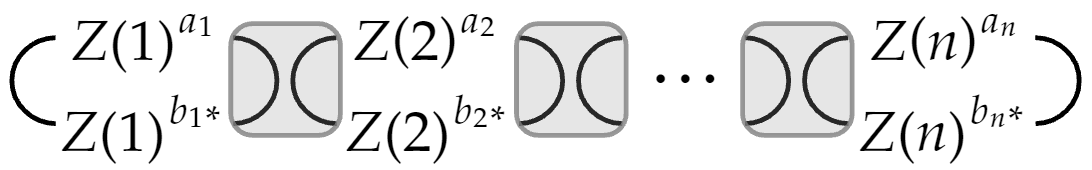}}.
    \end{equation}
The right-hand side checks that at each decoration layer $t$, $a_t=b_t$, this is precisely the same as the decoration delta constraint. A site $r\leq 0$ with decoration $\Gamma^r$ contributes $\bra{+}\Gamma^r\ket{+}$ in the decoration expansion, suppose that we demanded that this contribution contained only trivial correlators ($\bra{+}\Gamma^r\ket{+}=\langle \mathbb{1}\rangle^2$). The decorations that meet this condition are those that satisfy the constraint $\delta^{\Gamma^r_1,\Gamma^r_{\overline{1}}}\delta^{\Gamma^r_2,\Gamma^r_{\overline{2}}}$. The projector insertion technique described above selects precisely these relevant decorations as follows
\begin{equation}
    \delta^{\Gamma^r_1,\Gamma^r_{\overline{1}}}\delta^{\Gamma^r_2,\Gamma^r_{\overline{2}}}=\bra{+}\Gamma^r(1)\ket{+}\bra{+}\Gamma^r(2)\ket{+}\cdots\bra{+}\Gamma^r(n)\ket{+}.
\end{equation}
For sites $r>x+1$, where the contributions take the form $\bra{-}\Gamma^r\ket{-}$, the decorations that contribute trivial correlators are identified in the same way but by projecting with $\ket{-}\bra{-}$.

\subsection{The constraint \texorpdfstring{$\delta^{\Gamma_1\Gamma_{\overline{2}}^\dagger\Gamma_2\Gamma_{\overline{1}}^\dagger,\mathbb{1}}$}{Lg}}
\begin{figure}[H]
    \centering
    \includegraphics[height=3.5cm]{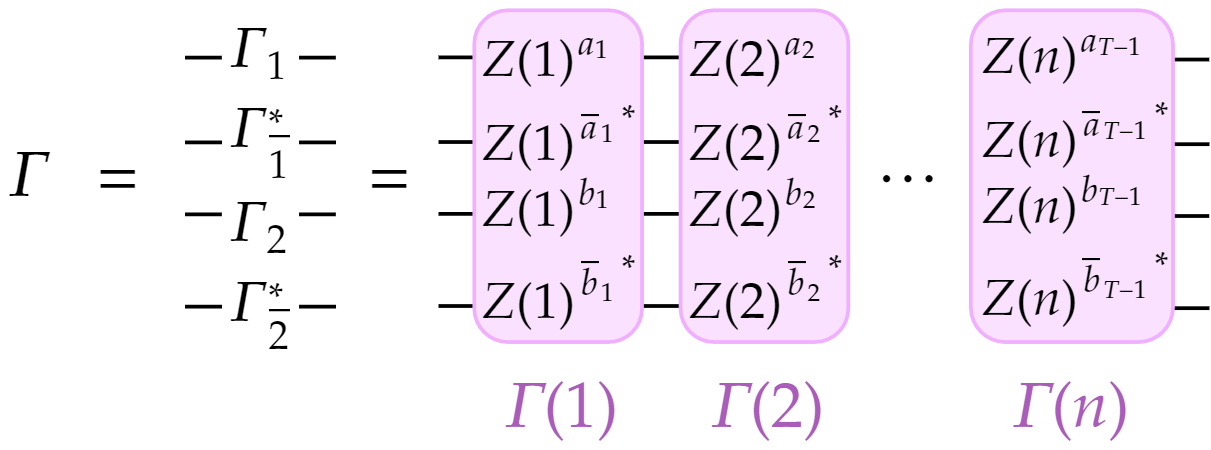}
    \caption{A decoration $\Gamma$ on the four legs $1,\overline{1},2$ and $\overline{2}$ with $n$ decoration layers.}
    \label{Gamma_dec}
\end{figure}

Consider next, a decoration $\Gamma=\Gamma_1\otimes\Gamma_{\overline{1}}^*\otimes\Gamma_2\otimes\Gamma_{\overline{2}}^*$ with $n$ decoration layers ($\Gamma=\Gamma(1)\cdots \Gamma(n)$) given in \figref{Gamma_dec}, with $\Gamma_1 = Z(1)^{a_1}\cdots Z(n)^{a_{n}}$, $\Gamma_{\overline{1}} = Z(1)^{\overline{a}_1}\cdots Z(n)^{\overline{a}_{n}}$, $\Gamma_2 = Z(1)^{b_1}\cdots Z(n)^{b_{n}}$ and $\Gamma_{\overline{2}} = Z(1)^{\overline{b}_1}\cdots Z(n)^{\overline{b}_{n}}$ for binary strings $\boldsymbol{a}, \overline{\boldsymbol{a}}, \boldsymbol{b}$ and $\overline{\boldsymbol{b}}$.
As in the previous case, we will obtain a prescription for rewiring the legs of a contour at every time step, this prescription will identify decorations $\Gamma$ that satisfy the constraint. The delta constraint $\delta^{\Gamma^r_1\Gamma^{r\dagger}_{\overline{2}}\Gamma^r_2\Gamma^{r\dagger}_{\overline{1}},\mathbb{1}}$ appears whenever we demand that the decoration $\Gamma^r$ on some site $r$, with $1<r<x$, contributes only a trivial correlator, $q\bra{-}\Gamma^r\ket{+}=\langle \mathbb{1} \rangle = 1$. Finding the $\Gamma$ which contribute non-trivial correlators is equivalent to finding $\Gamma$ which satisfy the delta constraint.

Start then, with $\bra{-}\Gamma\ket{+}$. Assume that, working in from the left, at least one of the decoration layers non-trivially decorates the $\bra{-}$ wiring (so that the either one or both of the $(i,\overline{i})$ wirings in $\bra{-}$ carry non-identity operators). This excludes the case where $\bra{-}\Gamma=\bra{-}$, which we will examine last. Let $t_i$, $0<t_i\leq n$, be the first decoration layer in from the left that non-trivially decorates the $\bra{-}$ wiring. Likewise, let $t_f$ be the first decoration layer in from the right that non-trivially decorates the $\ket{+}$ wiring. If $t_i\neq t_f$, the resulting correlator is certainly non-trivial and therefore $X\neq \mathbb{1}$ (the delta constraint is not satisfied). Otherwise, if $t_i=t_f$, we have $\bra{-}\Gamma\ket{+}=\bra{-}\Gamma(t_i)\ket{+}$. In order for this to be a trivial correlator, $\Gamma(t_i)$ must decorate $\bra{-}$ such that $\bra{-}\Gamma(t_i)=\bra{-}Z(t_i)^{\otimes2}$ (see \eqnref{Kdef_Zsquareddef} for definition of $Z^{\otimes 2}$). We can select this case by sandwiching every decoration layer $t<t_i$ by $\bra{-}$ and $\ket{-}$, every layer $t>t_i$ by $\bra{+}$ and $\ket{+}$ and the layer $t_i$ by $\bra{-}$ on the left and on right by $q\ket{0}$. Finally, the cases where no decoration layer decorates the $\bra{-}$ (an obvious example where $X=\mathbb{1}$) wiring can be selected by sandwiching every layer with $\bra{-}$ and $\ket{-}$. Therefore, the decoration delta constraint can be rewritten as
    \begin{align}\label{OTOCdelta}
        \delta^{\Gamma^r_1\Gamma^{r\dagger}_{\overline{2}}\Gamma^r_2\Gamma^{r\dagger}_{\overline{1}},\mathbb{1}} = \sum_{m=1}^n & \left(\prod_{t=1}^{m-1}\bra{-}\Gamma^r(t)\ket{-}\right)q\bra{-}\Gamma^r(m)\ket{0}\left(\prod_{t=m+1}^{n}\bra{+}\Gamma^r(t)\ket{+}\right)\nonumber\\ 
        & + \prod_{t=1}^{n}\bra{-}\Gamma^r(t)\ket{-}.
    \end{align}
Or diagrammatically as
\begin{align}\label{OTOCGamma_dec}
\delta^{\Gamma^r_1\Gamma^{r\dagger}_{\overline{2}}\Gamma^r_2\Gamma^{r\dagger}_{\overline{1}},\mathbb{1}}=\sum_{m=1}^n \ &\raisebox{-0.45\totalheight}{\includegraphics[height=1.8cm]{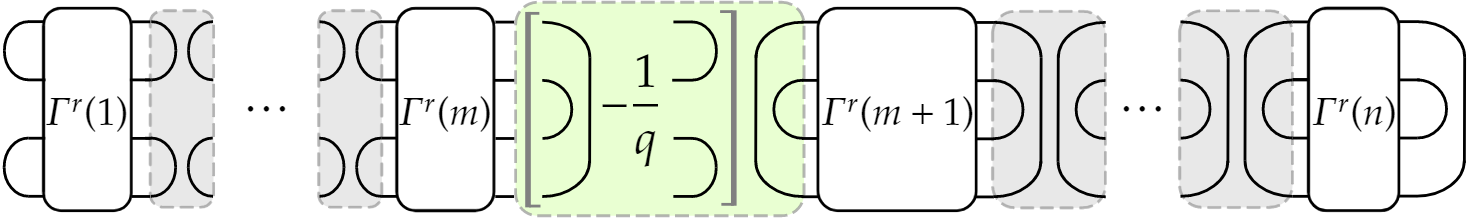}}\nonumber\\
&\quad +\raisebox{-0.45\totalheight}{\includegraphics[height=1.8cm]{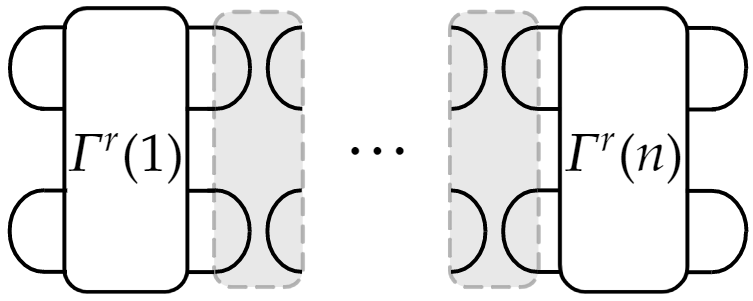}}
\end{align}

\subsection{Connection to the OTOC Haar average}
In this section we use the results of \cite{mcculloch2021haar} to re-express the Haar average of a physical OTOC in terms of decoration delta constraints, arriving at the form of this theorem presented in theorem \ref{OTOCtheorem} of \secref{theorems}.
The Haar average of an OTOC given in \cite{mcculloch2021haar} is quoted below
\begin{align}
    \int dV \langle Z \Gamma_1 Z(T) &\Gamma_{\overline{1}}^\dagger Z \Gamma_2 Z(T) \Gamma_{\overline{2}}^\dagger \rangle\nonumber \\
    &= \frac{1}{q^2}\sum_{m=1}^n \left(\prod_{t=1}^{m-1}\bra{-}\Gamma(t)\ket{-}\right)q\bra{-}\Gamma(m)\ket{0}\left(\prod_{t=m+1}^{n}\bra{+}\Gamma(t)\ket{+}\right)\nonumber\\
    &\quad - \frac{1}{q^2}\prod_{t=1}^{n}\bra{-}\Gamma(t)\ket{-} + \order{1/q^3}.
\end{align}
Comparing this to \eqnref{OTOCdelta}, we have the following,
\begin{equation}\label{OTOC theorem derived}
    \int dV \langle Z \Gamma_1 Z(T) \Gamma_{\overline{1}}^\dagger Z \Gamma_2 Z(T) \Gamma_{\overline{2}}^\dagger \rangle = \frac{1}{q^2}\left(\delta^{\Gamma_1\Gamma_{\overline{2}}^\dagger\Gamma_2\Gamma_{\overline{1}}^\dagger,\mathbb{1}} - \delta^{\Gamma_1,\Gamma_{\overline{1}}}\delta^{\Gamma_2,\Gamma_{\overline{2}}}-\delta^{\Gamma_1,\Gamma_{\overline{2}}}\delta^{\Gamma_2,\Gamma_{\overline{1}}}\right)+\order{1/q^3}.
\end{equation}
\section{\texorpdfstring{$(a,b)=(2,1)$}{Lg}}\label{(2,1) appendix}
In this appendix, we evaluate the Haar average of the $(a,b)=(2,1)$ contribution to $\mathcal{D}^{a,b}(x>0,T)$.
\subsection{\texorpdfstring{$x=0$}{Lg}}
    \begin{equation}
        \mathcal{D}^{2,1}_\Gamma(x=0,T)=g^2\times \left[ \ \begin{matrix}
        \textcolor{gray}{\textrm{site } 0}\\ 
        \textcolor{gray}{\textrm{site } 1}
        \end{matrix} \ \ 
        \begin{matrix}
        \ \vdots\\
        \bra{+}\\ 
        \hspace{-5pt}\bra{\phi_-}\\
        \ \vdots
        \end{matrix}
        \
        \fbox{ $\begin{matrix}
        \\
        \hspace{-5pt}\Gamma\\
        \\
        \end{matrix}$}\
        \begin{matrix}
        \hspace{-20pt}\vdots\\
        \ket{\phi_+(T)}\\
        \hspace{-20pt}\ket{-}\\
        \hspace{-20pt}\vdots
        \end{matrix}\right]
    \end{equation}
    Sites $0$ and $1$ each contribute a product of non-trivial correlation functions plus terms of size $1/q^2$. The Haar average of this is $\order{1/q^4}$.
    
\subsection{\texorpdfstring{$x = 1$}{Lg}}
    \begin{equation}
        \mathcal{D}^{2,1}_\Gamma(x=1,T) = g^2 \times \left[ \ \begin{matrix}
        \\
        \textcolor{gray}{\textrm{site } 1} \\
        \\
        \end{matrix} \ \ 
        \begin{matrix}
        \ \ \vdots\\
        \hspace{8pt} \bra{+}\\
        \hspace{-3pt}q\bra{\phi_-}\\
        \hspace{8pt} \bra{-}\\
        \ \ \vdots
        \end{matrix}
        \
        \fbox{ $\begin{matrix}
        \\
        \\
        \hspace{-5pt}\Gamma\\
        \\
        \\
        \end{matrix}$}\
        \begin{matrix}
        \hspace{-20pt}\vdots\\
        \hspace{-20pt} \ket{+}\\
        \ket{\phi_+(T)}\\
        \hspace{-20pt} \ket{-}\\
        \hspace{-20pt}\vdots
        \end{matrix}\right]
    \end{equation}
    Splitting the decoration site by site, we find
    \begin{equation}
    \mathcal{D}^{2,1}_\Gamma(x=1,T) = qg^2\bra{\phi_-}\Gamma^1\ket{\phi_+(T)}
    \left(\prod_{r\leq 0}\bra{+}\Gamma^r 
    \ket{+}\right) \left(\prod_{r>1}\bra{-}\Gamma^r\ket{-}\right). 
    \end{equation}

    The contribution from site $1$ is given in full below,
    \begin{align}\label{(2,1) site 1}
        \begin{matrix}
        \hspace{-3pt}q\bra{\phi_-}
        \end{matrix}
        \ \Gamma^1 \ 
        \begin{matrix}
        \ket{\phi_+(T)}
        \end{matrix}
        &=\langle Z \Gamma^1_1 Z(T) \Gamma^{1\dagger}_{\overline{2}} Z \Gamma^1_2 Z(T) \Gamma^{1\dagger}_{\overline{1}}\rangle - \frac{1}{1-q^{-2}}\langle Z\Gamma^1_1\Gamma^{1\dagger}_{\overline{1}}\rangle\langle Z\Gamma^1_2\Gamma^{1\dagger}_{\overline{2}}\rangle\nonumber \\
        &-\frac{1}{1-q^{-2}}\langle \Gamma^{1\dagger}_{\overline{2}}\Gamma^1_1 Z(T)\rangle\langle \Gamma^{1\dagger}_{\overline{1}}\Gamma^1_2 Z(T)\rangle
        +\frac{1}{q^2-1}\langle Z \Gamma^1_1\Gamma^{1\dagger}_{\overline{2}} Z \Gamma^1_2\Gamma^{1\dagger}_{\overline{1}} \rangle\nonumber\\
        &+\frac{1}{q^2-1}\langle \Gamma^{1\dagger}_{\overline{1}} \Gamma^1_1 Z(T) \Gamma^{1\dagger}_{\overline{2}} \Gamma^1_2 Z(T) \rangle\nonumber
        +\frac{1}{q^2(1-q^{-2})^2}\langle \Gamma^1_1\Gamma^{1\dagger}_{\overline{1}} \Gamma^1_2\Gamma^{1\dagger}_{\overline{2}} \rangle\nonumber\\
        &-\frac{1}{q^2(1-q^{-2})^2}\langle \Gamma^1_1\Gamma^{1\dagger}_{\overline{2}}\rangle \langle \Gamma^1_2\Gamma^{1\dagger}_{\overline{1}}\rangle
        -\frac{1}{q^2(1-q^{-2})^2}\langle \Gamma^1_1\Gamma^{1\dagger}_{\overline{1}}\rangle \langle \Gamma^1_2\Gamma^{1\dagger}_{\overline{2}} \rangle\nonumber\\
        &+\frac{1}{(q^2-1)^2}\langle \Gamma^1_1\Gamma^{1\dagger}_{\overline{2}}\Gamma^1_2\Gamma^{1\dagger}_{\overline{1}} \rangle
    \end{align}
    Every term is either an OTOC, a product of two non-trivial correlators, or is manifestly $\order{1/q^2}$. The decorations on each site $r\neq 1$ may result in contributions that are either: (1) a trivial correlator; (2) a single non-trivial correlator; (3) a product of two non-trivial correlators. Note that none of these non-trivial correlators are OTOCs because they live on a contour with only a single forward and backward segments. Therefore, if any decoration on sites $r\neq 1$ does anything other than contribute trivial correlators, we have $\int dV \mathcal{D}^{2,1}_\Gamma(x=1,T)=\order{1/q^3}$. Keeping only $\order{1/q^2}$ contributions forces every site $r\neq 1$ to contribute trivial correlators only. This allows us to take the Haar average of \eqnref{(2,1) site 1} in isolation. To do this we find it useful to write the follows results (consequences of  theorem \ref{productofcorrelators}),
    \begin{align*}
        &\int dU \langle \Gamma\rangle\langle \Gamma'^\dagger\rangle
       =\delta^{\Gamma,\mathbb{1}}\delta^{\Gamma',\mathbb{1}} + \order{1/q^2},\\
        &\int dU \langle \Gamma\rangle = \delta^{\Gamma,\mathbb{1}} + \order{1/q^2},\\
        &\int dU \langle Z \Gamma Z \Gamma' \rangle = \delta^{\Gamma,\mathbb{1}}\delta^{\Gamma',\mathbb{1}} + \order{1/q^2}.
    \end{align*}
    Where all $\Gamma$ are products $Z(1)^{\alpha_1} \cdots Z(T-1)^{\alpha_{T-1}}$ for some binary string $\alpha = (\alpha^i_1,\cdots,\alpha^i_{T-1})$.
    Using theorem \ref{two-correlators} we find the useful result
    \begin{equation}
        \int dU \langle Z \Gamma\rangle\langle Z \Gamma'^\dagger\rangle =\frac{1}{q^2}\delta^{\Gamma,\Gamma'}(1-\delta^{\Gamma,\mathbb{1}}) + \order{1/q^4}.
    \end{equation}
    Using these results and theorem \ref{OTOCtheorem} for the Haar average of a physical OTOC, we find that at $\order{1/q^2}$, every term in \ref{(2,1) site 1} cancels, 
    \begin{align}
        \begin{matrix}
        \hspace{-3pt}q\bra{\phi_-}
        \end{matrix}
        \ \Gamma^1 \ 
        \begin{matrix}
        \ket{\phi_+(T)}
        \end{matrix}_{\textrm{Haar}} =& \  \frac{1}{q^2}\left[ \delta^{\Gamma^1_1 \Gamma^{1\dagger}_{\overline{1}}\Gamma^1_2 \Gamma^{1\dagger}_{\overline{2}},\mathbb{1}} - \delta^{\Gamma_1, \Gamma_{\overline{1}}}\delta^{\Gamma_2, \Gamma_{\overline{2}}}- \delta^{\Gamma_1, \Gamma_{\overline{2}}}\delta^{\Gamma_2, \Gamma_{\overline{1}}}\nonumber\right.\\
        &-\delta^{\Gamma^1_1\Gamma^{1\dagger}_{\overline{1}},\Gamma^1_{\overline{2}}\Gamma^{1\dagger}_2}(1-\delta^{\Gamma^1_1,\Gamma^1_{\overline{1}}})
        -\delta^{\Gamma^1_1\Gamma^{1\dagger}_{\overline{1}},\Gamma^1_{\overline{2}}\Gamma^{1\dagger}_2}(1-\delta^{\Gamma^1_1,\Gamma^1_{\overline{2}}})\nonumber\\
        &+\delta^{\Gamma^1_1,\Gamma^1_{\overline{2}}}\delta^{\Gamma^1_2,\Gamma^1_{\overline{1}}}
        +\delta^{\Gamma^1_1,\Gamma^1_{\overline{1}}}\delta^{\Gamma^1_2,\Gamma^1_{\overline{2}}}\nonumber+\delta^{\Gamma^1_1\Gamma^{1\dagger}_{\overline{1}} \Gamma^1_2\Gamma^{1\dagger}_{\overline{2}},\mathbb{1}}
        -\delta^{\Gamma^1_1,\Gamma^1_{\overline{2}}}\delta^{\Gamma^1_2,\Gamma^1_{\overline{1}}}\nonumber\\
        &\left.-\delta^{\Gamma^1_1,\Gamma^1_{\overline{1}}}\delta^{\Gamma^1_2,\Gamma^1_{\overline{2}}}\right] + \order{1/q^3}\nonumber\\
        =& \ \order{1/q^3}\label{7}.
    \end{align}
    and we find $\int dV \mathcal{D}^{2,1}(x=1,T)=\order{1/q^3}$.
    
\subsection{\texorpdfstring{$x\geq2$}{Lg}}\label{x>1}
    
    \begin{equation}\label{D21_x>2}
        \mathcal{D}_{\Gamma}^{2,1}(x>2,T)= g^2\times \left[ \ \begin{matrix}
        \vspace{-1mm}
        \\
    \textcolor{gray}{\textrm{site } 0} \ \ \\ 
    \textcolor{gray}{\textrm{site } 1}\ \ \\
    \\ \\ \\ \\
    \end{matrix} \ \ 
    \begin{matrix}
    \ \ \vdots\\
    \ \ \bra{+}\\ 
    \hspace{-5pt}q\bra{\phi_-}\\
    \ \ \vdots\\
    q\bra{-}\\
    \ \ \bra{-}\\
    \ \ \vdots
    \end{matrix}\quad
    \fbox{ $\begin{matrix}
    \\
    \\
    \\
    \hspace{-5pt}\Gamma\\
    \\
    \\
    \\
    \end{matrix}$}\
    \begin{matrix}
    \hspace{-20pt}\vdots\\
    \hspace{-20pt}\ket{+}\\
    \hspace{-20pt}\ket{+}\\
    \hspace{-20pt}\vdots\\
    \ket{\phi_+(T)}\\
    \hspace{-20pt}\ket{-}\\
    \hspace{-20pt}\vdots
    \end{matrix} \ \ 
    \begin{matrix}
    \\ \\ \\ \\
    \hspace{-10pt}\textcolor{gray}{\textrm{site } x}\\ 
    \hspace{-10pt}\textcolor{gray}{\textrm{site } x+1}\\
    \vspace{-2mm}
    \end{matrix}\right]
    \end{equation}

    Sites $1$ and $x$ each contribute factors of form $\textrm{OTOC}+\textrm{Corr}\times\textrm{Corr}'+\order{1/q^2}$ (see \eqnref{dec phi overlaps}). Our theorem for the Haar average of a product of correlators (theorem \ref{productofcorrelators}) implies that if any other site contributes a non-trivial correlator, the Haar average of the total contribution will be $\order{1/q^3}$ or smaller. Thus, working to  $\order{1/q^2}$, we will look for contributions where sites $r\neq 1,x$ give only trivial correlators. Moreover, theorem \ref{productofcorrelators} also implies that the leading order contribution comes from the $\textrm{OTOC}_1 \textrm{OTOC}_x$ cross term
    \begin{align}
        \int dU &\left(\textrm{OTOC}_1+\textrm{Corr}_1\times\textrm{Corr}'_1+\order{1/q^2}\right)\nonumber\left(\textrm{OTOC}_x+\textrm{Corr}_x\times\textrm{Corr}'_x+\order{1/q^2}\right)\nonumber\\
        = & \int dU \ \textrm{OTOC}_1\times \textrm{OTOC}_x + \order{1/q^3}
    \end{align}
    In summary, in evaluating the contributions for $x\geq 2$, we need only consider those terms in the decoration expansion corresponding to OTOCs on site $1,x$, and trivial correlators on all other sites. As in the $(a,b)=(4,4)$ calculation, we select decorations that leave the contours on sites $r>x$ ($r<1)$ undecorated by inserting the projector $\ket{+}\bra{+}$ ($\ket{-}\bra{-}$) between every Floquet layer. For sites $1<r<x$ the non-decoration condition is more delicate. For these sites, the input wiring configuration is of $-$ type and the output wiring configuration is of $+$ type giving an OTO type contour. The requirement that the OTO contour is undecorated (i.e., a trivial correlator) is equivalent to the decoration delta constraint $\delta^{\Gamma^r_1\Gamma^{r\dagger}_{\overline{2}}\Gamma^r_2\Gamma^{r\dagger}_{\overline{1}},\mathbb{1}}$. In \ref{decoration delta constraint appendix} we show that this decoration delta constraint can be rewritten as
    \begin{align}\label{OTO_delta_constraint}
        \delta^{\Gamma^r_1\Gamma^{r\dagger}_{\overline{2}}\Gamma^r_2\Gamma^{r\dagger}_{\overline{1}},\mathbb{1}} = \sum_{m=1}^n& \left(\prod_{t=1}^{m-1}\bra{-}\Gamma^r(t)\ket{-}\right)q\bra{-}\Gamma^r(m)\ket{0}\left(\prod_{t=m+1}^{n}\bra{+}\Gamma^r(t)\ket{+}\right)\nonumber\\
        &+ \prod_{t=1}^{n}\bra{-}\Gamma^r(t)\ket{-}.
    \end{align}
    
    The final term in \eqnref{OTO_delta_constraint} selects the decorations $\Gamma^r$ that never decorate the initial state $\bra{-}$, so that at each time step the $\bra{-}$ wirings never carry any non-identity operators. With these $\ket{-} \bra{-}$ projectors in place, let us
    sum over all decorations $\Gamma$ with the coefficients $C_\Gamma$ of \eqnref{decexpansion}, in doing so we replace each of the decoration layers with the full unitary layers $U(t)$. This is pictured below, where we have highlighted site $x$, with its terminating state $\ket\phi_+(T)$.
    \begin{equation}\label{append3_2_minusminus_proj_insertion}
    \raisebox{-0.78\totalheight}{\includegraphics[height=5cm]{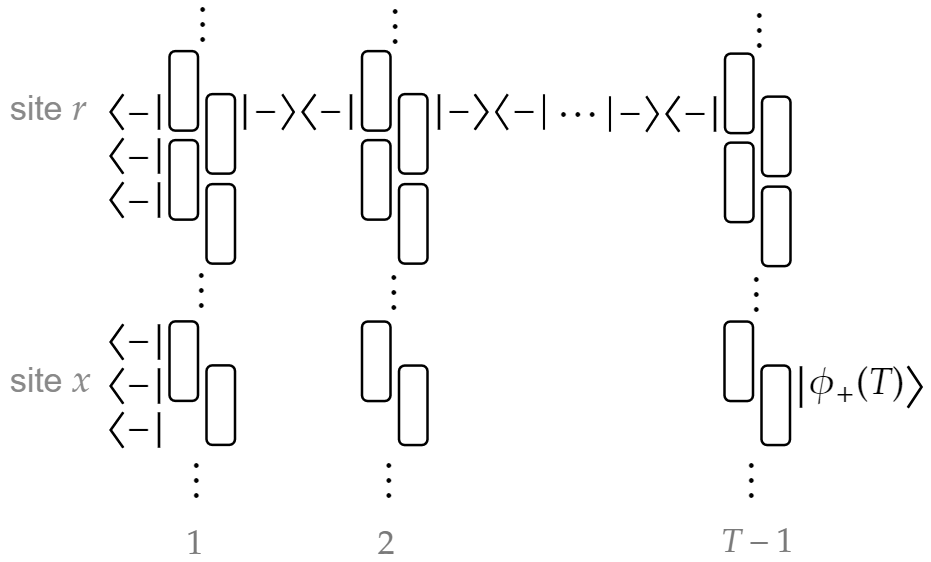}} =\ \raisebox{-0.6\totalheight}{\includegraphics[height=2.9cm]{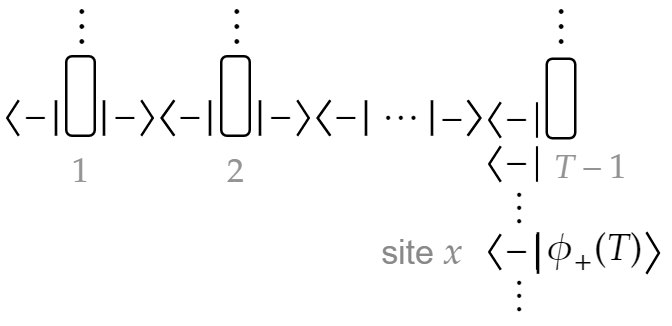}}
    \end{equation}
    Where, crucially, the brick property of \eqnref{rules_away_from_cut} can be used to remove every brick to the right of site $r$ (below site $r$ in the diagram above). This yields the right hand-side of the equation above. A consequence of which is that the terminating states $\ket{\phi_+(T)}$ on site $x$ is contracted directly with $\bra{-}$. Using \eqnref{nodecphi}, we see that this diagram vanishes. Therefore, the decorations selected by the final term of \eqnref{OTO_delta_constraint} cannot contribute to $\mathcal{D}^{2,1}(x>2,T)$. In what follows, we will consider only the decorations selected by the sum in \eqnref{OTO_delta_constraint}.
    
    As previously noted, only the $\textrm{OTOC}_1\times \textrm{OTOC}_x$ terms can contribute at $\order{1/q^2}$. This allows us to drop all but the $\bra{-}Z^{\otimes 2}$ term in the $\bra{\phi_{-}}$ state (see \eqnref{phi}) of site $1$ and the $Z(T)^{\otimes 2}\ket{+}$ term in the state $\ket{\phi_+(T)}$ of site $x$. The fact that $\textrm{OTOC}_1$ begins with $\bra{-}Z^{\otimes 2}$ and $\textrm{OTOC}_x$ ends with $Z(T)^{\otimes 2}\ket{+}$ and the condition that these OTOCs are complex conjugates of each other (using theorem \ref{two-correlators}) forces both OTOCs to be physical OTOCs of the same length (the length of physical OTOCs is the difference between the latest and earliest time appearing in the OTOC). We will sum over all possible OTOC lengths $\tau$, $1\leq\tau\leq T-1$.
    
    We now describe how we select only those decorations that produce a physical OTOC with length $\tau$. For $\textrm{OTOC}_1$, we must ensure that every unitary layer $t>\tau$ does not decorate the $\ket{+}$ wirings on site $1$. We do this by inserting the projector $\ket{+}\bra{+}$ to the left of each of these layers. Requiring then that the $\tau$-th unitary layer decorates the $\ket{+}$ wiring by leaving $Z(\tau)^{\otimes 2}\ket{+}$ is achieved by sandwiching the layer with $q\bra{\perp}$ and $\ket{+}$. Making all of these selections, the contribution from site $1$ is takes the form shown below,
    \begin{equation}
    \small
    q\bra{-}Z^{\otimes2}\Gamma^1(1)\cdots\Gamma^1(\tau-1)Z(\tau)^{\otimes 2}\ket{+}q\bra{\perp}\Gamma^1(\tau)\ket{+}\bra{+}\Gamma^1(\tau+1)\ket{+}\cdots\bra{+}\Gamma^1(T-1)\ket{+}.
    \end{equation}
    We use the same strategy to select decorations that contribute physical OTOCs of length $\tau$ on site $x$ as well. The resulting contribution takes the form shown below, where we have defined $\tau'=T-\tau$,
    \begin{equation}
    \small
    \bra{-}\Gamma^x(1)\ket{-}\cdots\bra{-}\Gamma^x(\tau'-1)\ket{-}q\bra{-}\Gamma^x(\tau')\ket{0}q\bra{-}Z(\tau')^{\otimes2}\Gamma^x(\tau'+1)\cdots\Gamma^x(T-1)Z(T)^{\otimes 2}\ket{+}.
    \end{equation}
    Rather than focus on a single decoration $\Gamma$, we are able to select every $\order{1/q^2}$ to $\mathcal{D}^{2,1}(x=2,T)$ simultaneous by summing over decorations $\Gamma$ with the appropriate coefficients $C_\Gamma$ (as introduced in the decoration expansion in \eqnref{decexpansion}),
    \begin{equation}
        \mathcal{D}^{2,1}(x=2,T)=\sum_{\Gamma} C_\Gamma \mathcal{D}^{2,1}_\Gamma(x=2,T).
    \end{equation}
    Each of these layers is contracted by various combinations of the $+,-,\perp$ and $0$ states. We introduce a short-hand for each of these contractions, this is given below,
    \begin{equation}
     \raisebox{-0\totalheight}{\includegraphics[height = 3.5cm]{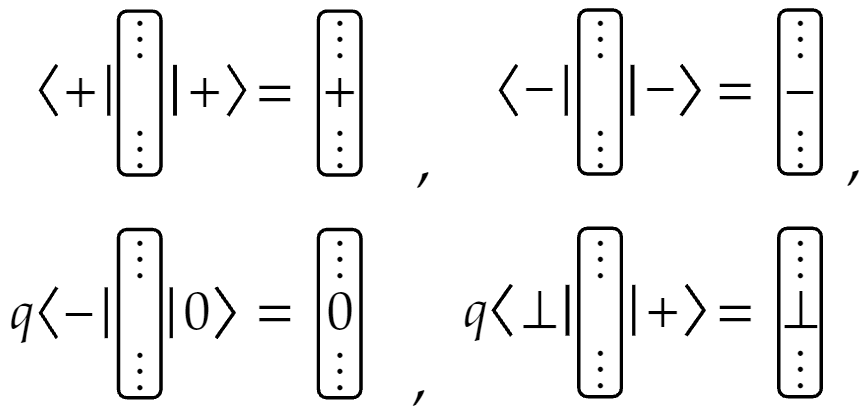}}.
    \end{equation}
    One further short-hand we use is $\bra{Z_t^-}\equiv\bra{-}Z(t)^{\otimes2}$ and $\ket{Z_t^+}\equiv Z(t)^{\otimes2}\bra{+}$. The contractions (of the unitary layers) at site $1$ now take the more readable form,
    \begin{equation}
        \raisebox{-0.45\totalheight}{\includegraphics[height=2cm]{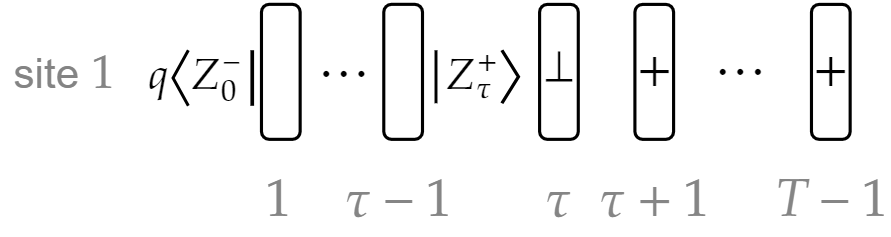}}.
    \end{equation}
    The short-hand version for site $x$ is found similarly, but with $T-\tau-1$ `$-$' contractions, followed by a `$0$' contraction on layer $T-\tau$, followed by the OTOC. We now also apply this short-hand to the contributions on site $r$, $1<r<x$, in particular, this yields
    \begin{equation}
        \sum_{t_r=1}^{T-1}\raisebox{-0.6\totalheight}{\includegraphics[height=2cm]{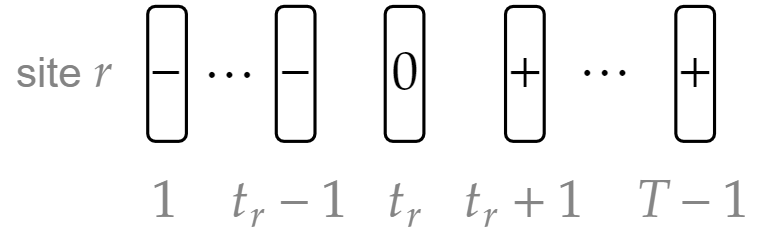}},
    \end{equation}
    where we have discarded the decoration that never decorates the initial state, as previously discussed.

    By keeping only the $\order{1/q^2}$ contributions to $\mathcal{D}^{2,1}(x\geq 2,T)$, have found a set of diagrams labelled by: $\tau$, the length of each of the physical OTOCs; $t_r$ for each site $r\in \{2,\cdots, x-1\}$, the positions the `$0$' contraction on site $r$. Setting $t_1=\tau$ and $t_x=T-\tau$, we label each diagram by a sequence $(t_1,t_2,\cdots,t_{x-1},t_x)$. An example diagram, labelled $(t_1,t_2,t_3,t_4,t_5,t_6)=(5,3,6,2,8,7)$, that contributes to $\mathcal{D}^{2,1}(x=6,T=12)$ is given below
    \begin{figure}[H]
    \centering
    \raisebox{-0.45\totalheight}{\includegraphics[height = 5cm]{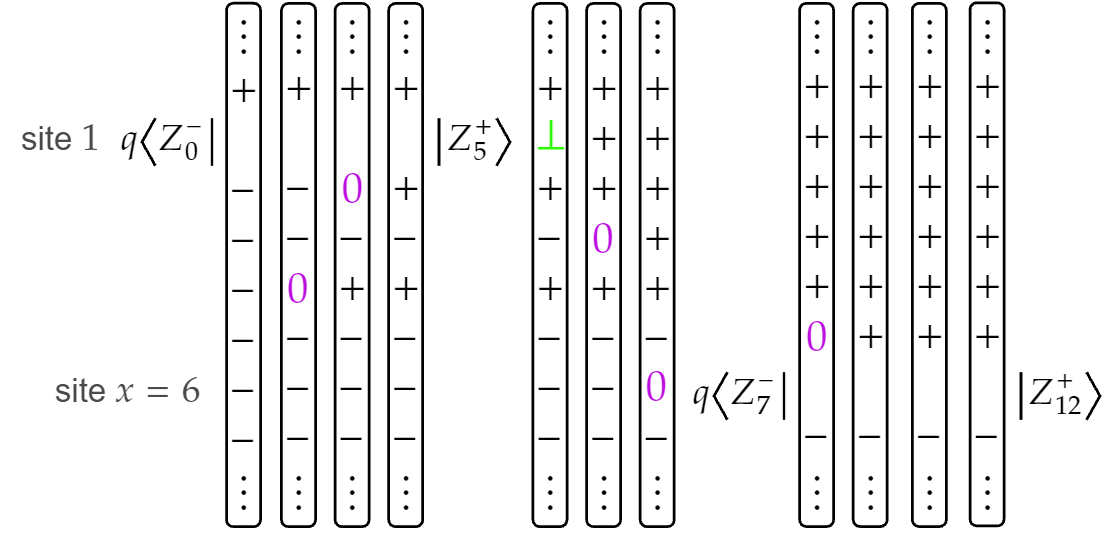}}\caption{The contribution $(t_1,t_2,t_3,t_4,t_5,t_6)=(5,3,6,2,8,7)$ to $\int dV \mathcal{D}^{(2,1)}(x=6,T=12)$}\label{tensor-contraction_diagram}
    \end{figure}

    In fact, for $x>2$, only contributions where $t_1\leq t_2<t_3<\cdots<t_x$ are non-zero. One of the following motifs must appear in any diagram not satisfying this property, each of which is zero,
    \begin{align}\label{layer contractions}
    \raisebox{-0.45\totalheight}{\includegraphics[height = 2cm]{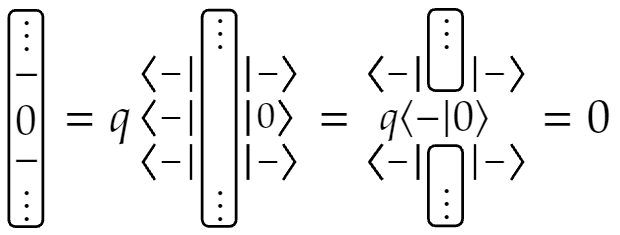}}, &\quad \raisebox{-0.45\totalheight}{\includegraphics[height = 2cm]{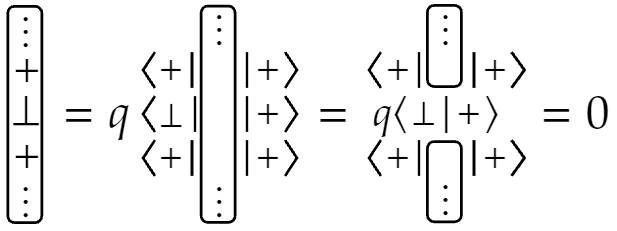}},\nonumber\\  &-\hspace{-3cm}\raisebox{-0.45\totalheight}{\includegraphics[height = 2cm]{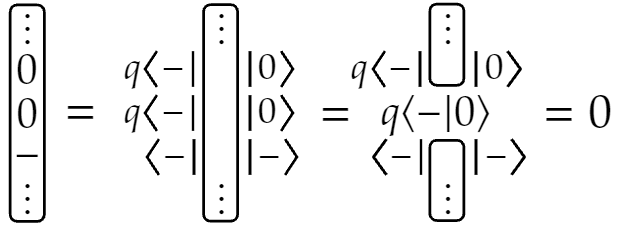}}.
    \end{align}
    In the non-zero diagrams (i.e., those for which $t_1\leq t_2<t_3<\cdots<t_x$) the contracted unitary layers collapse into contractions of only a short portion of the full layer this follows from the brick property \eqnref{rules_away_from_cut}. We demonstrate this process by collapsing a semi-infinite domain of `$+$' contractions below,
    \begin{equation}
        \raisebox{-0.45\totalheight}{\includegraphics[height=3.2cm]{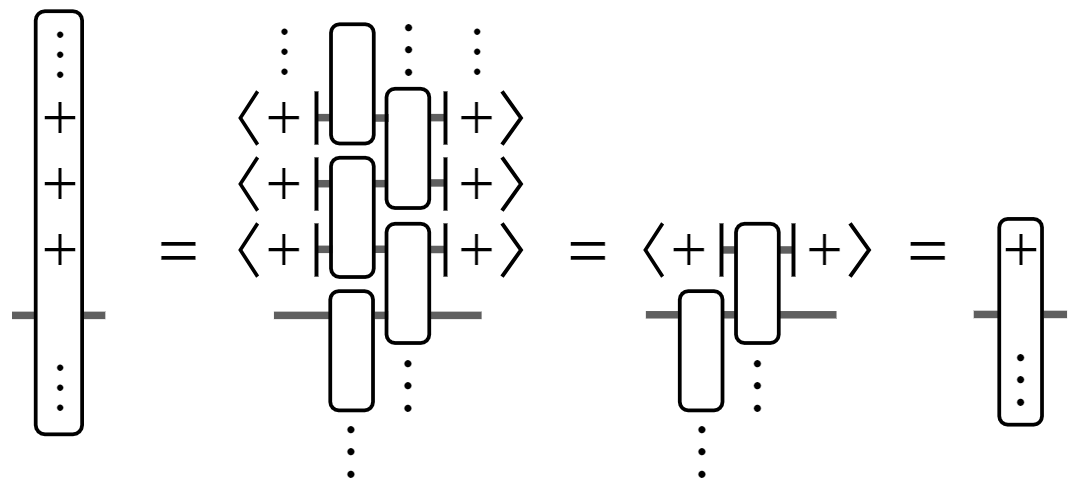}}.
    \end{equation}
    We collapse the `$-$' domain in the same way. Ultimately, every layer reduces to one of the motifs below, 
    \begin{equation}
    \raisebox{-0.45\totalheight}{\includegraphics[height=1.2cm]{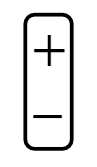}}=1+g, \quad
    \raisebox{-0.45\totalheight}{\includegraphics[height = 1.6cm]{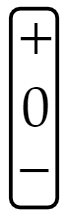}}=\raisebox{-0.45\totalheight}{\includegraphics[height = 1.6cm]{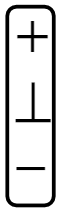}}=-g, \quad\raisebox{-0.45\totalheight}{\includegraphics[height=1.8cm]{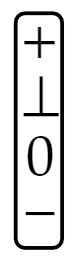}}=g, \quad  M(t) \equiv \raisebox{-0.55\totalheight}{\includegraphics[height = 1.8cm]{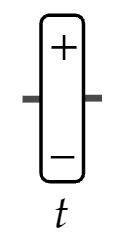}},\label{Mdef}
    \end{equation}
    where $M(t)$ is a single site operator. Each diagram as a whole decomposes into a products of the motifs below, which we use to introduce a compact diagrammatic notation. We also give the numerical value of each motif.
    \begin{align}\label{motif}
        & \ \raisebox{-0.45\totalheight}{\includegraphics[height=1.4cm]{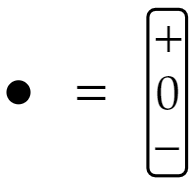}}=-g, \quad \raisebox{-0.2\totalheight}{\includegraphics[height=1.05cm]{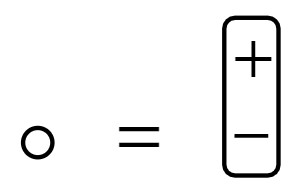}}=1+g,\nonumber\\
        &\raisebox{-0.45\totalheight}{\includegraphics[height=1.5cm]{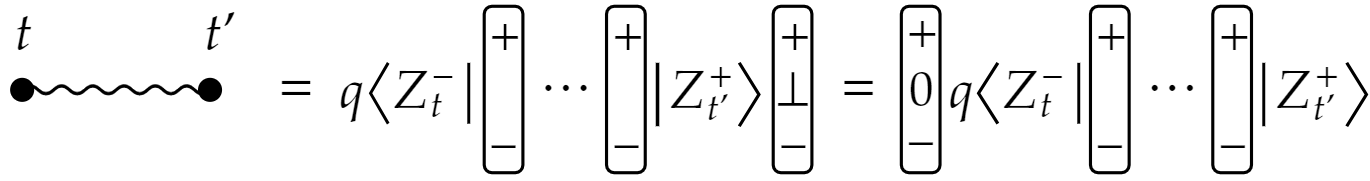}}=-g\lambda(t'-t)\nonumber,\\
        &\raisebox{-0.55\totalheight}{\includegraphics[height=2cm]{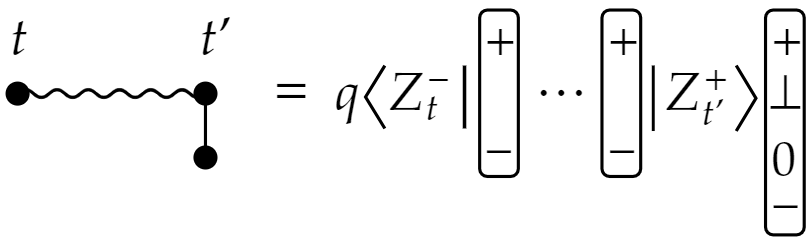}}=g\lambda(t'-t)=-\raisebox{-0.16\totalheight}{\includegraphics[height=1cm]{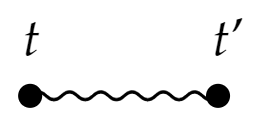}},
    \end{align}
    where we have defined $\lambda(t-t')=q\bra{Z^-_{t}}M(t+1)\cdots M(t'-1)\ket{Z^+_{t'}}$. To give an example of the correspondence between the contracted unitary layer diagrams and this new diagrammatic notation, consider the diagram corresponding to the sequence $(t_1,\cdots,t_4)=(3,3,6,7)$ below
    \begin{equation}\label{equivalency of notation}
    \raisebox{-0.45\totalheight}{\includegraphics[height = 3.7cm]{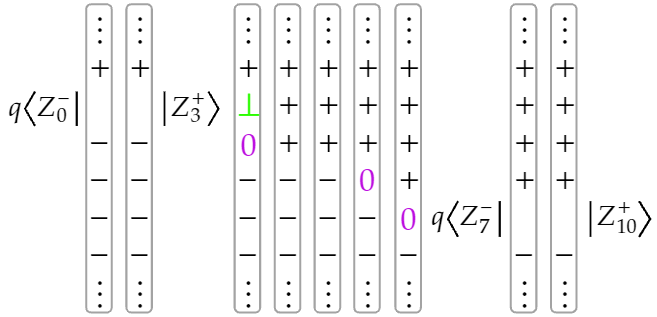}}=\raisebox{-0.45\totalheight}{\includegraphics[height = 2.2cm]{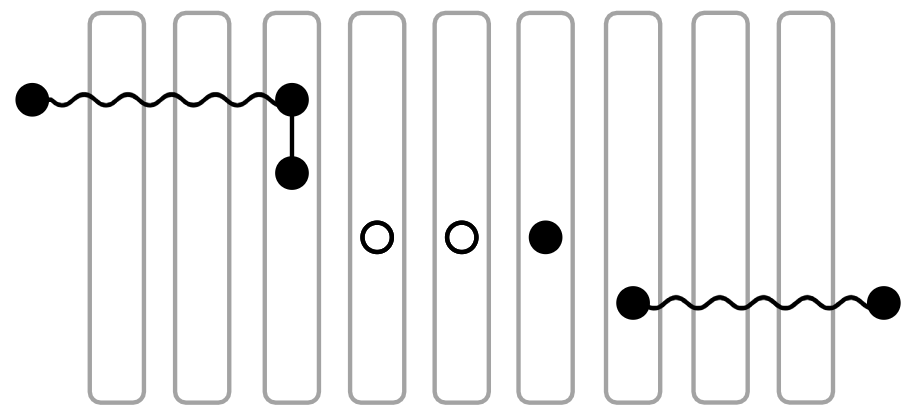}},
    \end{equation}
    
    For $x>2$, the diagrams come in two qualitatively different types: (1) $t_1<t_2$ and (2) $t_1=t_2$.  All type 1 diagrams have the following diagrammatic form
    \begin{equation}\label{type-1}
    \raisebox{-0.45\totalheight}{\includegraphics[height = 2.2cm]{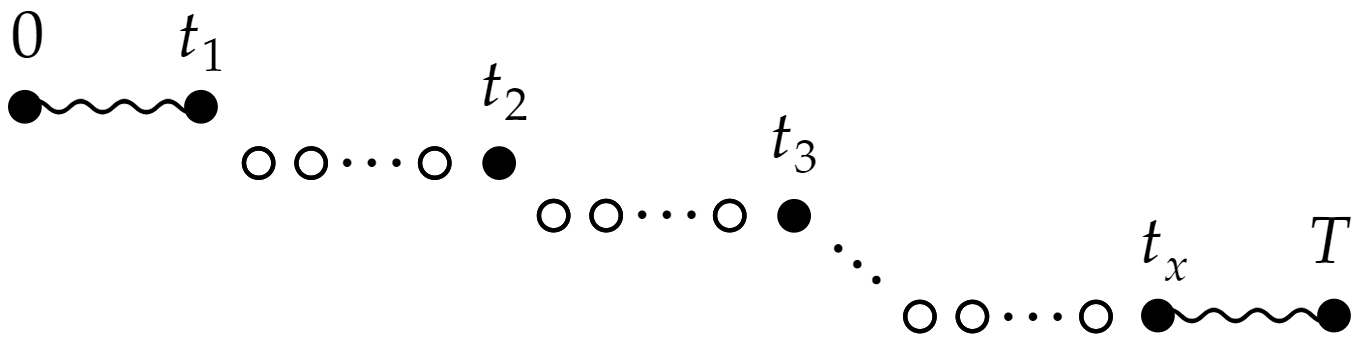}},
    \end{equation}
    where $T-t_x=t_1$. Whereas, type 2 diagrams have the form
    \begin{equation}\label{type-2}
    \raisebox{-0.45\totalheight}{\includegraphics[height = 2.2cm]{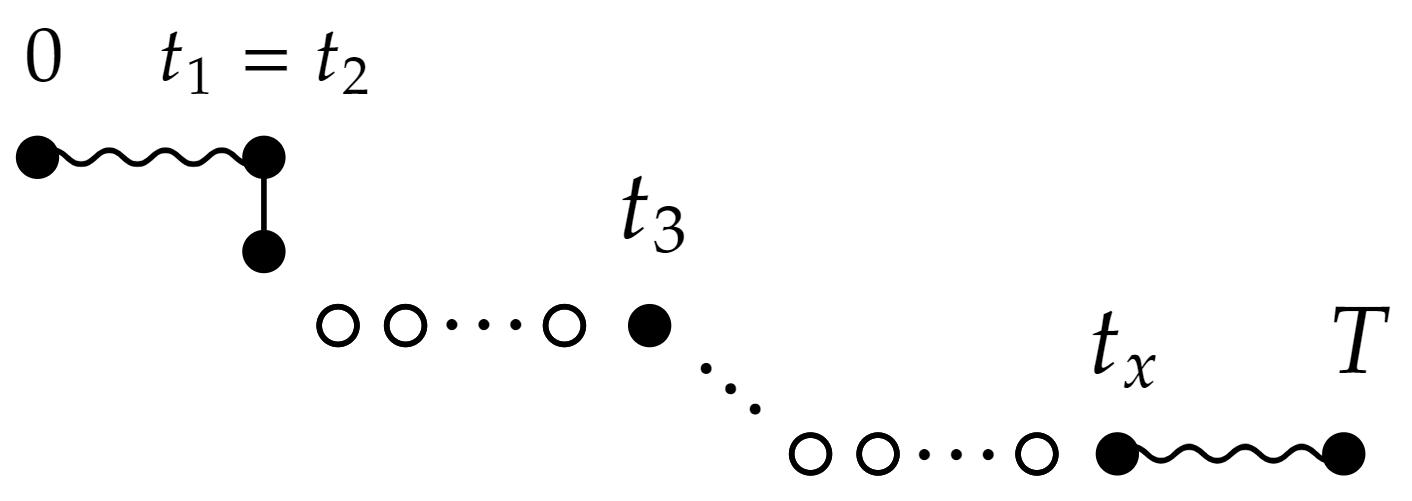}}.
    \end{equation}
    
    For $x=2$, we will find diagrams with similar motifs. We had found that for $x\geq 2$, the relevant diagrams are labelled by a sequence $(t_1,t_2,\cdots,t_x)$, with $t_x=T-t_1$ and where $t_1=\tau$ is the length of the OTOCs. For $x=2$ the diagrams are simply labelled by $(\tau,T-\tau)$. A complication for $x=2$ is the fact that the OTOCs may overlap in time. This is because no matter where we position the `$\perp$' contraction of site $1$ and `$0$' contraction of site $2$, we can never encounter any of the vanishing motifs of \eqnref{layer contractions}, which in the case of $x>2$ force $t_1\leq t_2$.
    
    To address this complication, we split $x=2$ in three types of diagram: (1) $T-\tau>\tau$, the OTOCs do not overlap -- the treatment of these diagrams is exactly the same the type-1 diagrams discussed for $x>2$; (2) $\tau=T-\tau$, the OTOC's `touch' -- this is similar to the type-2 diagrams in $x>2$; (3) $\tau>T-\tau$, the OTOCs overlap.
    
    The touching OTOC contributions ($\tau=T-\tau$) have the following form/motif,
    \begin{equation}\label{touchingOTOCs}
        \raisebox{-0.45\totalheight}{\includegraphics[height=1.9cm]{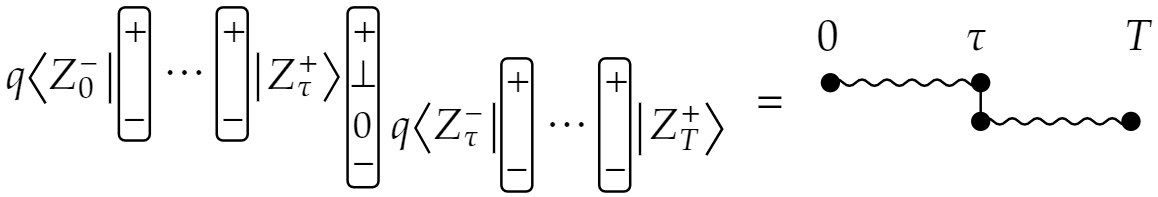}}=g\lambda(t''-t')\lambda(t'-t).
    \end{equation}
    We give the overlapping OTOC contributions ($\tau>T-\tau$) the following compact notation,
    \begin{equation}
        \raisebox{-0.15\totalheight}{\includegraphics[height=1.3cm]{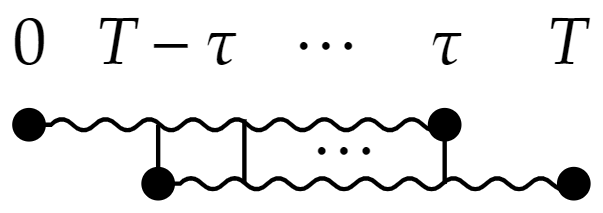}}\ .
    \end{equation}
    We study these contributions in detail in the next section, \ref{overlapping OTOCs}. 
    
    We are interested in computing $\int dV \sum_x\mathcal{D}^{(2,1)}(x,T)$, we have seen earlier in this appendix that the $x=0,1$ contributions are $\order{1/q^3}$ (as are the $x<0$ contributions, \eqnref{x<0}). Therefore, at $\order{1/q^2}$, we need only sum over $x\geq 2$. Fortunately, there is an abundance of cancellation between these diagrams. We will cover some examples and then give the general result. Starting with the simplest, we compute $\int dV \sum_x\mathcal{D}^{(2,1)}(x,T)$ for $T=2,3$ and $4$ explicitly (where the Haar average is implied, but not written below). The additional factor of $g^2$ in \eqnref{D21_x>2} has been divided through in the equations below.
    
    \begin{align}
    \sum_x\mathcal{D}^{(2,1)}(x,T=2)/g^2&=
    \raisebox{-0.3\totalheight}{\includegraphics[height = 1.1cm]{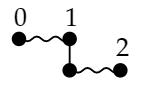}}\nonumber\\
    \sum_x\mathcal{D}^{(2,1)}(x,T=3)/g^2&=
    \raisebox{-0.38\totalheight}{\includegraphics[height = 1.4cm]{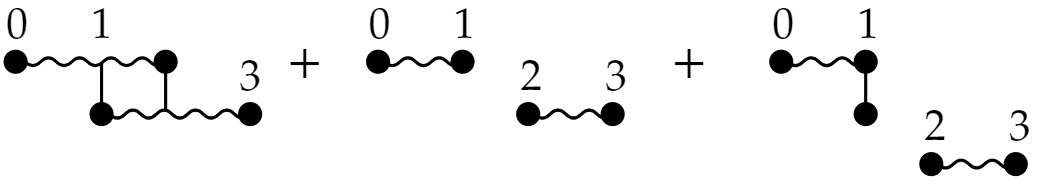}}\nonumber\\
    &=\raisebox{-0.38\totalheight}{\includegraphics[height = 1cm]{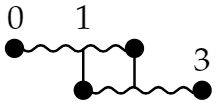}}\nonumber\\
    \sum_x\mathcal{D}^{(2,1)}(x,T=4)/g^2&=
    \raisebox{-0.7\totalheight}{\includegraphics[height = 2.8cm]{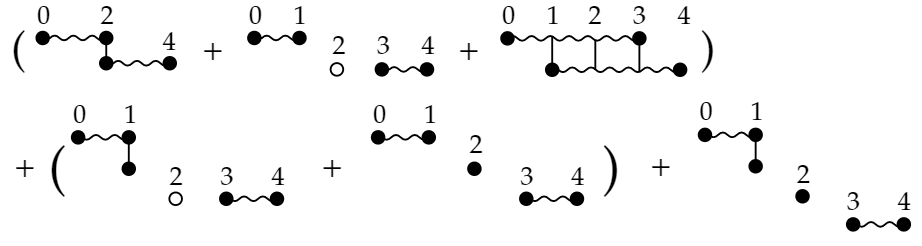}}\nonumber\\&=\raisebox{-0.28\totalheight}{\includegraphics[height = 1cm]{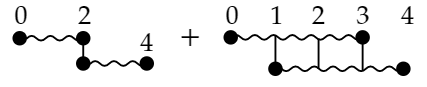}}.\nonumber
    \end{align}
    We have used the third rule of \eqnref{motif} to cancel the two terms (associated with $x=2$ and $x=3$) for $T=2$ and to cancel the second and fourth terms and the fifth and sixth terms for $T=4$. Notice that the terms that remain after cancellation are all connected diagrams (i.e., the OTOCs either touch or overlap), all the diagrams with `gaps' (i.e., where a vertical line can be drawn through them without intersecting a wobbly line, representing an OTOC) have conspired to cancel. This is no coincidence, it is a consequence of the fact that processes that contribute to $\Sigma$ (and hence the corrections to $v_B$) must explore only the fast space, this is due to the $Q$ projectors that project out all slow components at every time-step in $\Sigma$. Diagrams with a gap represent processes that take a detour to the slow space. We can see this by returning to the contracted unitary layer picture; take, for example, the $T=3$ diagram associated with $(t_1=1,t_2=2)$, this is the first diagram in the $T=3$ sum above. It is equivalently given by
    \begin{equation}
    \raisebox{-0.45\totalheight}{\includegraphics[height = 3.4cm]{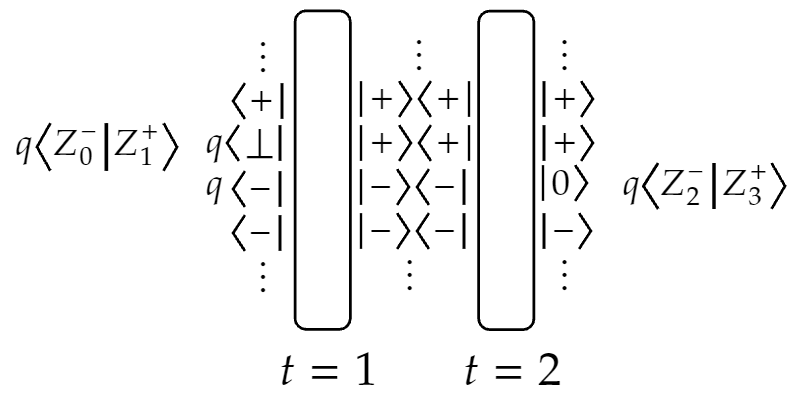}}
    \label{detour_to_P}
    \end{equation}
    Between the two unitary layers we have vectors that are clearly in $\mathcal{P}$. The only diagrams that have no gaps are those for $x=1$ (which we have seen all vanish at $\order{1/q^2}$) and the overlapping or touching OTOC diagrams for $x=2$. We will next evaluate the overlapping OTOC diagram contributions, before finally calculating the touching OTOC diagram contributions.
    
\subsection{Overlapping OTOC diagrams.}\label{overlapping OTOCs}

    In this section we investigate the contributions to $\mathcal{D}^{(2,1)}(x=2,T)$ that take the form of overlapping OTOCs. We name this contribution $\mathcal{D}^{(2,1)}_{\textrm{O}}(x=2,T)$. These OTOC overlap diagrams are given in detail below for OTOC length $\tau$ and total diagram length $T$.
    \begin{equation}
        \int dV \mathcal{D}^{(2,1)}_{\textrm{O}}(x=2,T) \approx g^2 \int dV \sum_{\tau=1}^{T-1} \  \raisebox{-0.55\totalheight}{\includegraphics[height = 2cm]{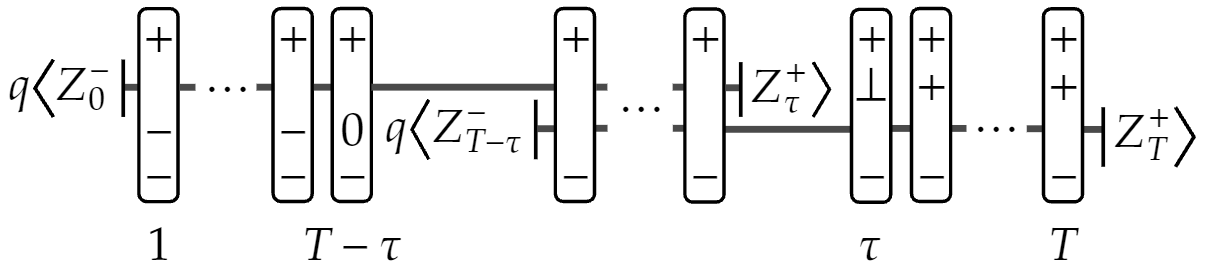}}
    \end{equation}
    It will be useful to define the single site operators $M_+(\tau)$ and $M_-(\tau')$ and repeat the definition for $M(t)$ seen in \eqnref{Mdef}.
    \begin{equation}
    M_-(\tau) \equiv \raisebox{-0.55\totalheight}{\includegraphics[height = 2.4cm]{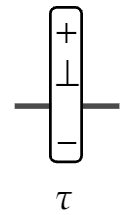}},\quad M_+(\tau') \equiv \raisebox{-0.55\totalheight}{\includegraphics[height = 2.4cm]{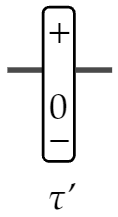}}, \quad
    M(t) \equiv \raisebox{-0.55\totalheight}{\includegraphics[height = 2.0cm]{Diagrams/Mdef.png}}.
    \label{Transfer_matrix_contraction_2}
    \end{equation}
    The Floquet layers $T-\tau+1$ through to $\tau-1$ have been reduced to two site operators which has already been introduced in \eqnref{calTdef} and named $\mathcal{T}(t)$ for Floquet layer $t$. With these definitions we have
    \begin{equation}\label{Transfer_matrix_tensor_1}
         \int dV \mathcal{D}^{(2,1)}_{\textrm{O}}(2,T) \approx g^2 \int dV \sum_{\tau=1}^{T-1} \  \raisebox{-0.55\totalheight}{\includegraphics[height = 2cm]{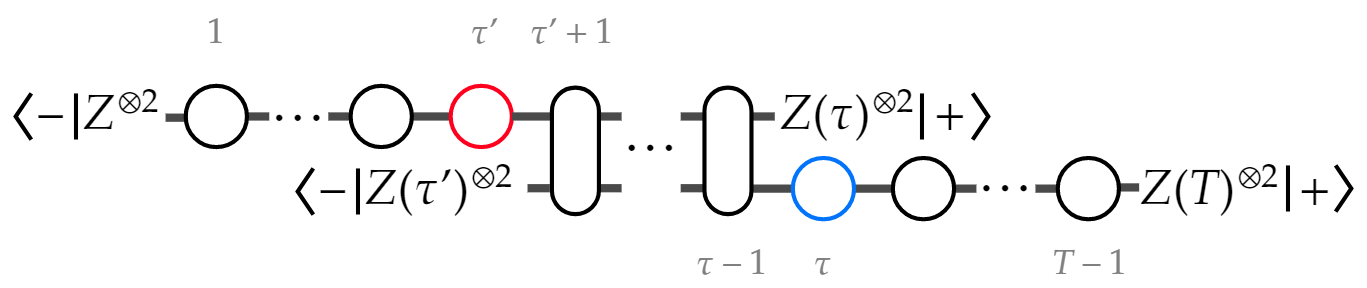}},
    \end{equation}
    where we have denoted the following,
    \vspace{-2mm}
    \begin{figure}[H]
    \centering
    \includegraphics[height = 1.5cm]{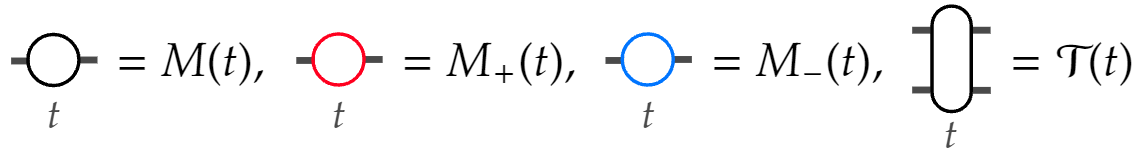}.
    \label{Transfer_matrix_definitions}
    \end{figure}
    \vspace{-4mm}
    Using theorem \ref{two-correlators} of \secref{theorems}, the Haar average of a product of two physical OTOCs is found to $\order{1/q^2}$ to be
    \begin{align}
        \int dV \begin{matrix}
        \langle Z \Gamma^1_1 Z(T) \Gamma^{1\dagger}_{\overline{2}} Z \Gamma^1_2 Z(T) \Gamma^{1\dagger}_{\overline{1}} \rangle\\ 
        \langle Z \Gamma^2_1 Z(T) \Gamma^{2\dagger}_{\overline{2}} Z \Gamma^2_2 Z(T) \Gamma^{2\dagger}_{\overline{1}} \rangle
        \end{matrix} =
        \frac{1}{q^2}
        &\left(\delta^{\Gamma^1_1,\Gamma^2_{\overline{1}}}\delta^{\Gamma^1_{\overline{2}},\Gamma^2_2}\delta^{\Gamma^1_2,\Gamma^2_{\overline{2}}}\delta^{\Gamma^1_{\overline{1}},\Gamma^2_1}\right.\nonumber\\
        &\quad +\left.\delta^{\Gamma^1_1,\Gamma^2_{\overline{2}}}\delta^{\Gamma^1_{\overline{2}},\Gamma^2_1}\delta^{\Gamma^1_2,\Gamma^2_{\overline{1}}}\delta^{\Gamma^1_{\overline{1}},\Gamma^2_2}\right) + \order{1/q^3}.\label{Av_two_OTOCs}
    \end{align}
    Each of the decoration delta constraints can be implemented as described in \ref{decoration delta constraint appendix}. In doing so, we sandwich each decoration layer with a wiring configuration labelled $A$ for the first term in \eqnref{Av_two_OTOCs} and by a configuration labelled $B$ for the second term. We then use the shorthand below.
    \begin{equation}
    \int dV \raisebox{-0.55\totalheight}{\includegraphics[height = 1.8cm]{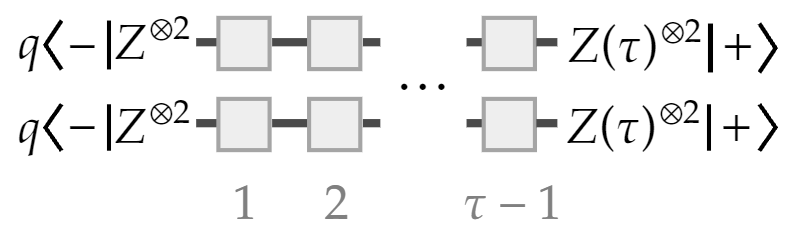}} = \frac{1}{q^2} \sum_{a=A,B} \raisebox{-0.55\totalheight}{\includegraphics[height = 1.7cm]{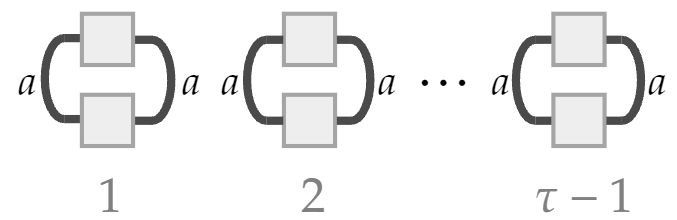}},
    \label{Haar_average_OTOC_product}
    \end{equation}
    where the grey boxes are placeholders for the possible decorations at each layer. A grey box (labeled $t$) may decorated each of the incoming legs $1$ and $2$ with $Z(t)$ and each of the legs $\overline{1}$ and $\overline{2}$ with $Z(t)^*$. On the right-hand side, every super-leg carries a label $a$ which labels one of the permutations $A, B$ of the legs $1,\overline{1},2,\overline{2}$ given in \eqnref{superleg} below,
    \begin{equation}\label{superleg}
    \centering
    \includegraphics[height = 0.9cm]{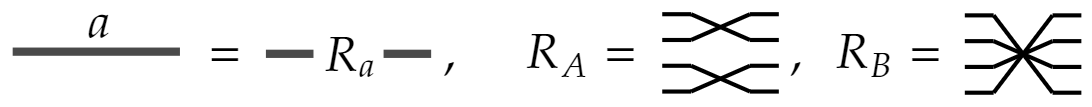}.
    \end{equation}
    Each super-leg is implicitly carrying a factor $1/q^2$. The contraction between decorations within a column is given more explicitly below
    \begin{equation}
    \centering
    \raisebox{-0.42\totalheight}{\includegraphics[height = 1.2cm]{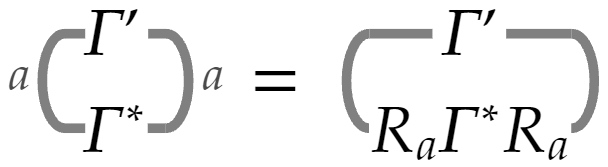}}=\Tr(R^a \Gamma^\dagger R^a \Gamma')/q^4=\bra{R^a \Gamma R^a}\ket{\Gamma'}.
    \end{equation}
    All of the onsite scrambling evolution can be dropped on the right-hand-side of \eqnref{Haar_average_OTOC_product} as the leg contractions are between decorations at the same time. In the case of \eqnref{Transfer_matrix_tensor_1}, the OTOCs are off-set by $\tau'=T-\tau$ Floquet time-steps. These OTOCs can be brought into alignment by globally shift the time arguments in the OTOCs, then we can use \eqnref{Haar_average_OTOC_product}. After Haar averaging, and shifting the OTOCs back to their original positions, the leg contractions will stretch over $\tau'$ steps, as shown below.
    \begin{equation}
    \frac{g^2}{q^2}\sum_{s\geq 0}\sum_{t\geq 0}\delta^{s+2t+3,T}\sum_{a=A,B} \raisebox{-0.55\totalheight}{\includegraphics[height = 2cm]{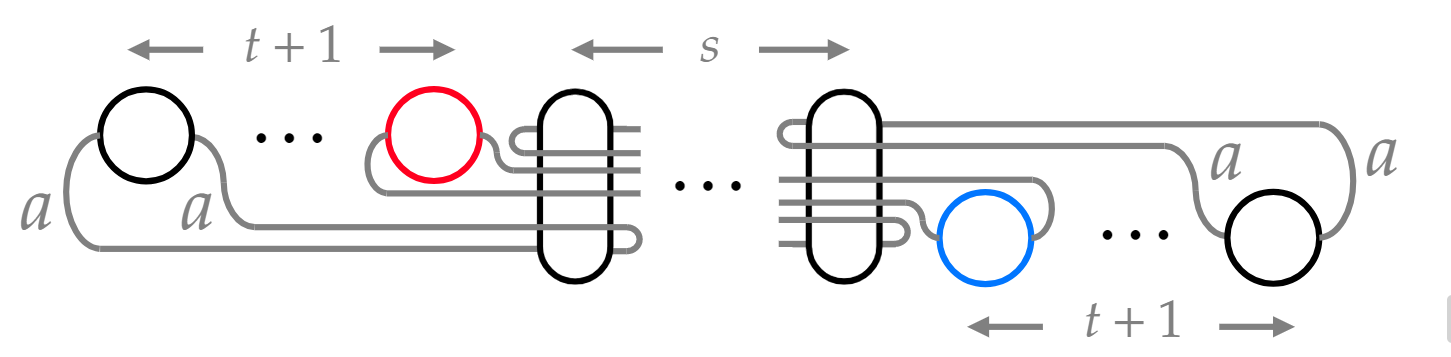}},
    \label{Transfer_matrix_contraction_5}
    \end{equation}
    where all scrambling evolution time arguments have been dropped. The new variables are related to those in \eqnref{Transfer_matrix_tensor_1} by $t=T-\tau-1\geq 0$ and $s=2\tau-T-1\geq 0$. As an example, we show the $t=1$, $s=2$ case,
    \begin{equation}
    \raisebox{-0.4\totalheight}{\includegraphics[height = 1.4cm]{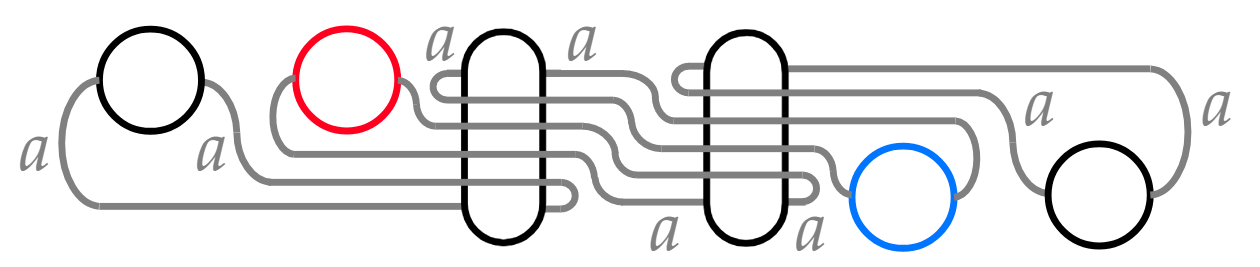}} = \raisebox{-0.4\totalheight}{\includegraphics[height = 1.3cm]{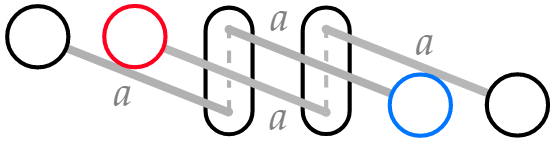}},
    \end{equation}
    where on the right hand side we have introduced a shorthand which makes obvious the decomposition into two chains of tensor contractions. In general, the Haar average decomposes into $t+1$ chains. By summing over $T\geq 1$, we are able to drop the delta constraint above, doing this sum is equivalent to calculating the Laplace transformed $\int dV \mathcal{D}^{2,1}(x=2,T)$ at $z=0$, i,e $\int dV \mathcal{D}^{2,1}(x=2,z=0)$. This is sufficient for calculating $v_B$. Before we determine the decomposition for a general $s\geq 0$ and $t\geq 0$, we first define the following chains,
    \begin{align}
    &\raisebox{-0.4\totalheight}{\includegraphics[height = 1.6cm]{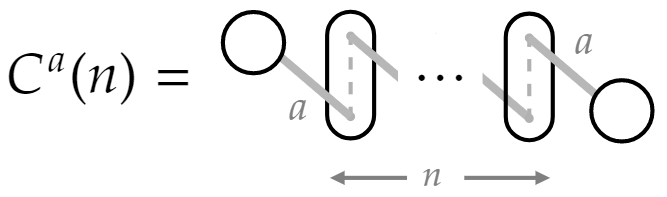}}, \quad \raisebox{-0.4\totalheight}{\includegraphics[height = 1.6cm]{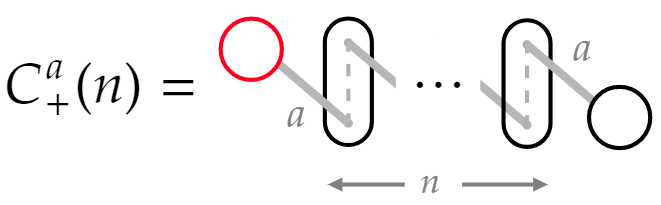}},\nonumber\\
    &\raisebox{-0.4\totalheight}{\includegraphics[height = 1.6cm]{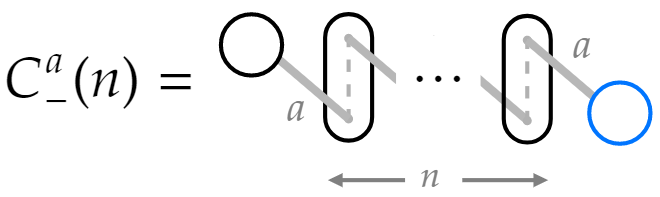}}, \quad \raisebox{-0.4\totalheight}{\includegraphics[height = 1.6cm]{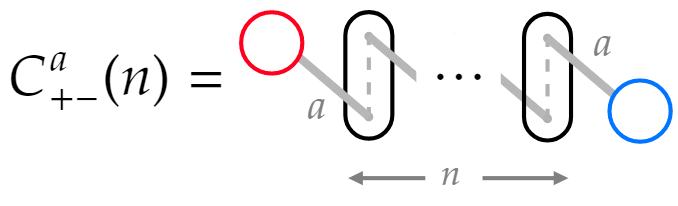}}.
    \end{align}
    For any $(s,t)$ There are two qualitatively distinct types of contribution to \eqnref{Transfer_matrix_contraction_5}: (1) $t+1 \equiv 0 \pmod{s+1}$, a product of $t$ $C^a$ chains and one $C^a_{+-}$ chain; (2) otherwise, a product of $t-1$ $C^a$ chains, one $C^a_+$ chain and one $C^a_-$ chain. For case 1, the $C^a_{+-}$ chain is $n=\frac{s+1}{t+1}-1\geq0$ transfer matrices long, i.e., $C^a_{+-}(n)$, while all $t$ of $C^a$ chains are $n+1$ matrices long, $C^a(n+1)$. For case 2 with $s+1=n(t+1)+k$ for $1\leq k\leq t$ and starting from the left, the first $k-1$ $M$'s are on chains of length $n+1$ and terminate on an $M$. The $k$-th $M$ sits on a chain of length $n$ and terminates on $M_-$, the following $t-k$ $M$'s sit on chains length $n$ that terminate on an $M$. Finally, the $M_+$ sits on a chain of length $n$ and terminates on an $M$. All together, the contribution is $C^a_+(n)C^a_-(n)C^a(n+1)^{k-1}C^a(n)^{t-k}$. This is summarised below,
    \begin{equation}
    \frac{g^2}{q^2}\sum_{a=A,B}\sum_{n\geq0}\left[\sum_{t\geq 0}C^a_{+-}(n)C^a(n+1)^t + \sum_{t\geq 1}\sum_{k=1}^t C^a_+(n)C^a_-(n)C^a(n+1)^{k-1}C^a(n)^{t-k}\right].
    \end{equation}
    In both cases, all but the $n$ sums can be evaluated to give,
    \begin{align}
    \int dV \mathcal{D}^{(2,1)}_{\textrm{O}}&(x=2,z=0)=\nonumber\\
    &\frac{g^2}{q^2}\sum_{a=A,B}\sum_{n\geq0}\left[\frac{C^a_{+-}(n)}{1-C^a(n+1)} + \frac{C^a_+(n)C^a_-(n)}{(1-C^a(n))(1-C^a(n+1))}\right].
    \end{align}
    It remains to calculate the different chains. To do this we approach the problem as a transfer matrix problem, where $\mathcal{T}$ is the transfer matrix. Using the definition of $\mathcal{T}$ in \eqnref{calTdef}, the decoration decomposition of a two-site brick in \eqnref{fullbrickdecomposition} and definitions of $M$ and $M^{\pm}$ in \eqnref{Transfer_matrix_contraction_2},
    $\mathcal{T}$ and each of $M$, $M_+$ and $M_-$ have the following properties,
    \begin{equation}
    \raisebox{-0.4\totalheight}{\includegraphics[height = 1.05cm]{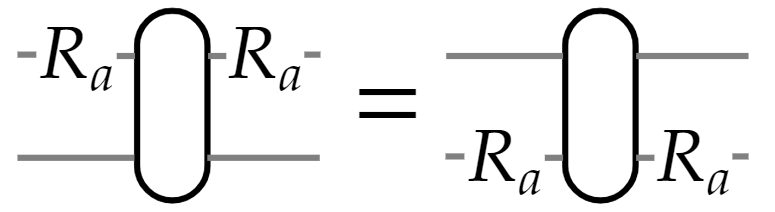}}, \quad \raisebox{-0.4\totalheight}{\includegraphics[height = 0.6cm]{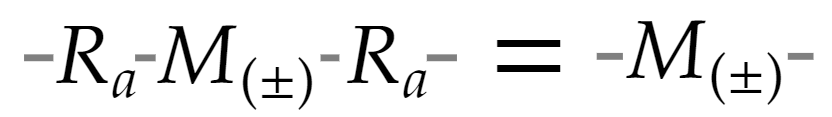}}.
    \end{equation}
    Then, using the definition of the labelled super legs in \eqnref{superleg}, the chain $C^a(n)$ simplifies to the expression below,
    \begin{equation}
    C^a(n) = \raisebox{-0.4\totalheight}{\includegraphics[height = 1.3cm]{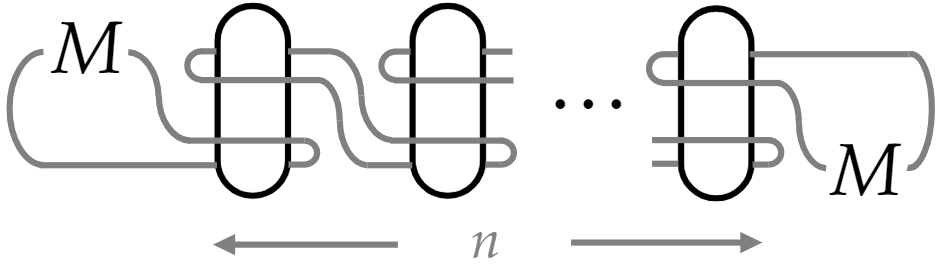}}.
    \end{equation}
    Notice that the dependence on the label $a$ vanished. This is true of all chains. Algebraically, these chains (now without leg labels) are equivalently given by
    \begin{align}
        C(n) = \bra{M}\mathcal{T}^n\ket{M}&, \quad C_+(n)= \bra{M_+}\mathcal{T}^n\ket{M},\nonumber\\
        C_+(n) = \bra{M}\mathcal{T}^n\ket{M_-}&, \quad C_{+-}(n)= \bra{M_+}\mathcal{T}^n\ket{M_-},
    \end{align}
    where these angles braces reflect the trace inner product for tensors with input and output super legs $l=(1,\overline{1},2,\overline{2})$.
    \begin{equation}
    \bra{B}\ket{A} = \Tr(B^\dagger A)/q^4 = \raisebox{-0.4\totalheight}{\includegraphics[height = 1cm]{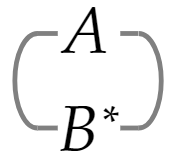}}.
    \end{equation}
    Equipped with this, we can now write, for the contribution due to overlapping OTOC diagrams, the following
    \begin{equation}
    \int dV \mathcal{D}^{(2,1)}_{\textrm{O}}(x=2,z=0)=\frac{2g^2}{q^2}f(\varepsilon),
    \end{equation}
    where $f(\varepsilon)$ is given by
    \begin{equation}\label{Chain sum}
    f(\varepsilon)= \sum_{n\geq0}\left[\frac{C_{+-}(n)}{1-C(n+1)} + \frac{C_+(n)C_-(n)}{(1-C(n))(1-C(n+1))}\right],
    \end{equation}
    where the chains are implicitly dependent on $\varepsilon$. To evaluate this sum, we must understand the space that the transfer matrix acts on. Each of the legs $1$, $\overline{1}$, $2$ and $\overline{2}$ may be either undecorated or carry a $Z$ decoration. This means that the state our state space is dimension $2^4$ and $\mathcal{T}$ is a $16\times16$ matrix. Because all the $M_{(\pm)}$ are even in the number of $Z$ decorations, and $\mathcal{T}$ preserves decoration parity, we are able to reduce the state space to those states with an even number of $Z$ decorations only, i.e., $8$ states. It will be useful to use the basis below,
    \begin{equation}
    \raisebox{-0.4\totalheight}{\includegraphics[height = 2.3cm]{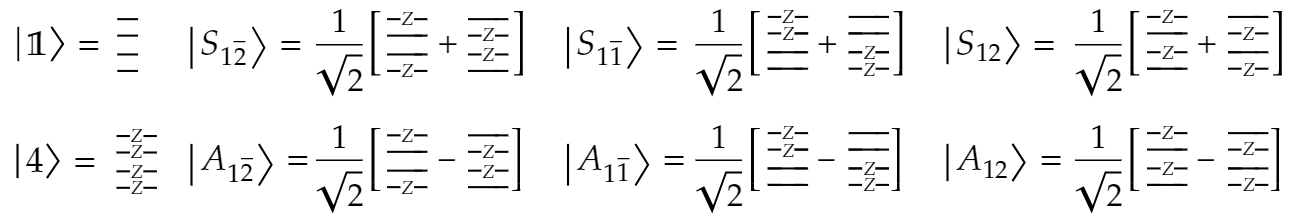}}.
    \end{equation}
    Let $\mathcal{S}$ be the space spanning $\{\ket{\mathbb{1}},\ket{S_{1\overline{1}}},\ket{S_{1\overline{2}}},\ket{S_{12}},\ket{4}\}$ and $\mathcal{A}$ be the space spanning $\{\ket{A_{1\overline{1}}},\ket{A_{1\overline{2}}},\ket{A_{12}}\}$ One can easily check that $\ket{M}$, $\ket{M_+}$ and $\ket{M_-}$ are all in $\mathcal{S}$. Explicitly, with $u(\varepsilon)=\sin(\varepsilon)^2$
    \begin{align}
        &\ket{M} = (u^4+(1-u)^4)\ket{\mathbb{1}} + 2u^2(1-u)^2\ket{4} - (1+g)\frac{g}{\sqrt{2}}\left(\ket{S_{1\overline{1}}}+\ket{S_{1\overline{2}}}\right) + \frac{g^2}{\sqrt{2}}\ket{S_{12}},\\
        &\ket{M_+} = -\frac{g(1+g)}{\sqrt{2}}\left(\ket{S_{12}} -\ket{S_{1\overline{2}}}\right) + \frac{g^2}{2}\left(\sqrt{2}\ket{S_{1\overline{2}}} -\ket{\mathbb{1}} -\ket{4}\right),\\
        &\ket{M_-} = -\frac{g(1+g)}{\sqrt{2}}\left(\ket{S_{12}} -\ket{S_{1\overline{1}}}\right) +\frac{g^2}{2}\left(\sqrt{2}\ket{S_{1\overline{1}}} -\ket{\mathbb{1}} -\ket{4}\right).
    \end{align}
    $\mathcal{T}$ is block diagonal in the subspaces $\mathcal{S}$ and $\mathcal{A}$, and since all $\ket{M_{(\pm)}}$ lie in $\mathcal{S}$, we are able to restrict our considerations to this five dimensional space only. In this restricted space with basis order $(\mathbb{1},4,S_{1,\overline{1}},S_{1,\overline{2}},S_{1,2})$, $\mathcal{T}$ is given by the product $\mathcal{T} = \tilde{T}_-\tilde{U}\tilde{T}_+$, where
    \begin{equation}
    \tilde{U}=\left(
    \begin{array}{ccccc}
    (1-u)^2 & 0 & 0 & 0 & 0 \\
    0 & u^2 & 0 & 0 & 0 \\
    0 & 0 & -\frac{g}{2} & 0 & 0 \\
    0 & 0 & 0 & -\frac{g}{2} & 0 \\
    0 & 0 & 0 & 0 & \frac{g}{2} \\
    \end{array}
    \right)
    \end{equation}
    \begin{equation}
    \tilde{T}_-=\left(
    \begin{array}{ccccc}
    (1-u)^2 & u^2 & -\frac{g}{\sqrt{2}} & 0 & 0 \\
    u^2 & (1-u)^2 & -\frac{g}{\sqrt{2}} & 0 & 0 \\
    -\frac{g}{\sqrt{2}} & -\frac{g}{\sqrt{2}} & g+1 & 0 & 0 \\
    0 & 0 & 0 & g+1 & -g \\
    0 & 0 & 0 & -g & g+1 \\
    \end{array}
    \right)
    \end{equation}
    \begin{equation}
    \tilde{T}_+=\left(
    \begin{array}{ccccc}
    (1-u)^2 & u^2 & 0 & -\frac{g}{\sqrt{2}} & 0 \\
    u^2 & (1-u)^2 & 0 & -\frac{g}{\sqrt{2}} & 0 \\
    0 & 0 & g+1 & 0 & -g \\
    -\frac{g}{\sqrt{2}} & -\frac{g}{\sqrt{2}} & 0 & g+1 & 0 \\
    0 & 0 & -g & 0 & g+1 \\
    \end{array}
    \right)
    \end{equation}
     Four of the eigenvalues of $\mathcal{T}$ are bounded by $\abs{\lambda_{1,2,3,4}}\leq \abs{g(\varepsilon)}(1 - 2\abs{g(\varepsilon)})/2$. The largest eigenvalue (for all $\varepsilon$) is bounded by $\abs{\lambda_5}\leq (1-\abs{g(\varepsilon)})^3$.

     In \secref{Minimal model} we showed that $\langle v_B(\varepsilon)\rangle$ must have the symmetries $\varepsilon\to-\varepsilon$ and $\varepsilon\to \pi/2+\varepsilon$. All analytically computed $\order{1/q^2}$ contributions (the $(a,b)=(4,4)$ contribution and the touching OTOC contribution) are found to respect this symmetry, as does the $\order{1}$ contribution from $\Omega$. We therefore conclude that $f(\varepsilon)$ must also have this symmetry. Additionally, we know that $f(\varepsilon)$ is a function of $\sin(\varepsilon)^2$ only (the transfer matrix and the initial and final vectors in our transfer matrix calculation are functions of $\sin(\varepsilon)^2$ only). Together with the symmetry requirements, this means that $f(\varepsilon)$ is in fact a function of $s(\varepsilon)=\sin(\varepsilon)^2\cos(\varepsilon)^2$. 
     
     We evaluate the sum in \eqnref{Chain sum} analytically for small $\varepsilon$, finding $f(\varepsilon)\approx \varepsilon^2/7$. Knowing that $f$ is a function of $s(\varepsilon)$ only, we use this to factorise $f(\varepsilon)=\frac{1}{7}s(\varepsilon)w(s(\varepsilon))$, where $w(\varepsilon)$ is a function that approaches $1$ as $\varepsilon\to0$. The transfer matrix dramatically simplifies at the point $s(\varepsilon) = 1/4$. Where all but one eigenvalue is zero. We find analytically that $f(\varepsilon)\propto (1-4s)^2$ around this point. This suggests a further factorisation $f(\varepsilon)=\frac{1}{7}s(\varepsilon)(1-4s)^2p(s(\varepsilon))$. We find a very good quadratic polynomial approximation for $p(\varepsilon)=1+as+bs^2$, with $a=6.8$ and $b=16.1$.
    \begin{equation}
        f(\varepsilon) = \frac{1}{7}s(\varepsilon)(1 - 4s(\varepsilon))^2(1+as(\varepsilon)+bs(\varepsilon)^2)
    \end{equation}
    \begin{figure}[H]
    \centering
    \raisebox{-0.4\totalheight}{\includegraphics[height = 5.5cm]{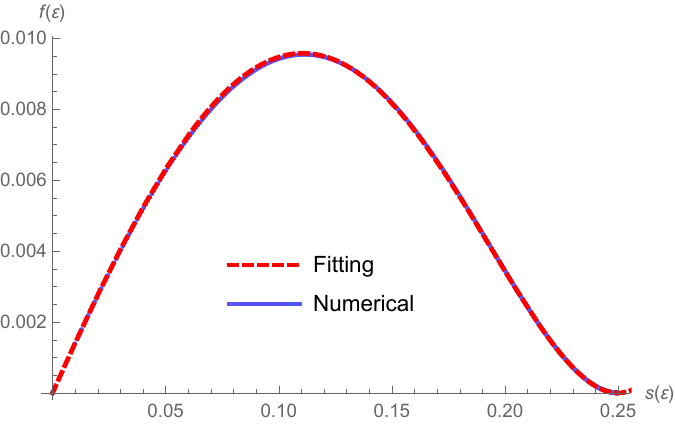}}.
    \caption{The contribution to $v_B$ from processes with overlapping OTOCs is given by $2g(\varepsilon)^2f(\varepsilon)/q^2$, $f(\varepsilon)$ is by numerically and plotted in blue. A fitting function is plotted in red.}\label{f_plot}
    \end{figure}
    We also find numerically that $\int dV \mathcal{D}^{2,1}_{\textrm{O}}(x=2,T)$ decays exponentially quickly, with a decay rate bounded below by $\gamma_{\textrm{O}}(\varepsilon)\geq 6s(\varepsilon)$.
\subsection{Touching OTOC diagrams}
We will now sum up all of the touching OTOC diagrams. These are given in \eqnref{touchingOTOCs} with $T=2\tau$ so that the two OTOC have the same length. We name this contribution $\mathcal{D}^{(2,1)}_{\textrm{Touch}}(x=2,T)$

We are only interested in the $z\to i0^+$ limit, this is given by
\begin{equation}
        \int dV \mathcal{D}^{(2,1)}_{\textrm{Touch}}(k=0,z=0) = g^2\int dV \sum_{t=1}^\infty g \lambda(t)^2 + \order{1/q^3}.
\end{equation}
The factor $g\lambda(t)^2$ is the contribution from the diagrams \eqnref{touchingOTOCs}. The Haar average of a product of OTOCs given in \eqnref{Haar_average_OTOC_product} is deployed again to say that
\begin{equation}
    \int dV \lambda(t)^2 = \frac{2}{q^2} \bra{M}\ket{M}^{t-1}.
\end{equation}
This gives 
\begin{equation}
        \int dV\sum_{t=1}^\infty g^3 \lambda(t)^2=\frac{2g^3}{q^2}\frac{1}{1-\bra{M}\ket{M}}=-\frac{2g^2}{q^2}\nu(\varepsilon),
\end{equation}
where $\bra{M}\ket{M}=1-8s(1-2s)(1-s(1-2s))$,  $s(\varepsilon)=\sin(\varepsilon)^2\cos(\varepsilon)^2$ and $\nu(\varepsilon)=-g/(1-\bra{M}\ket{M})$. Touching OTOC diagrams decay exponentially with a decay rate $\gamma_{\textrm{Touch}}(\varepsilon)=\log(\bra{M}\ket{M}^{-1})\geq 8s(\varepsilon)$.

\subsection{Summary}
Recalling \eqnref{x<0}, the contributions from $x<0$ are $\order{1/q^4}$ or smaller. Piecing together the $\order{1/q^2}$ corrections, we find $\int dV \mathcal{D}^{(2,1)}(k=0,z=0)$ is given by
\begin{equation}
    \int dV \mathcal{D}^{(2,1)}(k=0,z=0) = \frac{2g^2}{q^2}(f(\varepsilon) - \nu(\varepsilon)) + \order{1/q^3}.
\end{equation}

\section{\label{Remaining (a,b)}Remaining \texorpdfstring{$(a,b)$}{Lg}}
\subsection{\texorpdfstring{$(1,2)$}{Lg}}
    \begin{equation}
        \mathcal{D}^{1,2}_\Gamma(x\geq0,T)= g^2\times \left[ \ \begin{matrix}
        \vspace{-0.15cm}
        \\
    \textcolor{gray}{\textrm{site } 0}\\ 
    \textcolor{gray}{\textrm{site } 1}\\
    \\ \\ \\ \\
    \end{matrix} \ \ 
    \begin{matrix}
    \ \ \vdots\\
    \ \ \hspace{-5pt}\bra{\phi_+}\\ 
    q\bra{-}\\
    \ \ \vdots\\
    q\bra{-}\\
    \ \ \bra{-}\\
    \ \ \vdots
    \end{matrix}\quad
    \fbox{ $\begin{matrix}
    \\
    \\
    \\
    \hspace{-5pt}\Gamma\\
    \\
    \\
    \\
    \end{matrix}$}\quad 
    \begin{matrix}
    \hspace{-20pt}\vdots\\
    \hspace{-20pt}\ket{+}\\
    \hspace{-20pt}\ket{+}\\
    \hspace{-20pt}\vdots\\
    \hspace{-20pt}\ket{+}\\
    \ket{\phi_-(T)}\\
    \hspace{-20pt}\vdots
    \end{matrix} \ \ 
    \begin{matrix}
    \\ \\ \\ \\
    \vspace{-1.01cm}
    \\
    \textcolor{gray}{\textrm{site } x}\\ 
    \textcolor{gray}{\textrm{site } x+1}
    \end{matrix}\right]
    \end{equation}
    Here, we can simply use \eqnref{dec phi overlaps} to see that both sites $0$ and $x+1$ contribute a product of non-trivial correlation functions plus terms of size $1/q^2$. The Haar average of this is $\order{1/q^4}$.
    \begin{equation}
        \int dV \mathcal{D}^{1,2}_\Gamma(x\geq 0,T)=\order{1/q^4}.
    \end{equation}

\subsection{\texorpdfstring{$(3,3)$}{Lg}}
\begin{itemize}
    \item $x=0$:
    \begin{equation}
        \mathcal{D}^{3,3}_\Gamma(x=0,T)=g^2\times \left[ \ \begin{matrix}
        \textcolor{gray}{\textrm{site } 0}\\ 
        \textcolor{gray}{\textrm{site } 1}
        \end{matrix} \ \ 
        \begin{matrix}
        \bra{+}Z^{\otimes2}\\ 
        \bra{-}Z^{\otimes2}
        \end{matrix}
        \ \Gamma \ 
        \begin{matrix}
        Z(T)^{\otimes2}\ket{+}\\ 
        Z(T)^{\otimes2}\ket{-}
        \end{matrix}\right]
        \left(\prod_{r<0}\bra{+}\Gamma^r 
        \ket{+}\right) \left(\prod_{r>1}\bra{-}\Gamma^r\ket{-}\right)
    \end{equation}
    In this case, both contours on site $0$ and on site $1$ are non-trivially decorated, contributing a total of four non-trivial correlators. The Haar average is then $\order{1/q^4}$.
    \item $x\geq 1$:
    In this case, each site $0$ and $x+1$ contribute a product of non-trivial correlators and therefore, as above, are $\order{1/q^4}$ or smaller.
\end{itemize}
All together
\begin{equation}
        \int dV \mathcal{D}^{1,2}_\Gamma(x\geq 0,T)=\order{1/q^4}.
\end{equation}

\subsection{\texorpdfstring{$(3,4)$}{Lg} and \texorpdfstring{$(4,3)$}{Lg}}
\begin{itemize}
    \item $x=0$:
    \begin{equation}
        \mathcal{D}^{3,4}_\Gamma(0,T)=-ihg\times \left[ \ \begin{matrix}
        \textcolor{gray}{\textrm{site } 0}\\ 
        \textcolor{gray}{\textrm{site } 1}
        \end{matrix} \ \ 
        \begin{matrix}
        \bra{+}Z^{\otimes 2}\Gamma^0 K(T)\ket{+}\\ 
        \bra{-}Z^{\otimes 2}\Gamma^1 K(T)\ket{-}
        \end{matrix}\right]\left(\prod_{r<0}\bra{+}\Gamma^r 
        \ket{+}\right) \left(\prod_{r>1}\bra{-}\Gamma^r\ket{-}\right)
    \end{equation}
    Both site $0$ and site $1$ contribute two non-trivial correlators. The Haar average of this term is $\order{1/q^4}$ or smaller.
    \item $x\geq 1$:
    Each of the sites $0$ and $x+1$ contribute a product of two non-trivial correlators. Each of the sites $1$ and $x$ contribute a non-trivial correlator. If $x=1$, then this is only five non-trivial correlators, otherwise it is six. The Haar average is then $\order{1/q^5}$
 \end{itemize}  
 All together,
 \begin{equation}
        \int dV \mathcal{D}^{4,4}_\Gamma(x\geq 0,T)=\order{1/q^4}.
 \end{equation}

\subsection{\texorpdfstring{$(a,b)$}{Lg}, where either \texorpdfstring{$a\in {1,2}$}{Lg} and \texorpdfstring{$b\in{3,4}$}{Lg} or the converse}
    
    In these case, \eqnref{dec phi overlaps} is again enough to show that $\int dV \mathcal{D}^{a,b}_\Gamma(x,T)=\order{1/q^3}$.

\section{Model variation: Independent scramblers \texorpdfstring{$V_{x,t}$}{Lg}}\label{independent scramblers}
We can carry out a similar calculation for a version of the model with independently distributed scramblers $V_{x,t}$ for each site and at each layer of unitaries $U_t$ (analogous to the Floquet layer in the Floquet model), i.e., we break both spatial and temporal translation symmetry. We start with, for large $t$, the fact $\sum_x x \rho(x,t) = v_B t$ to write
\begin{align}\label{alt v def}
    v_B &= \lim_{t\to\infty} \sum_x x \left[\rho(x,t) - \rho(x,t-1)\right] = \lim_{t\to\infty}\lim_{k\to 0} \sum_x \mathrm{i}\partial_k e^{-\mathrm{i}kx} \left[\rho(x,t) - \rho(x,t-1)\right] \nonumber\\
    &= \lim_{t\to\infty}\lim_{k\to 0} \mathrm{i}\partial_k \left[\rho(k,t) - \rho(k,t-1)\right].
\end{align}
Using the definition of $\rho(k,n)$ we write the following
\begin{align}\label{rho}
    \rho(k,n) &= \left( W^k|U_0(P+Q)U_1\cdots U_n|W^k\right) = \left(W^k|U_0|W^k\right)\rho(k,n-1) + H(k,n)\nonumber\\
    &= (1+\Omega(k))\rho(k,n-1) + H(k,n)
\end{align}
where $H(k,n)$ is given by
\begin{align}\label{H def}
    H(k,n) &= \left(W^k|U_0QU_1\cdots U_n|W^k\right)\nonumber \\
    &= \left(W^k|U_0QU_1\cdots U_{n-1} (P+Q) U_n|W^k\right)\nonumber \\
    &= (1+\Omega(k))H(k,n-1) + \sigma(k,n),
\end{align}
where $\sigma(k,n)$ is as previously defined, but with the independently distributed scramblers.
\begin{equation}
    \sigma(k,n)=\left(W^k|L_0QU_1\cdots U_{n-1}QL_n|W^k\right).
\end{equation}
Using \eqnref{H def} inductively and the fact that $H(k,n=0)=0$ we find 
\begin{equation}
    \rho(k,n)-\rho(k,n-1) = \Omega(k) \rho(k,n-1) + \sum_{m=1}^n (1+\Omega(k))^{n-m-1}\sigma(k,m).
\end{equation}
Using this in \eqnref{alt v def}, we find
\begin{equation}
    v_B = \lim_{n\to\infty}\lim_{k\to 0} \mathrm{i}\partial_k \left[\Omega(k) \rho(k,n-1) + \sum_{m=1}^n (1+\Omega(k))^{n-m-1}\sigma(k,m)\right].
\end{equation}
Then writing $\Omega(k)=v_0(1-e^{-\mathrm{i}k})$ where $v_0 = \abs{g(\varepsilon)}$ is the $q\to\infty$ butterfly velocity, and noticing that $\lim_{k\to 0}\partial_k \left(\Omega(k)\rho(k,n)\right)= \mathrm{i} v_0$, we find
\begin{equation}\label{delta v}
    \delta v_B \equiv v_B - v_0 = \lim_{n\to\infty}\lim_{k\to 0} \mathrm{i}\partial_k \sum_{m=1}^n (1+\Omega(k))^{n-m-1}\sigma(k,m).
\end{equation}
Analogous to the definition in \eqnref{Ddef} for the Floquet model, we define the quantity,
\begin{equation}
    \mathcal{D}(x\to y,n) = q^{y-x}\bra{F^x}L_0QU_1\cdots U_{n-1}QL_n\ket{F^y}
\end{equation}
then $\sigma(k,n)$ is given by
\begin{align}
    \sigma(k,n) = \frac{1}{L(1-q^{-2})}\sum_{x,y}&e^{\mathrm{i}k(x-y)}\bigg[\mathcal{D}(x\to y,n) - \mathcal{D}(x\to y-1,n)\nonumber \\
    &\left. -\frac{1}{q^2}\left(\mathcal{D}(x-1\to y,n)-\mathcal{D}(x-1\to y-1,n)\right)\right]
\end{align}
The circuit averaged $\mathcal{D}(x\to y,n)$ is translationally invariant $\langle \mathcal{D}(x\to y,n) \rangle = \langle \mathcal{D}(y-x ,n) \rangle$ where $\mathcal{D}(x,n)=\mathcal{D}(0\to x,n)$. The circuit average of $\sigma(k,n)$ is given by
\begin{equation}
    \langle \sigma(k,n) \rangle = \eta(k)(1-e^{-\mathrm{i}k})\sum_x e^{-\mathrm{i}kx}\langle \mathcal{D}(x,n) \rangle = \eta(k)(1-e^{-\mathrm{i}k})\langle \mathcal{D}(k,n) \rangle.
\end{equation}
The circuit averaging of Eq. \ref {delta v} is given by
\begin{align}
    \langle \delta v_B \rangle &= \lim_{n\to\infty}\lim_{k\to 0} \mathrm{i}\partial_k \sum_{m=1}^n (1+\Omega(k))^{n-m-1}\eta(k)(1-e^{-\mathrm{i}k})\langle \mathcal{D}(k,m) \rangle\nonumber\\ 
    &= -\sum_{m=1}^\infty \langle \mathcal{D}(k=0,m) \rangle = -\langle \mathcal{D}(k=0,z=0) \rangle,
\end{align}
This is same expression for the circuit averaged correction to $v_B$ that we found using the memory matrix formalism in \eqnref{butterfly_velocity_correction}. We calculate $\langle \mathcal{D}(k=0,z=0) \rangle$ in the same way as in \secref{MMF calc}. However, the two $\order{1/q^2}$ contributions found for the Floquet model (i.e., the so-called $(2,1)$ and $(4,4)$ terms) relied on correlations between different sites and on the discrete time translation symmetry. With independently distributed scramblers, such terms are smaller than $\order{1/q^2}$. This gives the result $\langle \delta v_B \rangle  = \order{1/q^3}$ which leads to
\begin{equation}
    \langle v_B \rangle = \frac{1-\cos(4\varepsilon)}{4} + \order{1/q^3}.
\end{equation}

We can employ the same calculation above for the variant of the model with spatial translation symmetry but no time translation symmetry (i.e., independently random scramblers between time-steps) in order to find $\langle \delta v_B \rangle = \delta v_S(\varepsilon)  + \order{1/q^3}$ where $\delta v_S(\varepsilon)$ is as given in \eqnref{v contributions}.
\begin{equation}
    \langle v_B \rangle = \frac{1-\cos(4\varepsilon)}{4} + \delta v_S(\varepsilon) + \order{1/q^3}, \quad \delta v_S(\varepsilon) =\frac{1}{q^2}\frac{1+5s-4s^2}{1-s-3s^2}.
\end{equation}

Comparing these result to the result of \eqnref{butterfly_velocity_correction} allows us to see that the $\order{1/q^2}$ corrections only arise when there is spatial translation symmetry. Moreover, this enables us to identify the $\delta v_F$ contribution (see \eqnref{v contributions}) with the presence of time translation symmetry.

\subsection{The \texorpdfstring{$\varepsilon\to 0$}{Lg} limit in the absence of time-translation symmetry}\label{independent small epsilon}
In \secref{(4,4)}, we identified all the contributions to $\sigma(x=0,t)$ that contained the product of two non-trivial correlators and used theorem \ref{two-correlators} to evaluate these contributions to $\order{1/q^2}$. We were unable to extract the correct $\varepsilon\to0$ behaviour of these contributions due to the appearance of $\varepsilon$ in the denominator (after summing over the time $t$). Luckily, for the model variant with independently random scramblers at each time-step, we are able to accurately sum over all `two-correlator' contributions, after circuit averaging. While this is ignores correlations between Floquet layers, it demonstrated the failure of naively expanding the memory matrix in $1/q$ before taking the $\varepsilon\to0$ limit and the resolution by summing over all contributions. 

We identify the contributions that have the form of two non-trivial correlators in the same way as done in \secref{(4,4)}. We repeat this below for ease of reading.
\begin{equation}\label{F14}
        \mathcal{D}^{4,4}(x=0,T)= - h^2 \sum_{i,j\in \{1,2\}}
        \begin{matrix}
        \bra{+}Z_i\\
        \bra{-}Z_j
        \end{matrix}
        \ \fbox{ $\begin{matrix}
        \vspace{-5mm}\\
        \hspace{-5pt}1\vspace{1mm}
        \end{matrix}$} \ \fbox{ $\begin{matrix}
        \vspace{-5mm}\\
        \hspace{-5pt}2\vspace{1mm}
        \end{matrix}$}\ \cdots \ \fbox{ $\begin{matrix}
        \vspace{-5mm}\\
        \hspace{-5pt}T-1\vspace{1mm}
        \end{matrix}$}\ \begin{matrix}
        Z(T)_i\ket{+}\\
        Z(T)_j\ket{-}
        \end{matrix}\ +\ \cdots,
    \end{equation}
    where, so as to ensure that only sites $0$ and $1$ contribute non-trivial correlators, we have sandwiched each layer by $\bra{+}$ ($\bra{-}$) and $\ket{+}$ ($\ket{-}$) for sites $x<0$ ($x>1$) to leave the two site operators $\mathcal{T}(t)$ introduced in \eqnref{calTdef} and represented above in \eqnref{F14} by the rectangle blocks. The second ellipsis represents terms that contain at least three non-trivial correlators. Using the same argument as was used in \secref{(4,4)}, we see that the $(i,j)=(1,1)$ and $(2,2)$ choices give identical contributions and the $(1,2)$ and $(2,1)$ contributions are the complex conjugate of the $(i,i)$ contributions, \eqnref{104}. Therefore, we are free to study only the $(i,j)=(1,1)$ case. We insert projectors ensure that the $2,\overline{1}$ contour of site $0$ and the $2,\overline{2}$ contour of site $1$ remain undecorated (so as to ensure that there are only two non-trivial correlators). The contribution is given below,
    \begin{equation}
        \raisebox{-0.45\totalheight}{\includegraphics[height=3.2cm]{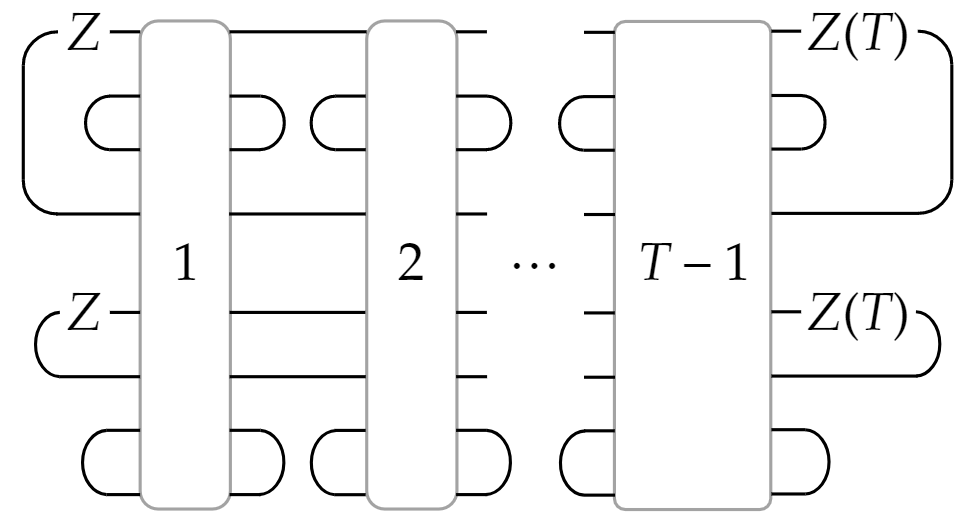}},
    \end{equation}
    where each closed loops is associated with a normalising factor of $1/q$ that is suppressed in the diagram. The contracted single time-step unitaries are now just operators on four indices. We denote the contribution to $\mathcal{D}^{4,4}(x=0,T)$ that includes only two non-trivial contours as $\mathcal{D}_{2}^{4,4}(x=0,T)$. Separating out the scrambling part of the unitaries we find
    \begin{equation}
    \mathcal{D}_{2}^{4,4}(x=0,T)= -\frac{2h^2}{q^2} \
        \raisebox{-0.45\totalheight}{\includegraphics[height=2cm]{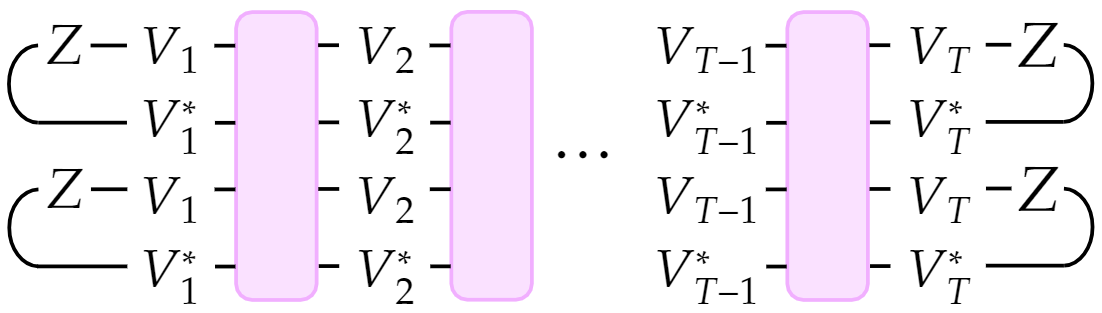}} + \textrm{c.c},
        \label{44-independent scramblers}
    \end{equation}
    where each of the pink bricks is given by the following contraction of $\mathcal{T}$.
    \begin{equation}
        Y = \raisebox{-0.45\totalheight}{\includegraphics[height=2.5cm]{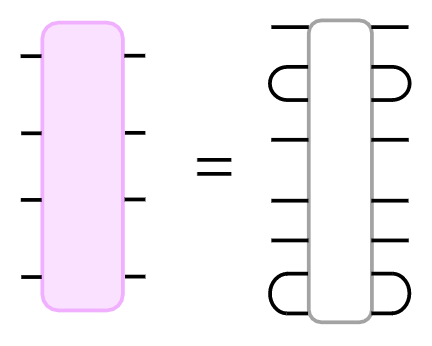}}\times \frac{1}{q^2}.
    \end{equation}
    Circuit averaging this tensor contraction is easy as each unitary only appears twice. We use the following result for the Haar average of two $V$'s and two $V^\dagger$'s. 
    \begin{equation}
        \int dV \raisebox{-0.5\totalheight}{\includegraphics[height=2cm]{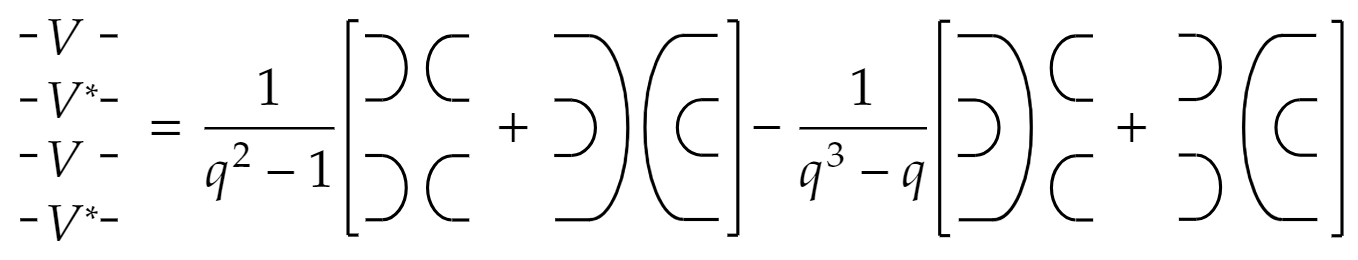}}=K,
    \end{equation}
    where $K$ is a projector. Using the definitions of $\ket{+}$, $\ket{-}$, $\ket{0}$ and $\ket{\perp}$, $K$ is given algebraically as
    \begin{equation}
        K \equiv \frac{\ket{-}\bra{\perp}}{1-q^{-2}}+\frac{\ket{+}\bra{0}}{1-q^{-2}}=\frac{\ket{\perp}\bra{-}}{1-q^{-2}}+\frac{\ket{0}\bra{+}}{1-q^{-2}}.
    \end{equation}
    Using this, we write
    \begin{align}
        \langle \mathcal{D}_{2}^{4,4}(x=0,T)\rangle &=-2h^2\bra{-}Z^{\otimes 2}(KYK)^{T-1}Z^{\otimes 2}\ket{-} + \textrm{c.c}\nonumber\\
        &=-\frac{2h^2}{q^2}\bra{0}(KYK)^{T-1}\ket{0}+ \textrm{c.c}.
    \end{align}
    The matrix $KYK$ is a $2\times2$ matrix acting on the space of wirings $\textrm{Span}\{\ket{+},\ket{-}\}$. Choosing the orthormal basis $\ket{S,A}=\frac{1}{\sqrt{2}}(\ket{+}\pm\ket{-})$. $KYK$ is given by
    \begin{equation}
    KYK = \frac{\xi}{2}\left(\begin{array}{cc}
    1 & 1 \\
    1 & 1 \end{array}\right) + \frac{1+g}{2} \left(\begin{array}{cc}
    1 & -1 \\
    -1 & 1 \end{array}\right) + \frac{1+g+ih}{q}\left(\begin{array}{cc}
    1 & 0 \\
    0 & -1 \end{array}\right).
    \end{equation}
    By diagonalising this, we find
    \begin{equation}
    \sum_{n=0}^\infty\frac{1}{q^2}\bra{0}(KYK)^n\ket{0}= \frac{2}{1+8q^2\varepsilon^2+\order{q^2\varepsilon^3,q\varepsilon^2}}.
    \end{equation}
    Using $h(\varepsilon)\approx \varepsilon$ for small $\varepsilon$, $\langle \mathcal{D}_{2}^{4,4}(x=0,z=0)\rangle $ is then given by
    \begin{equation}\label{small epsilon 44}
        \langle \mathcal{D}_{2}^{4,4}(x=0,z=0)\rangle =-\frac{8\varepsilon^2}{1+8q^2\varepsilon^2+\order{q^2\varepsilon^3,q\varepsilon^2}}.
    \end{equation}
    Importantly, this contribution, which keeps account of all orders in $q$, vanishes as $\varepsilon\to 0$, unlike the $\order{1/q^2}$ contribution (naively) calculated in \secref{(4,4)}, which (incorrectly) predicts $\lim_{\varepsilon\to 0}\langle \mathcal{D}^{4,4}(x=0,z=0) \rangle=-1/q^2$. However, this is still far from a complete account of contributions to $\langle \mathcal{D}(k=0,z=0)\rangle $ (for instance, contributions involving more than two correlators), but it does demonstrate that the apparent pathological behaviour as $\varepsilon\to 0$ is the result of a naive (and incorrect at small $\varepsilon$) counting of the $\order{1/q^2}$ contributions. Notice that for $q\varepsilon \gg 1$, the small $\varepsilon$ results of \secref{(4,4)} (which predict $\langle \mathcal{D}^{4,4}(x=0,z=0) \rangle\approx-1/q^2$) are in agreement with the analysis presented here.
    
\end{document}